\documentclass[fleqn,usenatbib]{mnras}

\usepackage{newtxtext,newtxmath}
\usepackage{subfigure}

\usepackage[T1]{fontenc}

\DeclareRobustCommand{\VAN}[3]{#2}
\let\VANthebibliography\thebibliography
\def\thebibliography{\DeclareRobustCommand{\VAN}[3]{##3}\VANthebibliography}

\usepackage{graphicx}	
\usepackage{amsmath}	
\usepackage{hyperref}
\usepackage{relsize}

\newcommand{\Msun}{\mbox{\,M$_\odot$} }

\newcommand{\Teff}{\mbox{\,T$_{\text{eff}}$ }}

\title[A Neural Network for Li-based Ages]{Using Neural Network Models to Estimate Stellar Ages from Lithium Equivalent Widths: An EAGLES Expansion}

\author[G. Weaver et al.]{
G. Weaver,$^{1}$\thanks{E-mail: g.weaver@keele.ac.uk}
R. D. Jeffries$^{1}$ and
R. J. Jackson$^{1}$
\\
$^{1}$Astrophysics Group, Keele University, Keele, ST5 5BG, UK
}

\date{September 2024}

\pubyear{2024}

\begin{document}
\label{firstpage}
\pagerange{\pageref{firstpage}--\pageref{lastpage}}
\maketitle

\begin{abstract}

We present an Artificial Neural Network (ANN) model of photospheric lithium depletion in cool stars ($3000<T_{\rm eff}/{\rm K} < 6500$), producing estimates and probability distributions of age from $^7$Li~6708\AA\ equivalent width (LiEW) and effective temperature data inputs.
The model is trained on the same sample of 6200 stars from 52 open clusters, observed in the Gaia-ESO spectroscopic survey, and used to calibrate the previously published analytical {\sc eagles} model, with ages 2 -- 6000 Myr and $-0.3 <$ [Fe/H] $< 0.2$.
The additional flexibility of the ANN provides some improvements, including better modelling of the ``lithium dip" at ages $< 50$ Myr and $T_{\rm eff}\sim 3500$\,K, and of the intrinsic dispersion in LiEW at all ages.
Poor age discrimination is still an issue at ages > 1 Gyr, confirming that additional modelling flexibility is not sufficient to fully represent the LiEW - age - \Teff relationship, and suggesting the involvement of further astrophysical parameters. 
Expansion to include such parameters -- rotation, accretion, and surface gravity -- is discussed, and the use of an ANN means these can be more easily included in future iterations, alongside more flexible functional forms for the LiEW dispersion. Our methods and ANN model are provided in an updated version 2.0 of the {\sc eagles} software.
\end{abstract}





\section{Introduction}\label{Introduction}

Stellar age determination is an important facet of the exploration of our Galaxy's history.
Stars contain tracers of the chemistry and dynamics of the environment in which they formed and evolved. Investigating stellar formation and evolution itself requires knowledge of stellar ages, which can be (indirectly) inferred from a host of techniques, with an effectiveness that varies with age, mass, and metallicity (discussed in detail by \citealt{soderblom2010} and \citealt{soderblom2014}).

For low-mass stars, stellar evolution models are limited in their age estimation capabilities.
On the main sequence, the observables of luminosity, temperature and gravity change very slowly, and asteroseismological observations are not widely available for most stars \citep{Epstein2014}. Despite the more rapid evolution of pre-main sequence (PMS) stars, significant theoretical uncertainties are still present in the form of differing treatments of convection and the influence of magnetic fields, rotation and starspots \citep[e.g.,][]{Hillenbrand2009,tognelli2011,feiden2013,soderblom2014,Somers2020}. Observational uncertainties associated with accretion, variability, binarity and extinction are also frequently large enough to confound attempts at age estimation in PMS stars \citep{Hartmann2001,  Preibisch2012, Jeffries2017b}.
As a result, empirical techniques have been developed for low-mass stars that rely on age-dependent properties that can be measured and calibrated using stars of better-known age, such as those in stellar clusters. Examples include rotation \citep[``gyrochronology",][]{barnes2003, barnes2016, bouma23} the frequency of accretion discs \citep{haisch2001, Fedele2010} and the depletion of photospheric lithium.
Lithium has been used as an age indicator for decades \citep[reviewed in][]{jeffries2014, barrado2016, randichmagrini2021}, since the photospheric abundance of $^7$Li (hereafter referred to as simply Li) follows an age-dependent trajectory in low-mass stars.
Within young stars, the initial Li is destroyed in $p$, $\alpha$ reactions once interior temperatures exceed $\sim 2.5 \times 10^{6}$K \citep{Pinsonneault1997}.
Any star with an outer mixing region extending deep enough to reach that temperature will deplete its photospheric Li.
In contracting PMS stars, Li-burning begins at a mass-dependent age \citep{Ventura1998}.
The lowest mass stars (M $<$ 0.35\Msun) are fully convective, and as such, once Li-burning commences, their photospheric Li depletes very rapidly.
Above $0.35M_\odot$, radiative cores develop and expand outwards on a timescale that decreases with increasing mass.
If the convection zone base falls below the Li-burning temperature, then Li-depleted material is mixed to the photosphere much more slowly. This causes a complex mass- and \Teff-dependent relationship between Li depletion and age at the end of the PMS phase. This depletion is primarily of use as an age indicator for stars in the PMS and ZAMS phases when depletion is most rapid, but can still be of use on the main sequence, since Li depletion, caused by one or more competing slow mixing mechanisms (slower than convection), continues -- for example, the Li abundance in the Sun's photosphere and solar-type stars in older clusters is orders of magnitude lower than solar-type ZAMS stars \citep{sestito2005}, and alternative observables may change too slowly to be useful. 

Theoretical models can reproduce the general trends of Li depletion with age and mass in PMS and main sequence phases, but they are extremely sensitive to model ingredients like opacities, mixing lengths, convective overshooting and exactly what is responsible for additional, non-convective mixing; different models can make wildly different predictions of Li abundance at a given age. In addition, current models fail to explain the dispersion in Li that is often observed at a given age and \Teff in open clusters like the Pleiades, M35 or M67 \citep{Pace2012, Bouvier2018}, which is quite likely related to rotation and starspots, or possibly even the presence of exoplanets \citep{Bouvier2008, Jeffries2021}; or the systematic disagreement in the Li depletion of older clusters, even with similar ages and metallicities \citep{randich2010}.
These systematic uncertainties in models point towards the empirical calibration of Li-based ages as a way forward.

To fully exploit Li abundance as an empirical age indicator requires both a thorough assessment of its age dependence and how this depends on mass (or its observational proxy, \Teff), and also of how any dispersion at a given age and \Teff depends on age or on other parameters, since this causes uncertainties in any Li-based age estimate. 
Initial empirical modelling by \citet{baffles} used calibrating data from 10 stellar clusters with ages of 5-5000 Myrs. This was used to generate a model from which Bayesian age estimates were made, constructed from a fitted LiEW - B-V colour - \Teff relationship.
This approach was improved significantly by \citet{Rob2023}, with a homogeneous calibrating dataset of Li~{\sc i}~6708\AA\ line equivalent widths (hereafter, LiEW) and \Teff from the {\it Gaia}-ESO survey \citep[GES,][]{Gilmore2022, randich2022}, for a much larger sample of 6200 stars in 52 open clusters with ages 2--6000 Myr, and with more clusters populating the younger and older ends of this range. This larger dataset calibrated an analytic empirical model ({\sc eagles} - Estimating AGes from Lithium Equivalent widthS) for both LiEW and its dispersion as a function of age between 5 Myrs and 10 Gyrs and $3000< $\Teff/K $<6500$. {\sc eagles} provides ($\log$) age estimates for individual (PMS) stars with precisions as small as 0.1-0.2 dex, or even more precisely for coeval groups or clusters. Nevertheless there are indications that the analytic function used to predict LiEW may be overly simplistic, the LiEW dispersion is incompletely modelled and relies on an {\it ad hoc} inflation of the dispersion in certain age and \Teff ranges and the model performs poorly for older ($>1$\,Gyr) stars and clusters, where significant systematic errors still remain.

In this work we develop and improve upon the {\sc eagles} methodology, using the same calibration dataset but using a data-driven artificial neural network (ANN) to discover the relationship between Li, age, and \Teff in PMS and main sequence stars.
The increase in data availability from large scale optical surveys such as the GES, SDSS-V \citep{kollmeier2019} and the forthcoming WEAVE \citep{Dalton2012} and 4MOST \citep{dejong2019} surveys at medium and high resolution will provide enormous stellar samples with Li abundance, \Teff and other parameters. In exploiting these data
it is essential to optimise age estimation codes, explore data-driven architectures and methods of modelling, and to investigate further parameters besides age and \Teff which may influence Li depletion.

The data utilised by this model, alongside the construction, training and prediction methods are described in Section \ref{sec:training}. Section \ref{sec:results} details the results of the model and compares its performance to literature ages and the analytical {\sc eagles} model for both single stars and clusters. This is done using residuals to the training set and to validation datasets from clusters and moving groups not used in the training. In Section \ref{sec:discussion}, these results are discussed and analysed in comparison with the analytical model and literature ages, alongside a discussion of the areas for development of the model. A summary is given in Section \ref{sec:conclusions} and an implementation of the model prediction code, written in {\sc python} code\footnote{\url{https://github.com/robdjeff/eagles}} is described in Appendix~\ref{code}.

\section{Training Data and Model}\label{sec:training}

Machine learning tools offer a powerful alternative in modelling unknown or uncertain complex, non-linear relationships between stellar parameters and observational features.
There are several machine learning methods used across a widening range of astronomical data types, but this work focuses on a supervised learning method -- the feed forward ANN.
ANNs use layers of neurons, with weighted inputs (and a bias) from the outputs of previous layers and with their outputs connected to the inputs of subsequent layers. The weights and biases are perturbed and tuned between training epochs in order to minimise a ``loss function" between the modelled and observed features in a training dataset and hence converge on a best-fit relationship.
When compared with more traditional fitting techniques (like those used in \citealt{Rob2023}), there are several potential advantages in using ANNs. 
They are purely data-driven, do not require any analytic constraints on the form of the model, and can provide as much complexity as required or allowed by regularisation techniques designed to prevent over-fitting.
The lack of constraints can lead to the discovery of model features that were not considered or expected when choosing an analytic fitting function and possibly giving insight into physical processes -- in this case the physics of Li depletion.
Alongside this, a major motivation for the use of an ANN in this work is the potential to expand the model, and future work will aim to encompass further age-dependent variables such as rotation, activity, and photometric features.

\subsection{The Dataset}\label{dataset}

The training data in this paper are taken from \citet{jackson2022} and \citet{Rob2023}, consisting of LiEW, an observational error in LiEW and a \Teff for kinematically-selected members of open clusters obtained from the FLAMES-GIRAFFE (17000 < R < 19200) and FLAMES-UVES (R = 47000) spectrographs \citep{randich2022,Pasquini2002}, on the 8m UT2-\textit{Kueyen} telescope of the Very Large Telescope, alongside data from the GESiDR6 Parameter Catalogue \citep{hourihane2023}.

These are the same data used to fit an empirical, analytic model by \citet{Rob2023} and further details about the data selection, filtering, and measurement of LiEW can be found there. In brief:
\begin{itemize}
    \item The \Teff, LiEW, observed error in LiEW, and adopted cluster ages used in this work are the same as in {\sc eagles}.
\item The clusters have metallicities of -0.3 < [Fe/H] < 0.2, taken from \cite{randich2022} (with the exception of NGC 6649, where the value is taken from \citealt{Rob2023}), and age values in the range 2 Myr to 6 Gyr.
\item 6200 target members from 52 open clusters for the final training dataset were selected in the same manner as \citet{Rob2023}, having a kinematic cluster membership probability $> 0.9$ and 2900\,K$ < \Teff <$ 6600\,K. 
Probable giants (with \Teff > 4000K and $\log$ $g$ < 3.4), and poor data (LiEW < -300 m\AA, LiEW > 800 m\AA, or $\Delta$LiEW > 300 m\AA) were rejected.
\item An additional 1503 field stars, with membership probability < 0.01 were added to the dataset, with a given age of 5 Gyr.
\end{itemize}

Literature/training ages for the data were the geometric mean values of ages drawn from isochrone fitting in \citealt{Dias2021}, a compilation of isochronal ages cited in \citealt{jackson2022} (with updates from \citealt{Franciosini2022} for some clusters) and the ages cited in \citealt{randich2022}. We adopted these ages to provide a consistent comparison with the original {\sc eagles} model of \citealt{Rob2023}. There is of course the likelihood of some systematic error and (stellar evolutionary) model-dependence in this absolute age scale. The adopted young cluster ages ($\lesssim$ 150 Myr) are broadly consistent with their lithium depletion boundary ages, as discussed in \citet{Rob2023}.

\subsection{Model Architecture}\label{subsec:architecture}

\begin{figure}
    \centering
    \includegraphics[width=\linewidth]{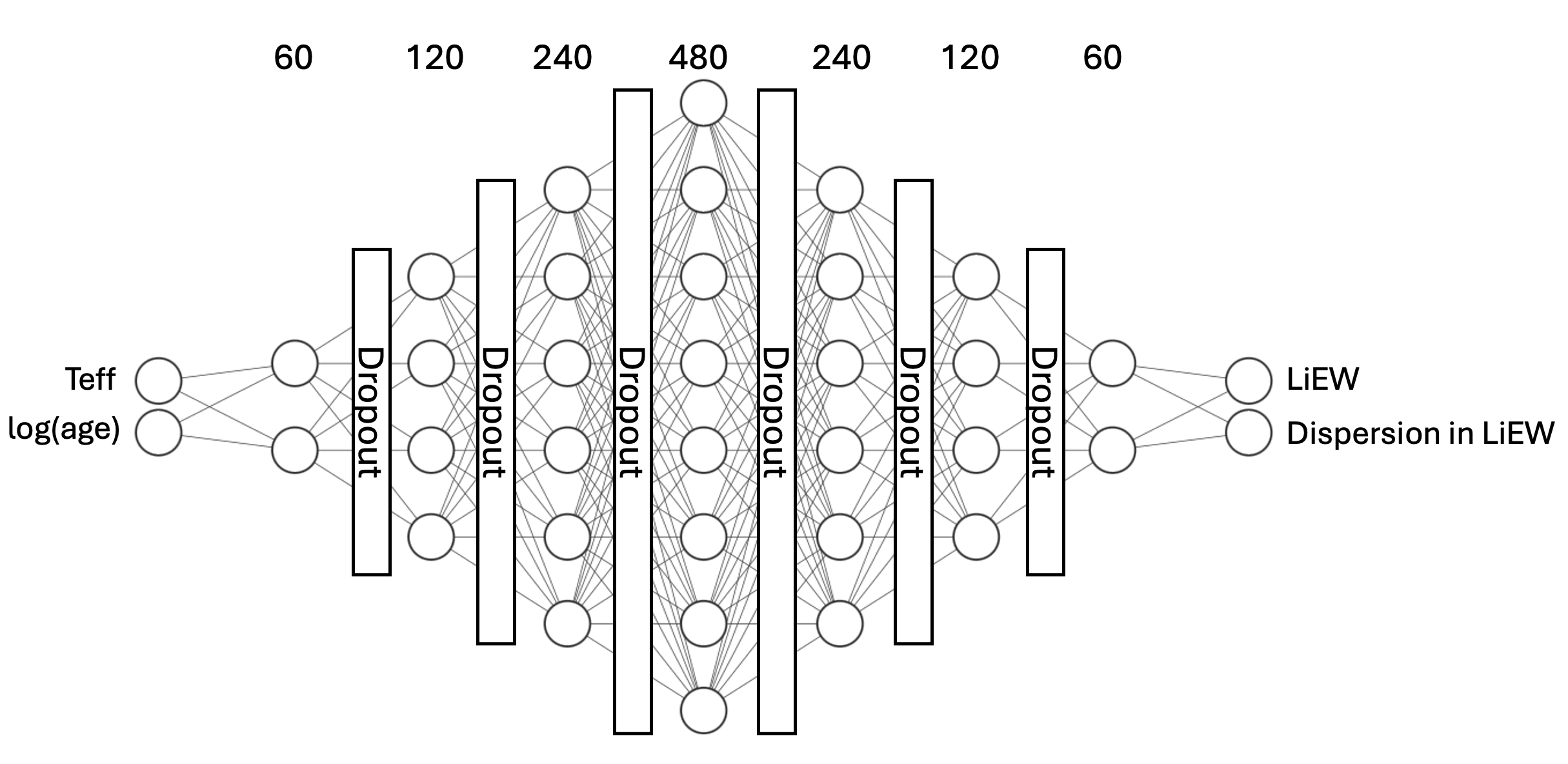}
    \caption{The structure of the ANN prediction model. The number of neurons in each hidden layer (i.e. not the input or output layers) is scaled down by a factor of 30 for this visualisation. The input layer takes scaled values of \Teff and $\log$(age), and the output layer produces similarly scaled values of LiEW and the dispersion in LiEW.
    Neurons are connected to the subsequent and previous layers' neurons, with dropout layers between each hidden layer dropping 20\% of the connections from the previous layer's neurons, as described in \S\ref{subsec:architecture}.}
    \label{fig:ann}
\end{figure}

The model is a simple ANN consisting of two input features, \Teff and $\log$(mean literature age), and two output features, LiEW and intrinsic dispersion in LiEW ($\sigma$LiEW, a dispersion beyond that explained by observational uncertainties), as shown in Fig. \ref{fig:ann}.
It was generated with the {\sc keras} application programming interface for {\sc tensorflow} \citep{keras,tensorflow}, using the sequential model function to create the ANN.
The model itself is structured with 9 dense layers, with the input and output layers each containing 2 neurons, and the output layer constrained by a ``ReLU" activation function.
In between these two are 7 hidden layers, of $n =$ 60, 120, 240, 480, 240, 120 and 60 neurons respectively, each with ``leaky ReLU" activation functions applied to avoid gradient explosion issues \citep{maas2013}.
After each hidden layer, there are ``dropout layers", set to drop 20 per cent of inputs from the previous layer. This provides a regularisation to help avoid overfitting, and also allows use of ``Monte Carlo dropout" error estimation.
This makes use of the dropout layers during prediction, so that the output from the model varies on each iteration. Over the course of many iterations (2000 in the final model), this is a means of assessing the uncertainty in the model predictions (see \S\ref{predictions}).

The compiled model used the {\sc adam} optimiser \citep{Kingma2014}, and a custom loss function.  
This takes in the predicted and observed values for LiEW from the model inputs and outputs, alongside the observational error and predicted $\sigma$LiEW, and calculates a negative log likelihood (see Appendix \ref{sec:appendixloss}).
This allows the model to make use of the observed error in LiEW in the training data to make a prediction of $\sigma$LiEW in the LiEW - \Teff - $\log$(age) parameter space. $\sigma$LiEW encapsulates the extent to which LiEW is not uniquely determined by age and \Teff; it is essential for accurate estimates of ages and age uncertainties.

\subsection{Training}
\label{training}

From the initial dataset described in \S\ref{dataset}, input features (\Teff and $\log$ (age/yr)) were scaled between 0 and 1 using a simple min-max scaler, alongside similarly scaled values for LiEW and observed error in LiEW.
The model hyperparameters were selected using a grid search, and the final model training took place over 3000 epochs, with early stopping procedures in place to prevent overfitting.
These procedures use a validation dataset made up of a randomly-selected 15 per cent of the shuffled training data. Training was halted when the rolling average loss from the previous 20 epochs, evaluated for the validation dataset no longer improved.

\subsection{Prediction and Final Model}
\label{predictions}

The trained model outputs predicted values for LiEW and $\sigma$LiEW for a given \Teff and $\log$ (age/yr).
In order to capture the epistemic uncertainty (i.e. those due to uncertainties in the ANN weghts and biases), predictions are run 2000 times with the dropout layers activated randomly.
This technique is equivalent to sampling from the posterior distribution of fits from the model and allows an approximate assessment of the model prediction uncertainties \citep{Gal2015}. 

To save computing time when using the model, a suite of predictions is made for $2900 < \Teff/{\rm K} < 6600$ in steps of 10\,K, and for  $6 \leq \log({\rm age}/{\rm yr}) \leq 10.1$ in steps of 0.005 dex. This generated a grid of mean values for LiEW and $\sigma$LiEW from 2000 predictions at each \Teff, $\log$(age) pair, alongside the standard deviation of those predictions.
This grid of values is used within a Bayesian age estimation scheme, to make a prediction of LiEW, $\sigma$LiEW and the epistemic uncertainties in those predictions, for any particular combination of \Teff and $\log$(age) by linear interpolation.

The rest of the age estimation proceeds almost identically to the original {\sc eagles} code.
Observed \Teff, LiEW and LiEW uncertainties are passed to the model.
The interpolated mean and standard deviation of predicted LiEW and $\sigma$LiEW are found for the observed \Teff across a range of 820 evenly spaced $\log$(age/yr) values between 6 and 10.1.
Each predicted LiEW is compared with the observed LiEW and its observational error to produce a $\log$(likelihood) for each $\log$(age) -- see eqn.~\ref{eq:modelloglikelihood}.
We assume that the observational uncertainty, predicted $\sigma$LiEW and the epistemic standard deviation of the predicted LiEW predictions are all normally distributed and they are added in quadrature.

\begin{equation}\label{eq:modelloglikelihood}
    \mathcal{L} = \prod \frac{1}{\sqrt{2\pi(\sigma_{s}^{2} + 
    \Delta^{2})}}\exp\bigg({{-\frac{({\rm LiEW}_{\rm obs} - {\rm LiEW}_{\rm pred})^2}{2(\sigma_{s}^{2} + 
    \Delta^{2})}}}\bigg)
\end{equation}
Where LIEW$_{\rm obs}$ and $\Delta$ are the observed LiEW and its error bar, whilst $\bar{\mu}$ is the predicted LiEW from the ANN model and $\sigma_{s}$ is the quadrature sum of the predicted intrinsic dispersion of LiEW and the standard deviation in LiEW from the 2000 model iterations.

Using the sum of predicted intrinsic dispersion and standard deviation in predicted LiEW in quadrature allows the model to take account of both observational aleatoric uncertainty of the data and the epistemic uncertainty of the model's fits. 
In almost all areas of the \Teff - Age plane, the epistemic uncertainty represents a value $\lesssim 60\%$ of the intrinsic dispersion.
The areas at which the epistemic uncertainty is highest in relation to the intrinsic dispersion tend to be those in which the training data is most sparse, and in areas with more training data it is $\lesssim 40\%$.
As such, it is important to include both uncertainties, particularly in the regions where calibration data is sparse.
While this aleatoric uncertainty could, in principle, be reduced with more precise LiEW data, and the epistemic uncertainty could be reduced with more training data, reducing the total uncertainty requires pinpointing the source of the dispersion and including those causes in the model.
If necessary, additional \Teff uncertainties can be incorporated by approximating the integrated likelihood over the \Teff probability distribution, by performing a weighted sum over 15 \Teff values at $\pm 3$ error bars from the mean.

The product of this likelihood function and an age prior probability function (either flat in age between 1 Myr and 12.4 Gyrs for single stars, or flat in $\log$ (age/yr) for coeval clusters or groups) then generates a posterior probability distribution of $\log$(age).
From this, the peak of the posterior is taken as the best-fitting (most probable) age, with asymmetric uncertainties estimated as $\pm 34$ per cent of the integrated posterior either side of the peak value or, if no significant peak is found, a 95 per cent upper or lower limit is quoted.
In one difference from the original {\sc eagles} however, we have decreased the lower bound of the likelihood calculations from $\log ({\rm age}/{\rm yr}) = 6.7$ to $\log ({\rm age}/{\rm yr}) = 6$. 
This is due to our belief that some age discrimination is now seen in the ANN model predictions at younger ages, which was suppressed by the analytic model used in {\sc eagles} (see \S~\ref{subsec:advantages}).
Further details of the treatment of upper and lower bounds, alongside the reasoning for the prior selection, can be found in \citet{Rob2023}.

To estimate the age for coeval clusters or groups of stars, the log likelihood distribution for each member star is summed before applying the prior. This then gives a combined posterior probability distribution in $\log$(age), with each star contributing an appropriate weight, which is analysed to provide a most probable age or an age limit as for individual stars.

\section{Results and Comparison with {\sc eagles}}\label{sec:results}

This artificial neural network (ANN) model is designed to replace the traditionally fitted analytical model in the {\sc eagles} code, aiming to improve or at least match the previous iteration in accuracy, whilst avoiding both overfitting or overly-constraining the form of the model.
To that end, we have conducted a series of tests in a similar manner to those performed for the original {\sc eagles} code by \citep{Rob2023}, comparing the performance of both approaches.

\subsection{LiEW And Dispersion Prediction Comparisons}\label{subsec:isochrones}

\begin{figure*}
  \centering
  \begin{minipage}{0.49\textwidth}
    \centering
    \includegraphics[width=\textwidth]{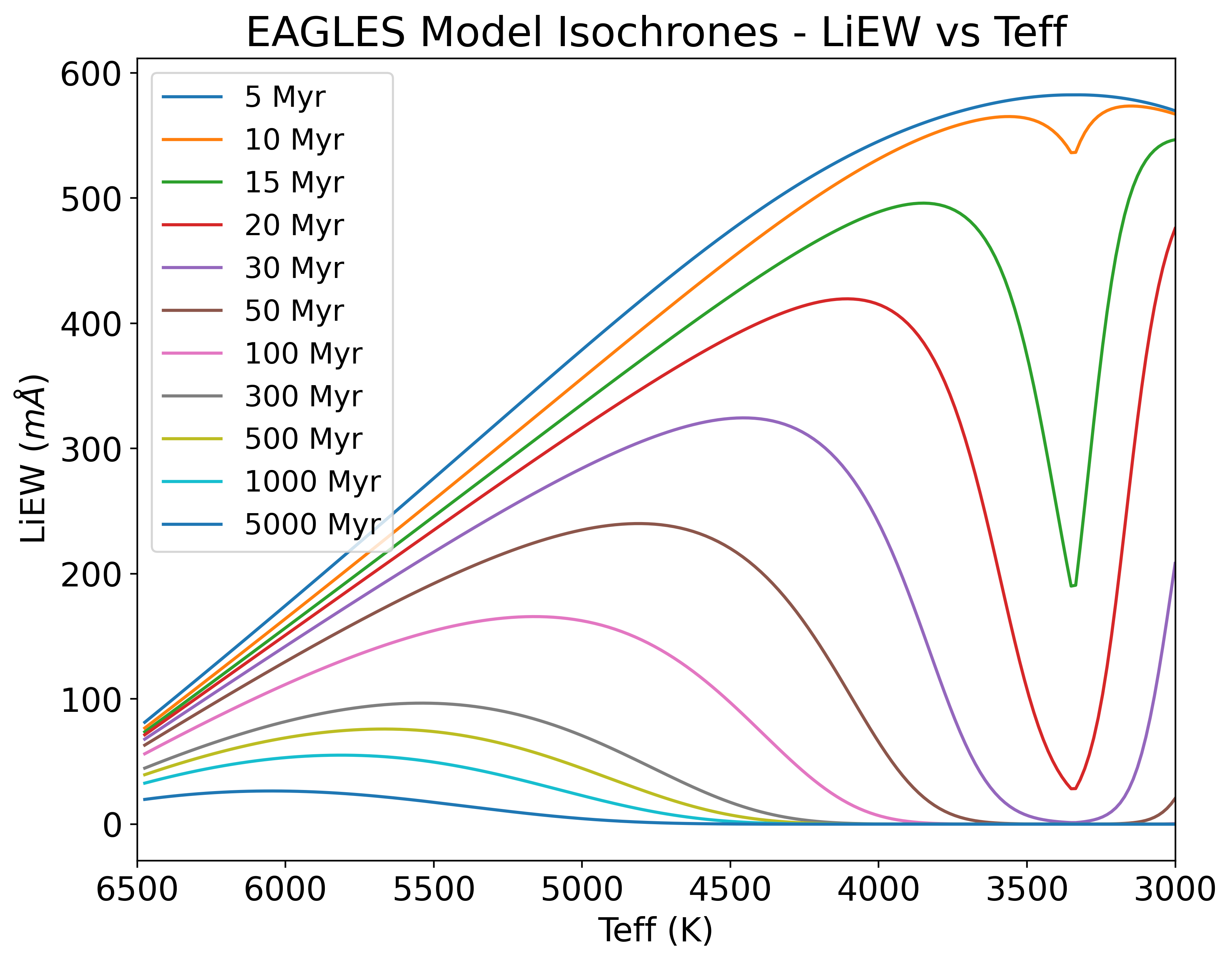}

  \end{minipage}
  \hfill
  \begin{minipage}{0.49\textwidth}
    \centering
    \includegraphics[width=\textwidth]{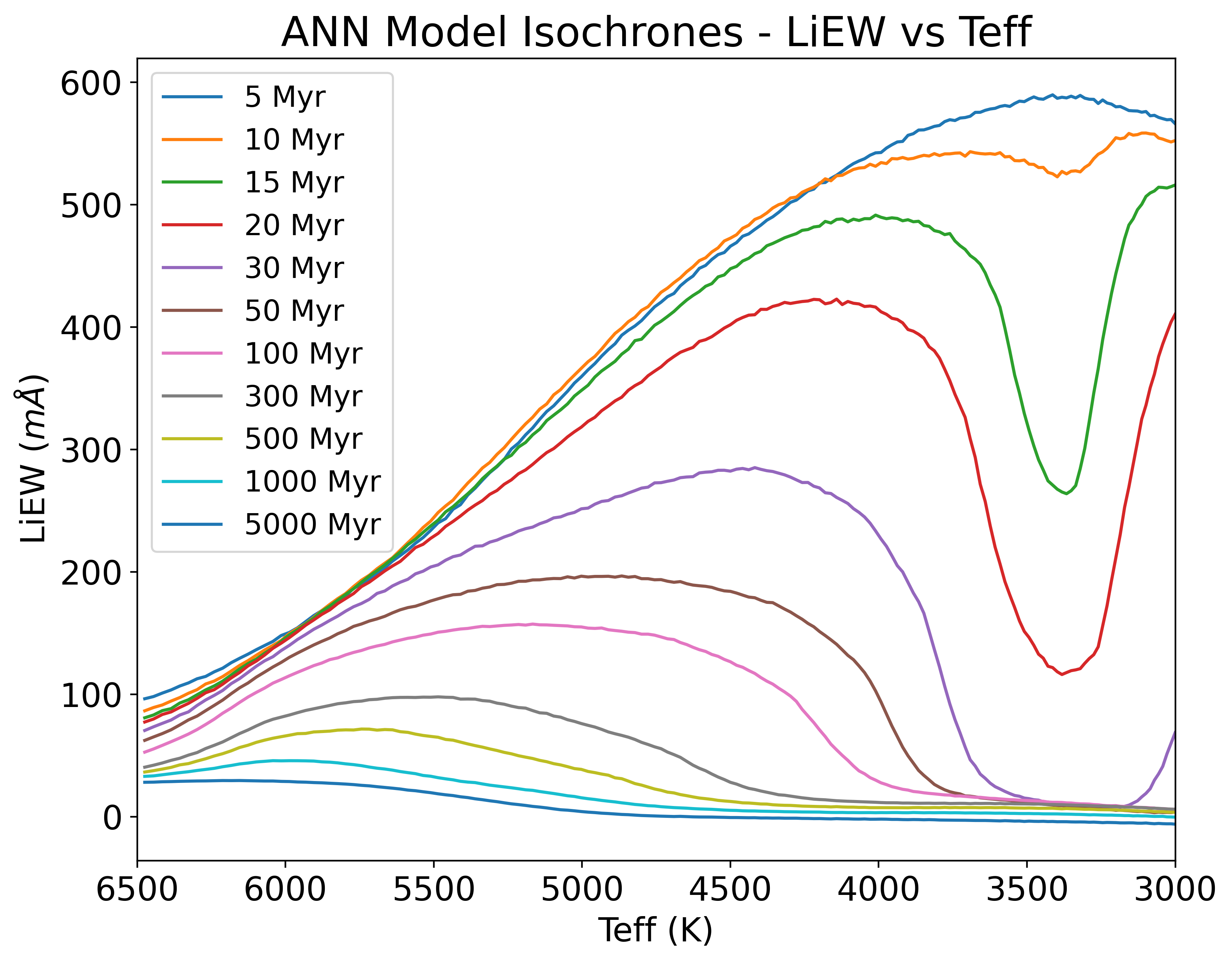}

  \end{minipage}
  
  \caption{Isochrones of predicted LiEW from {\sc eagles} (\textit{left}) and the ANN model (\textit{right}) on the LiEW - \Teff plane.}
  \label{fig:isochrones}
\end{figure*}

\begin{figure}

    \includegraphics[width = \linewidth]{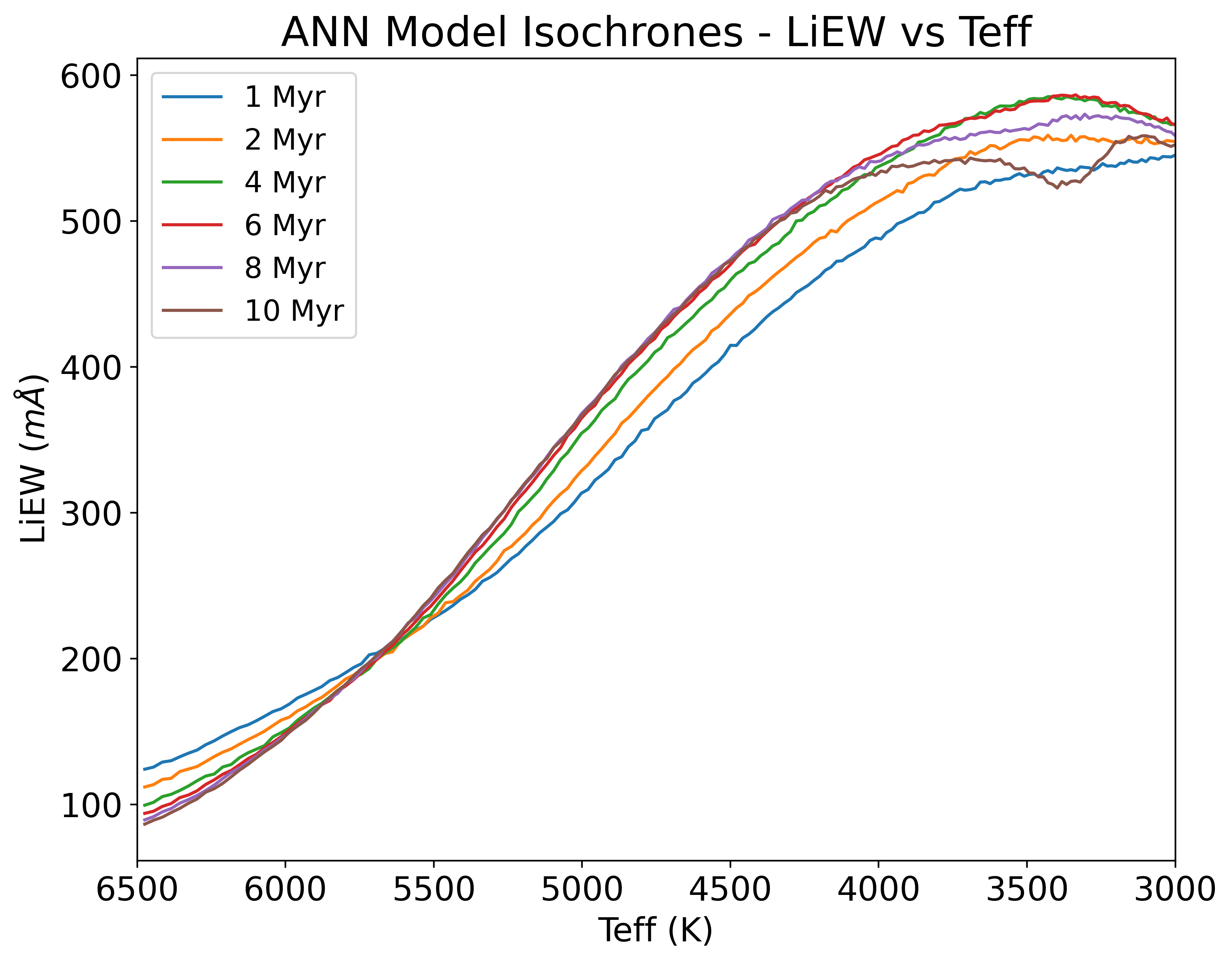}
    \caption{ANN model isochrones at the youngest ages (1-10 Myrs). For \Teff $<5500$\,K the 
    predicted LiEW increases before starting to decrease after about 5\,Myr.}
    \label{fig:overlapisochrones}
\end{figure}

The basic model isochrones from both {\sc eagles} and ANN models, shown in Fig.~\ref{fig:isochrones}, are superficially consistent in shape, but there are noticeable nuances in the predictions in specific areas of the LiEW-\Teff plane. 
The vastly increased number of parameters available in the ANN model, alongside the lack of mathematical constraints, allow the fit to find more subtleties that might be present in the data.
Whilst there is then a risk of overfitting, this was mitigated by the dropout layers and early stopping criteria discussed in \S\ref{training}. In Fig. \ref{fig:isochrones} we see the effects of this in the `rounding' of the `Lithium dip' area between 3200K - 3700K.
Additionally, the ANN predictions show non-monotonic behaviour with age, both in cool stars at the youngest ages where Li has not started, or has only just begun, to deplete and at the oldest ages where the Li has either gone or is almost gone. 
Whilst this behaviour may just be due to noise in the oldest stars, for the youngest stars (with isochrones shown in detail in Fig.~\ref{fig:overlapisochrones}), there is some evidence that the behaviour is resolved and may have a plausible physical explanation (see \S\ref{disc:constraints}). 
The ANN model has therefore picked out behaviour that was not allowed using the monotonically decreasing function mandated in {\sc eagles}.

\begin{figure}
    \centering
    \includegraphics[width = \linewidth]{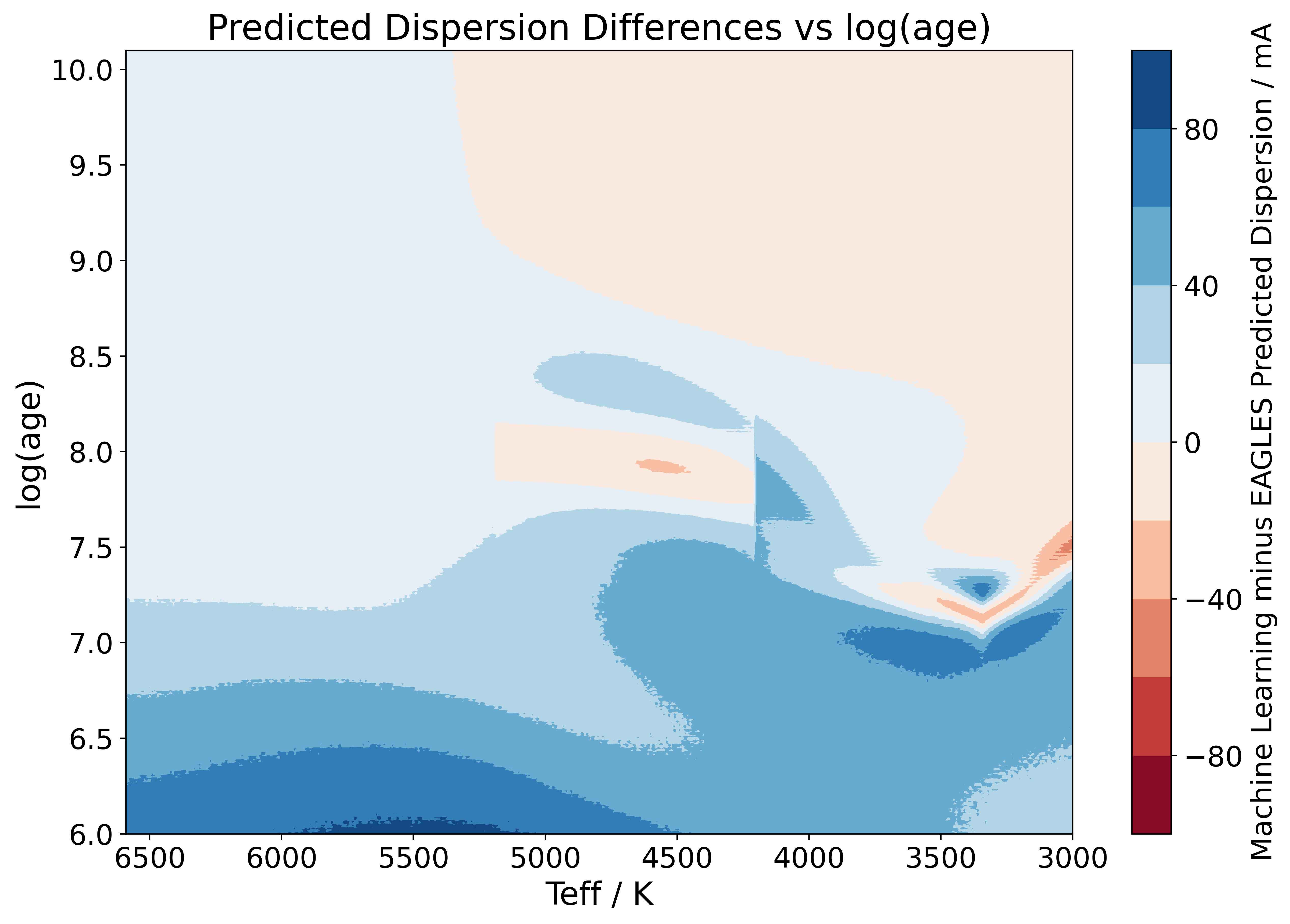}
    \caption{Predicted dispersion in LiEW from the ANN model minus the {\sc eagles} model on the $\log$ (age/yr) - \Teff plane. The ANN dispersion is higher in the majority of the plane. It is noted that, at $\log$ (age/yr) $\gtrsim$ 8.5, the small differences in predicted dispersion shown on this plot do not take into account the low LiEW prediction at these ages. This means that the relative dispersion compared to predicted LiEW is often very high despite the low raw predicted dispersion value. The linear features at 4200K and 5200K, between 7.5 < $\log$ (age/yr) < 8.3, are due to the manually added dispersion in the {\sc eagles} model.}
    \label{fig:dispersiondifferences}
\end{figure}

Figure \ref{fig:dispersiondifferences} shows the {\it difference} in the predicted $\sigma$LiEW (the intrinsic dispersion) between the ANN and {\sc eagles} model.
Below 30 Myr the {\sc eagles} model predicts a lower dispersion at all \Teff values, but at older ages there is a \Teff-dependent behaviour in which the ANN model predicts higher dispersion than {\sc eagles} for warmer stars and lower dispersion for cooler stars.

Across the whole LiEW - \Teff plane, the original {\sc eagles} dispersion model appears not to contain enough complexity to fully describe the intrinsic dispersion of the data.
At ages where the LiEW is changing slowly or not at all, there is no dependence on \Teff in the {\sc eagles} dispersion model.
As a result, at old ages we see underestimated dispersion for stars $\gtrsim$ 5300K and overestimated dispersions below this temperature in the {\sc eagles} model.
The lower dispersion in the {\sc eagles} model for the youngest stars is due to the slowly-decreasing LiEW with a small derivative, and as such the ANN model dispersion is larger for all \Teff values.

In {\sc eagles}, an {\it ad hoc} extra term in the dispersion model was added to describe an increased dispersion between 4200K < \Teff < 5200K that begun at ages $\sim 30$ Myr and disappeared at $\sim 300$ Myr. 
The ANN model has naturally identified this inflated dispersion, although without the sharp artificial \Teff or age boundaries, which are the reasons for the linear feature in Fig.\ref{fig:dispersiondifferences} and the slightly different age profile.

\subsection{Age estimates for individual stars}\label{subsec:individual}

The predictions of LiEW and its intrinsic dispersion are used as described in \S\ref{predictions} to make age estimates (with uncertainties) for stars with measured LiEW and \Teff. To compare these estimates with {\sc eagles}, each model was used to estimate ages for a grid of individual exemplar stars over the full range of the \Teff - LiEW plane, in steps of 10\,K and 5\,m\AA{} respectively, to compare single-star age estimations.

\begin{figure*}
    \centering
    \subfigure{\includegraphics[width=0.24\textwidth]{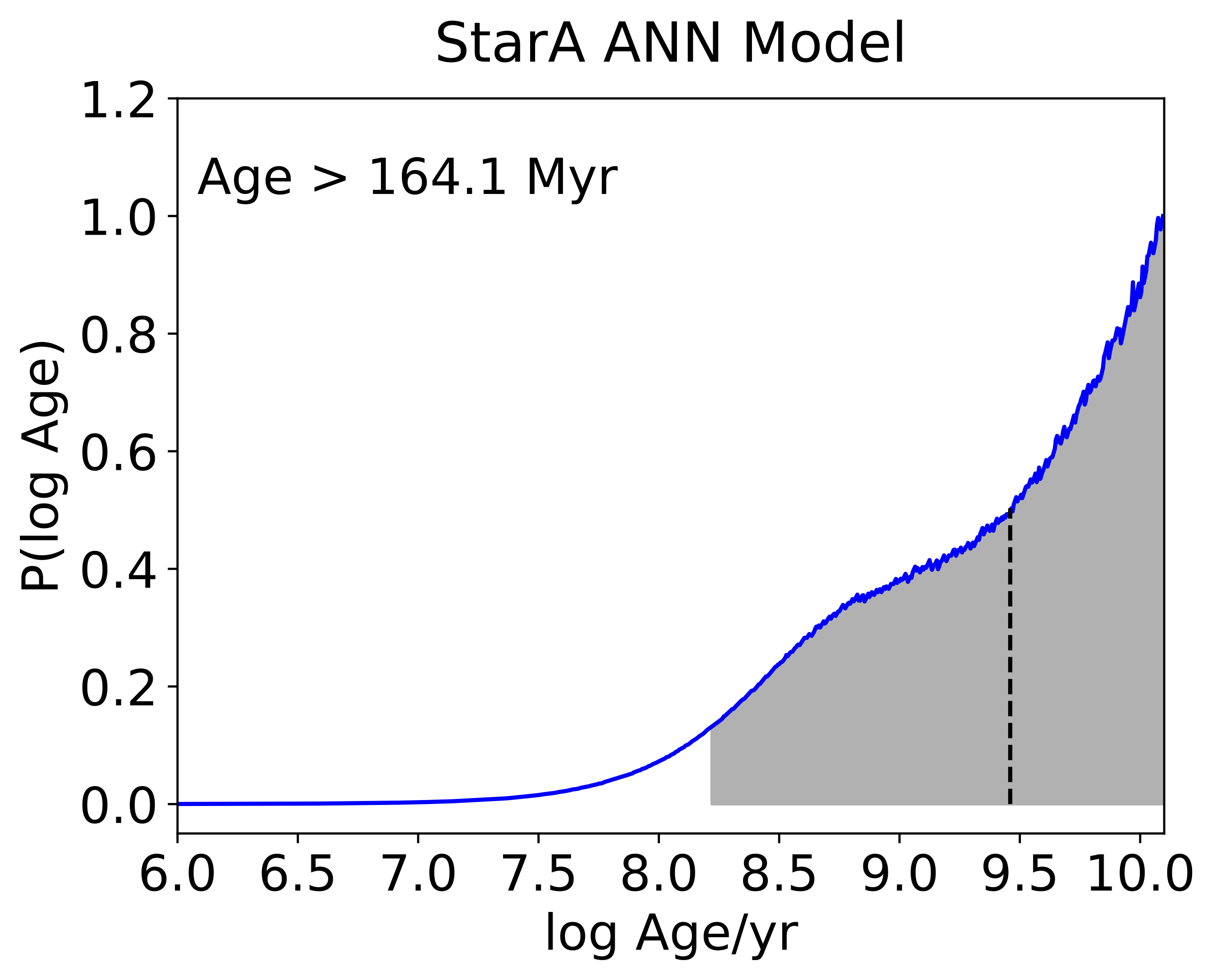}}
    \subfigure{\includegraphics[width=0.24\textwidth]{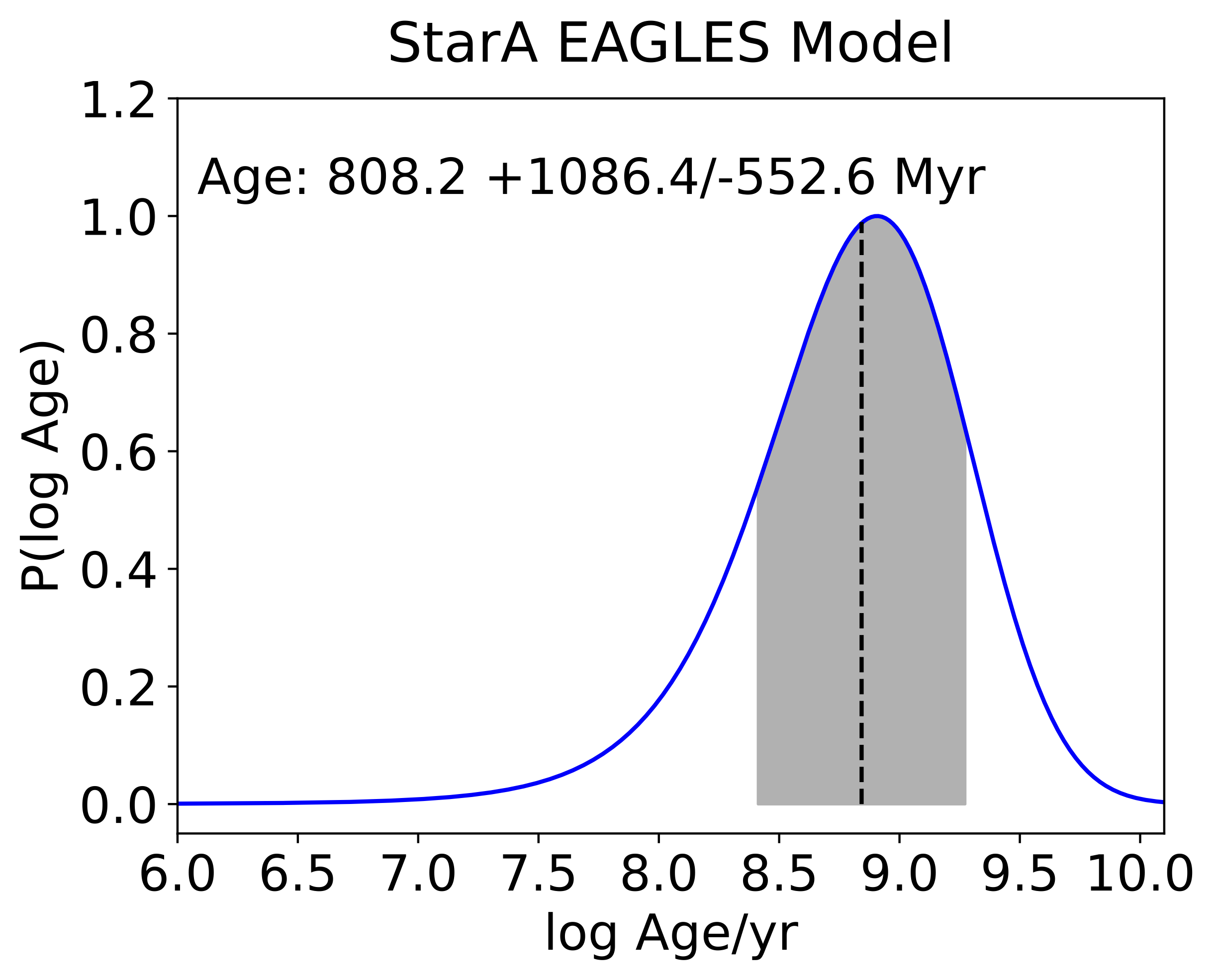}}
    \subfigure{\includegraphics[width=0.24\textwidth]{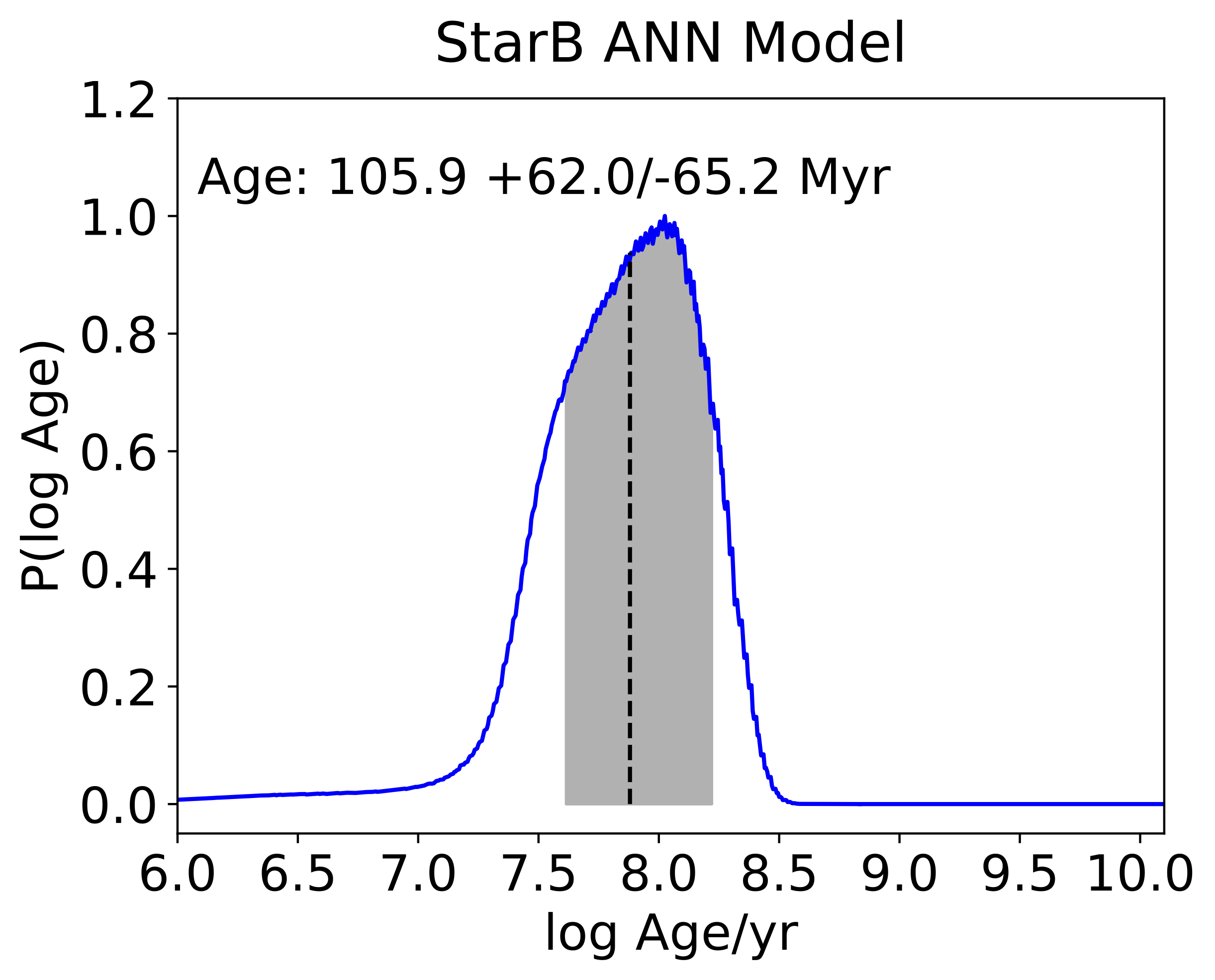}}
    \subfigure{\includegraphics[width=0.24\textwidth]{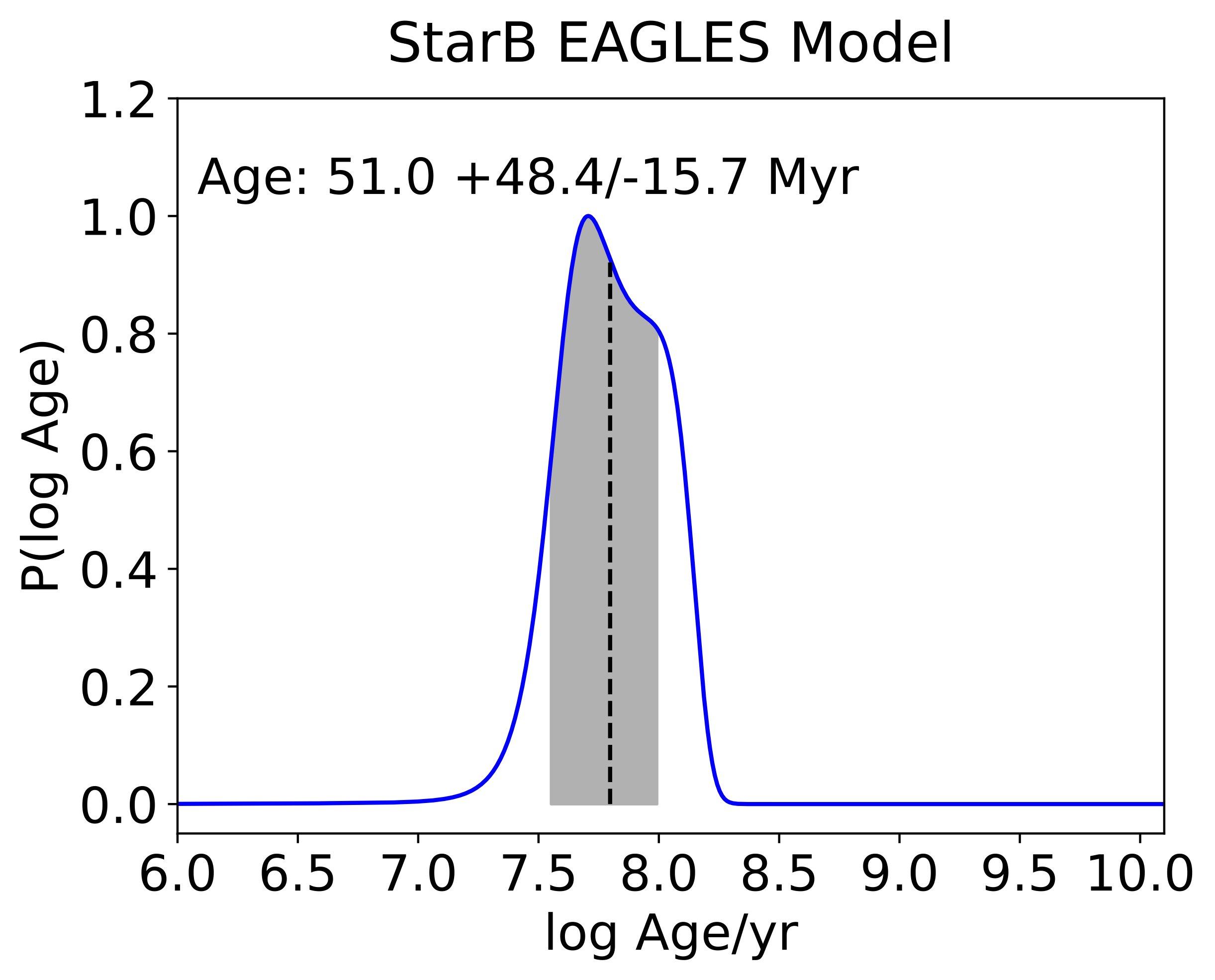}}

    \subfigure{\includegraphics[width=0.49\textwidth]{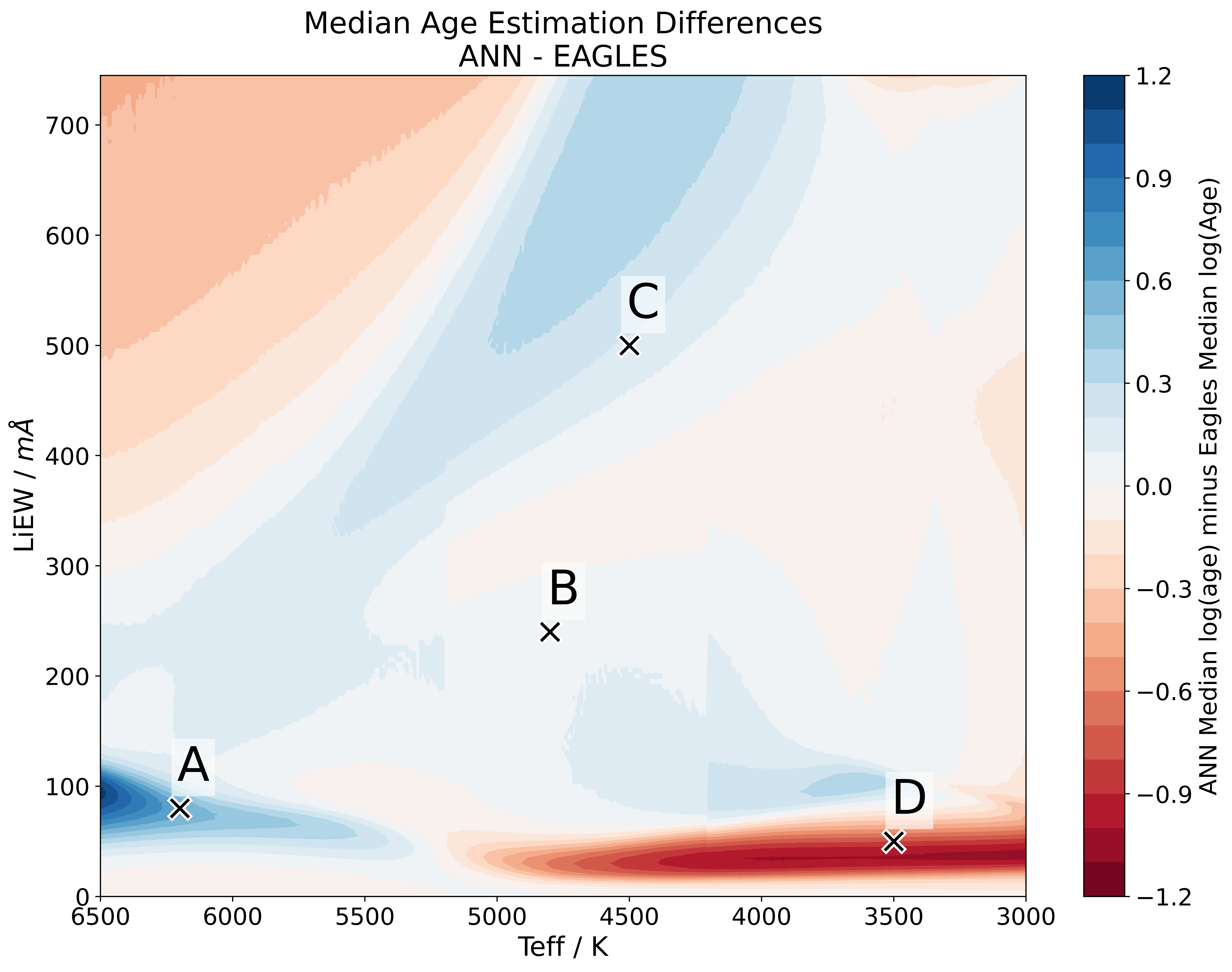}}
    \subfigure{\includegraphics[width=0.49\textwidth]{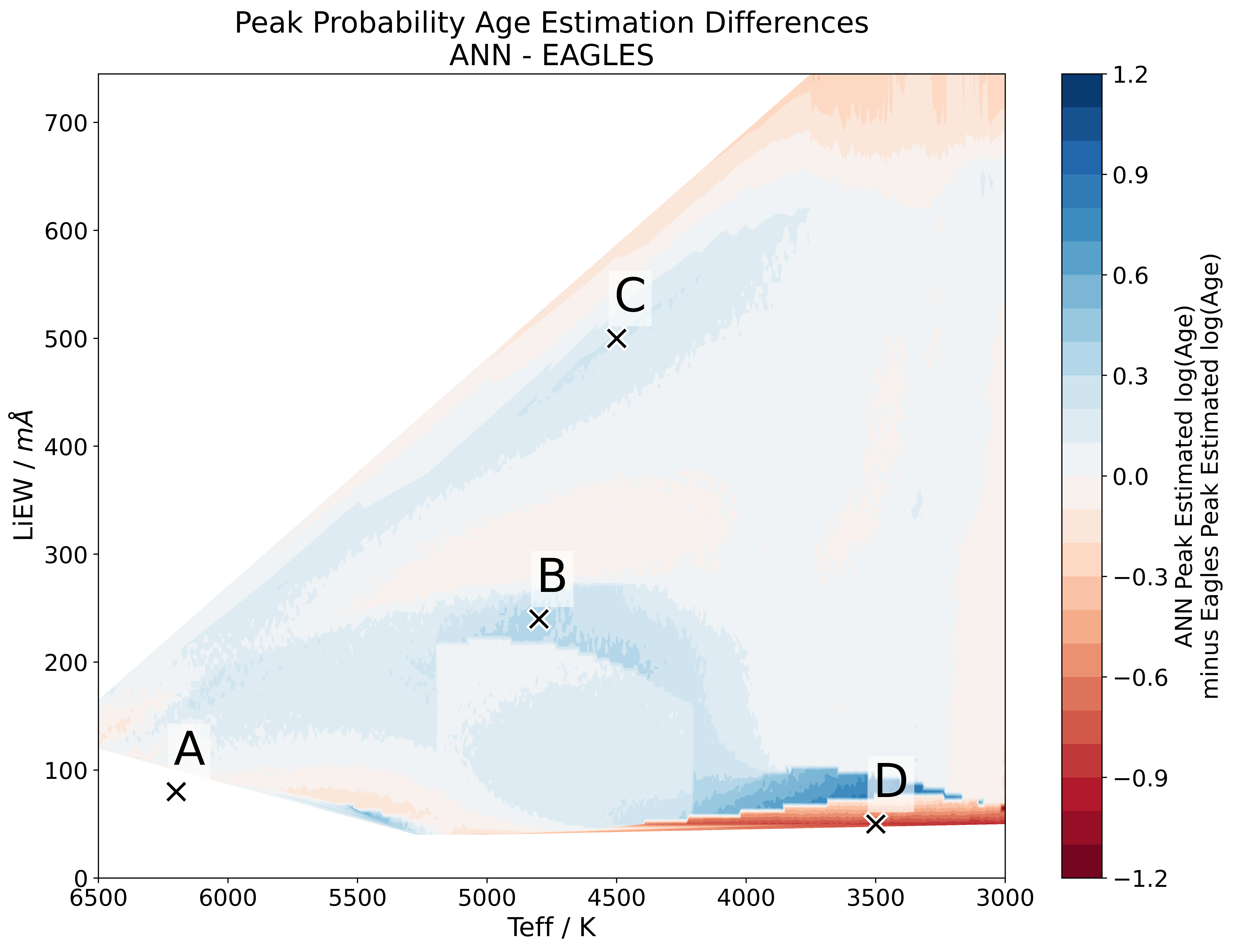}}
    
    \subfigure{\includegraphics[width=0.24\textwidth]{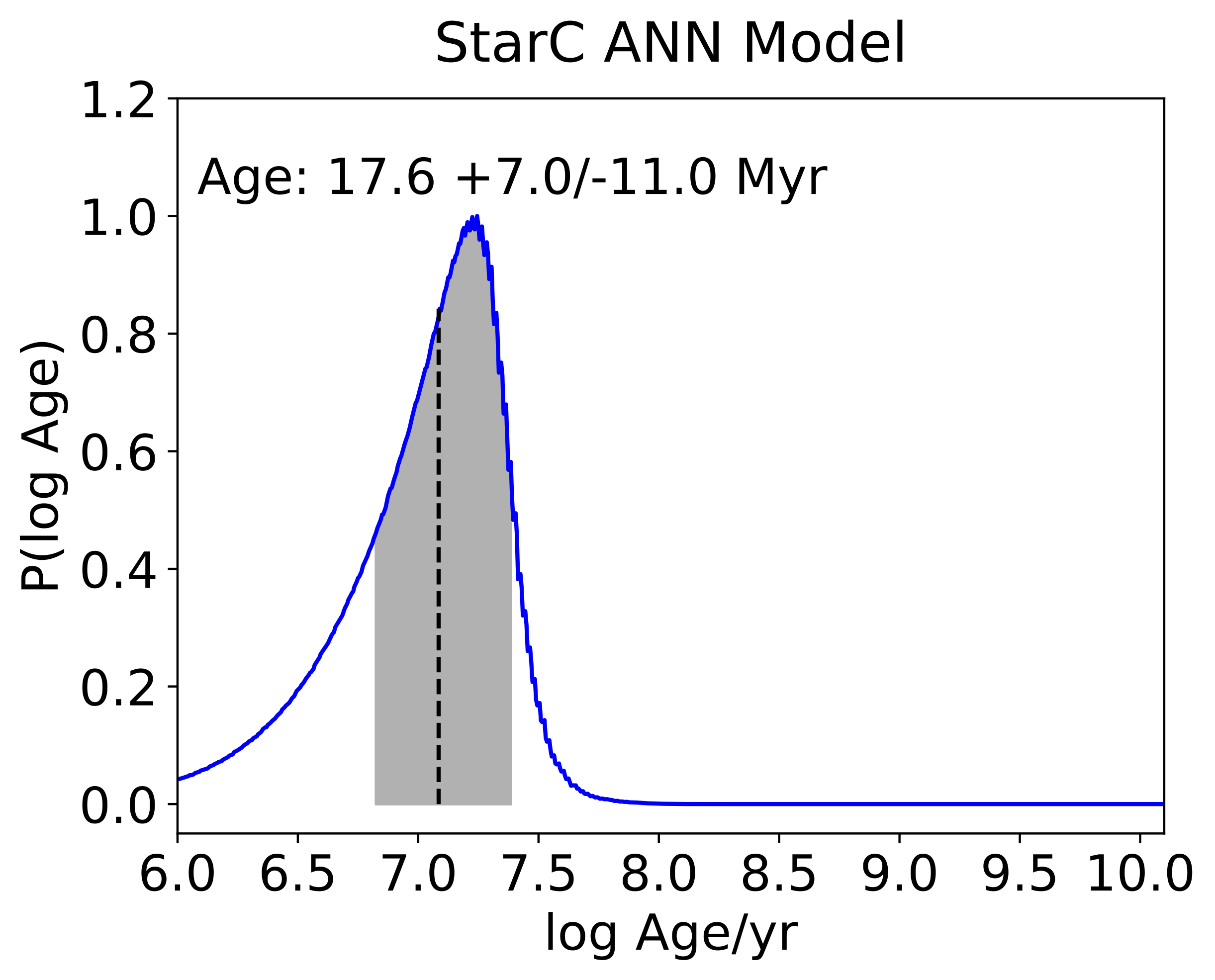}}
    \subfigure{\includegraphics[width=0.24\textwidth]{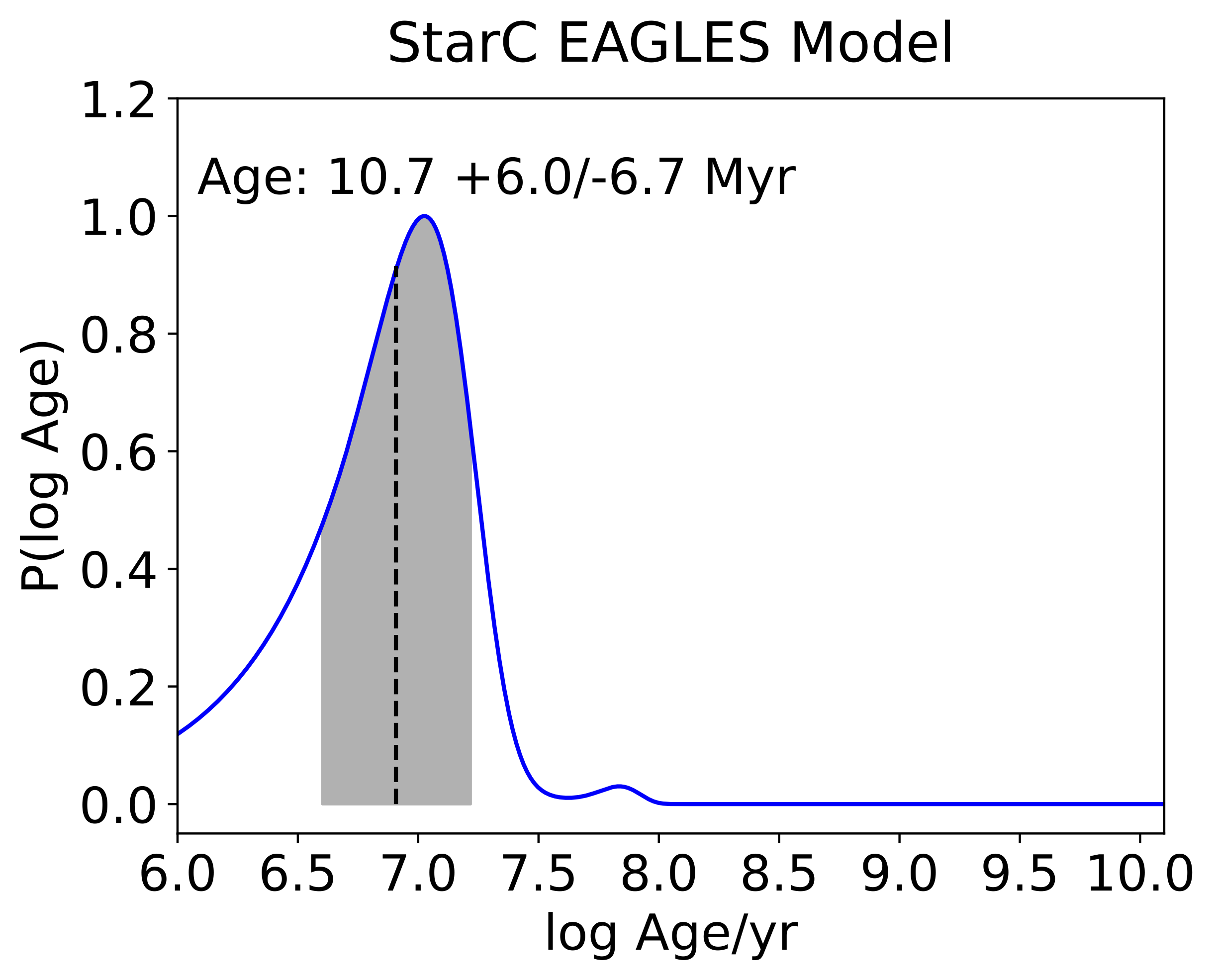}}
    \subfigure{\includegraphics[width=0.24\textwidth]{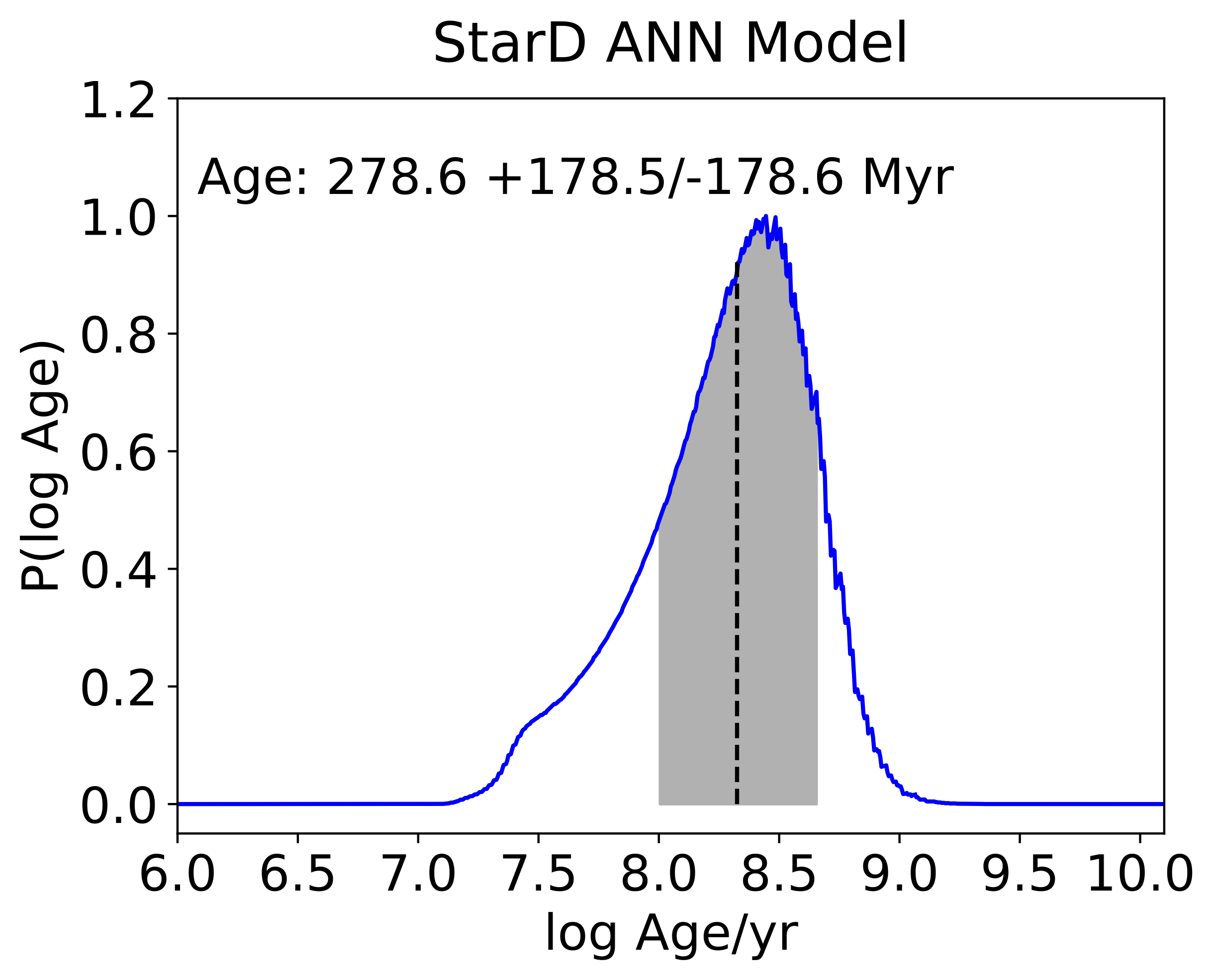}}
    \subfigure{\includegraphics[width=0.24\textwidth]{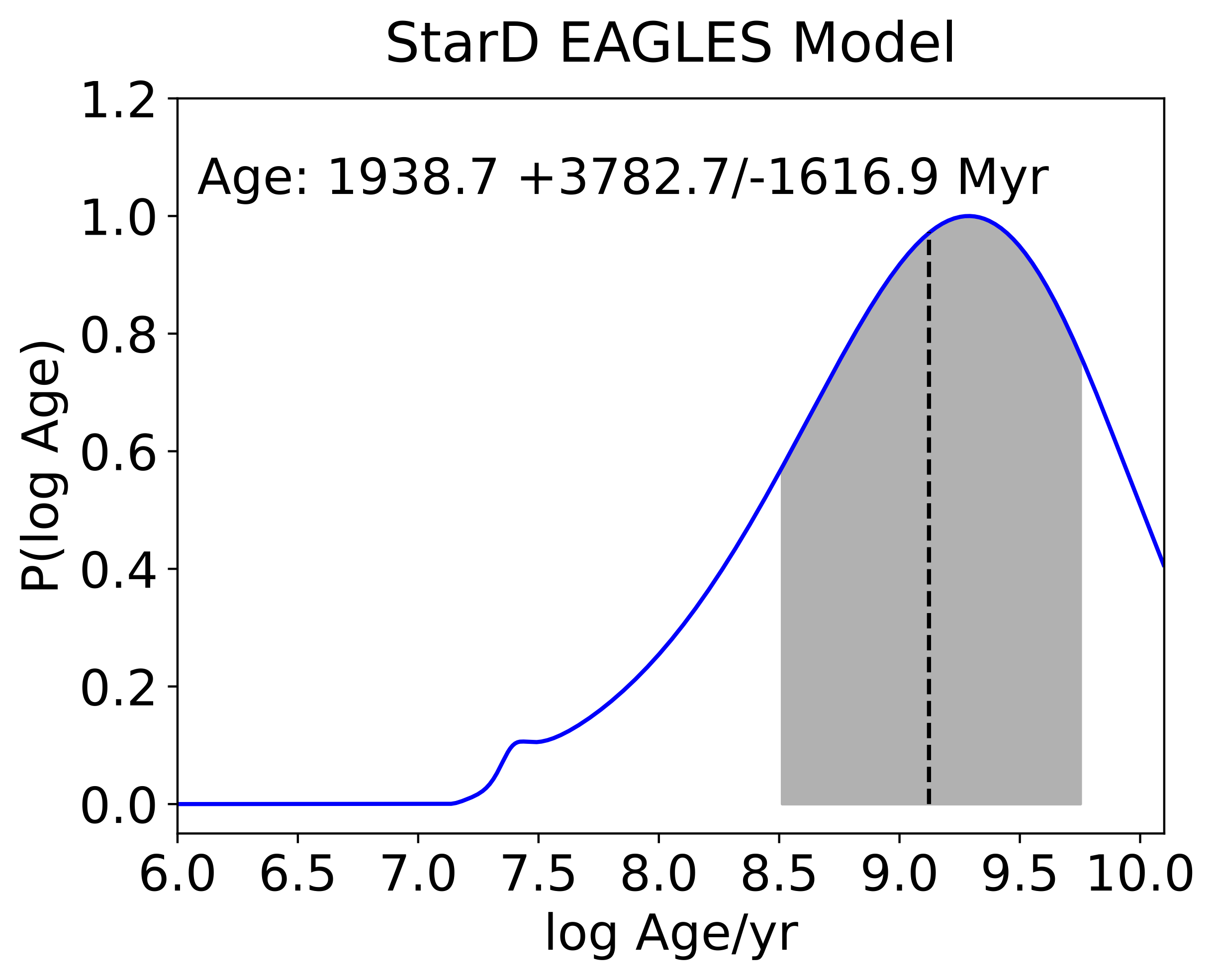}}
    
    \caption{\textit{Centre Left:} Differences in the median $\log$(age) estimates between the ANN and {\sc eagles} models, colour-coded and in dex units. Red areas denote the ANN model estimating higher ages than {\sc eagles}, and blue areas denote the opposite. A flat age prior is assumed in both cases.
    \textit{Centre Right:} Differences in the most probable $\log$(age) estimates shown with the same colour scheme. Areas in white are regions of the LiEW/\Teff plane where a significant probability peak is not found in one or both models (e.g., see the probability distributions for star A).
    \textit{Above/Below:} The probability distributions of $\log$(age) for four hypothetical stars: A, B, C and D, with no observational LiEW error, at the positions labelled in the central plots.}
    \label{fig:medvspeak}
\end{figure*}

Figure~\ref{fig:medvspeak} summarises the differences in estimated $\log$(age) in the LiEW vs \Teff plane, assuming there are no observational uncertainties and a flat prior probability in age. The left and right plots show the differences in median age (the 50th percentile in the posterior age probability distribution) and the most probable age respectively. The two models make similar predictions over most of the grid but four example stars are labelled and examined in detail to highlight differences in the age estimates and their posterior probability distributions (also shown). 

In stars A and C, the effects of a larger predicted LiEW dispersion for older warm stars and for very young stars with high LiEW in the ANN model are seen. This causes the flat prior in age to push the posterior distribution to include older ages. The effect is modest for star C, but much more important in star A, to the extent that there is no peak in the posterior, because the increased dispersion is a large fraction of the LiEW.
In the region of star B, the LiEW isochrones at older ages are closer together in the ANN model than in {\sc eagles}.
This, combined with the slightly larger dispersion in the ANN model, allows the prior to exert a greater influence and push the peak to older ages without greatly influencing the median age.
Finally, in star D, the ANN model produces a younger age estimate than {\sc eagles}.
Here, at ages $\gtrsim$ 500 Myrs, the dispersion predicted by the ANN model is much smaller than that of {\sc eagles}, and the Li is detected with enough significance (compared with the intrinsic dispersion) to rule out older ages, despite the rising prior.

\subsection{GES Cluster Comparisons}\label{subsec:ges}

Comparisons were made with selected GES clusters that were used in the testing of the original {\sc eagles} models (see \S \ref{predictions}). 
It is noted that some of these clusters have member stars included in the training dataset, and as such, the age predictions below are solely for the comparison of model residuals between the training ages and age estimates in both the ANN and {\sc eagles} models.
This is to ascertain whether the model is performing well in terms of accurately modelling the training data, in direct comparison with the same residuals from {\sc eagles}.
These residuals can highlight where there may be unexpected systematic departures or dispersion that may tell us that there is not enough complexity in the model, or possibly that other, unconsidered parameters, such as rotation, metallicity or a veiling continuum are affecting the Li abundance or measured LiEW.

Initially, comparisons were made in order to check whether the ANN model had improved the known shortcomings in the {\sc eagles} model -- the large residuals for clusters older than 1 Gyr, the need for an {\it ad hoc} additional intrinsic LiEW dispersion in some areas of the LiEW - \Teff plane, and the lack of any discrimination below ages of $\sim 10$ Myr. 

\begin{figure*}
  \centering
  \begin{minipage}{0.49\textwidth}
    \centering
    \includegraphics[width=\textwidth]{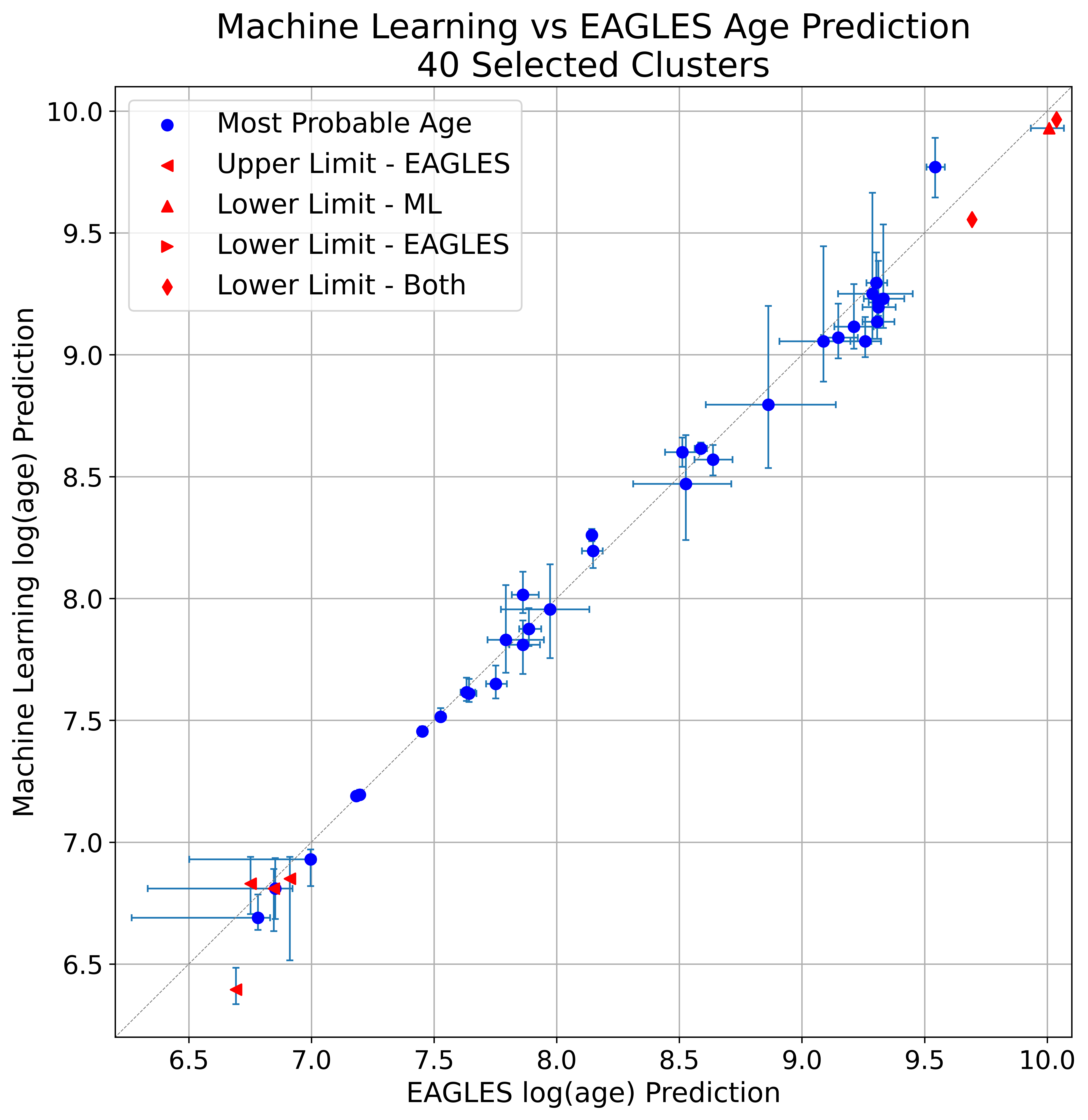}
    
  \end{minipage}
  \hfill
  \begin{minipage}{0.49\textwidth}
    \centering
    \includegraphics[width=\textwidth]{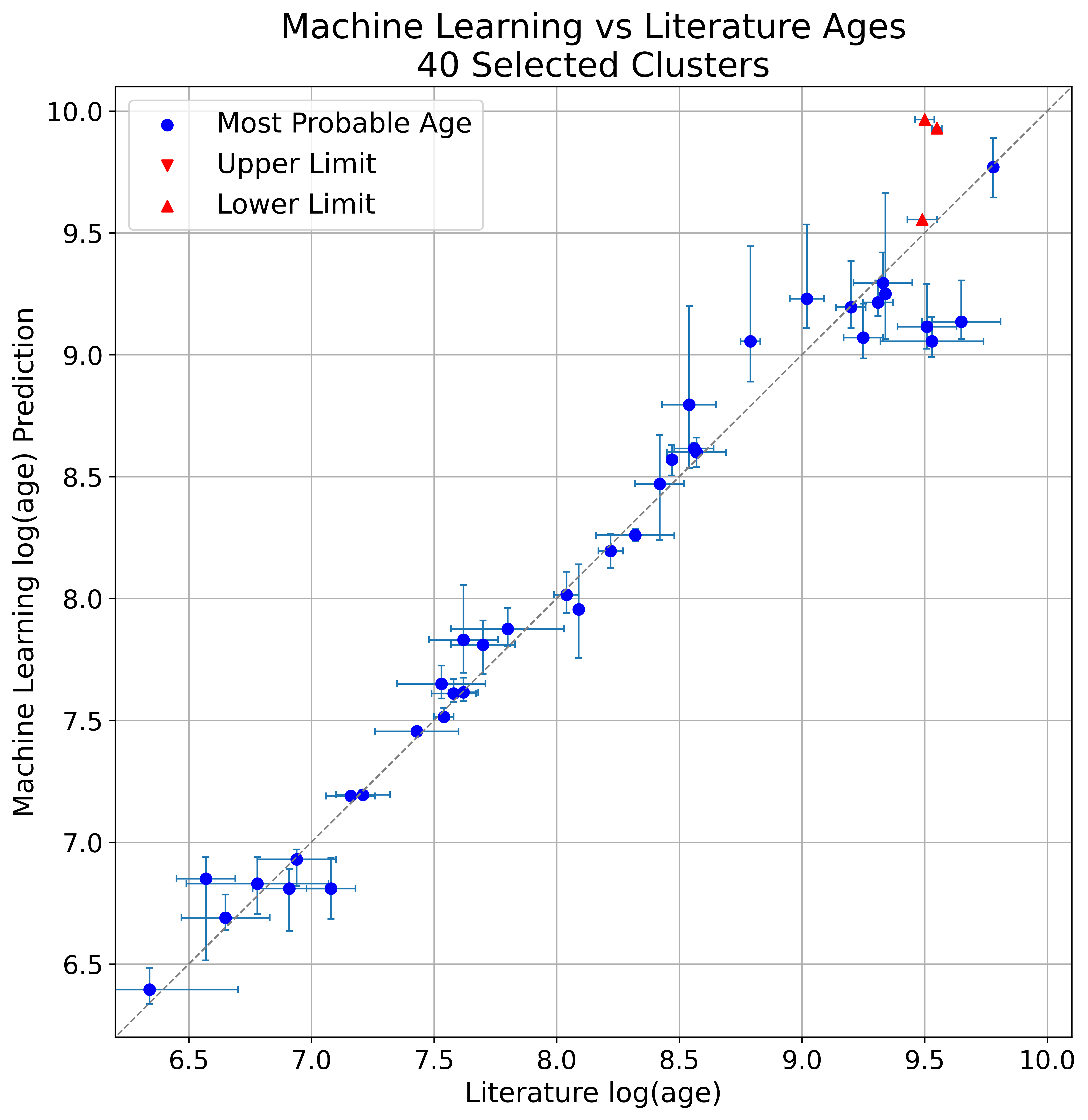}
    
  \end{minipage}
  
  \caption{\textit{Left:} Estimated $\log$(age) for 40 selected GES clusters using the ANN models versus that estimated using {\sc eagles}.
  \textit{Right:} Estimated $\log$(age) from the ANN model versus mean literature $\log$(age) \citep[the training ages from][]{Rob2023}.}
  \label{fig:GESagecomp}
\end{figure*}

Comparisons of the most probable age from the ANN model with those from {\sc eagles} and with the literature (training) ages are shown in Fig~\ref{fig:GESagecomp}. The  {\sc eagles} and ANN models yield very similar patterns of residual ages.
The majority of cluster age predictions, especially younger than $\lesssim$ 1 Gyr, are within 0.1 dex, as seen in the left plot of Fig \ref{fig:GESagecomp} however, there are some small, but interesting differences.

\begin{figure*}

    \centering
    \subfigure{
    \centering
    \includegraphics[width=0.32\linewidth]{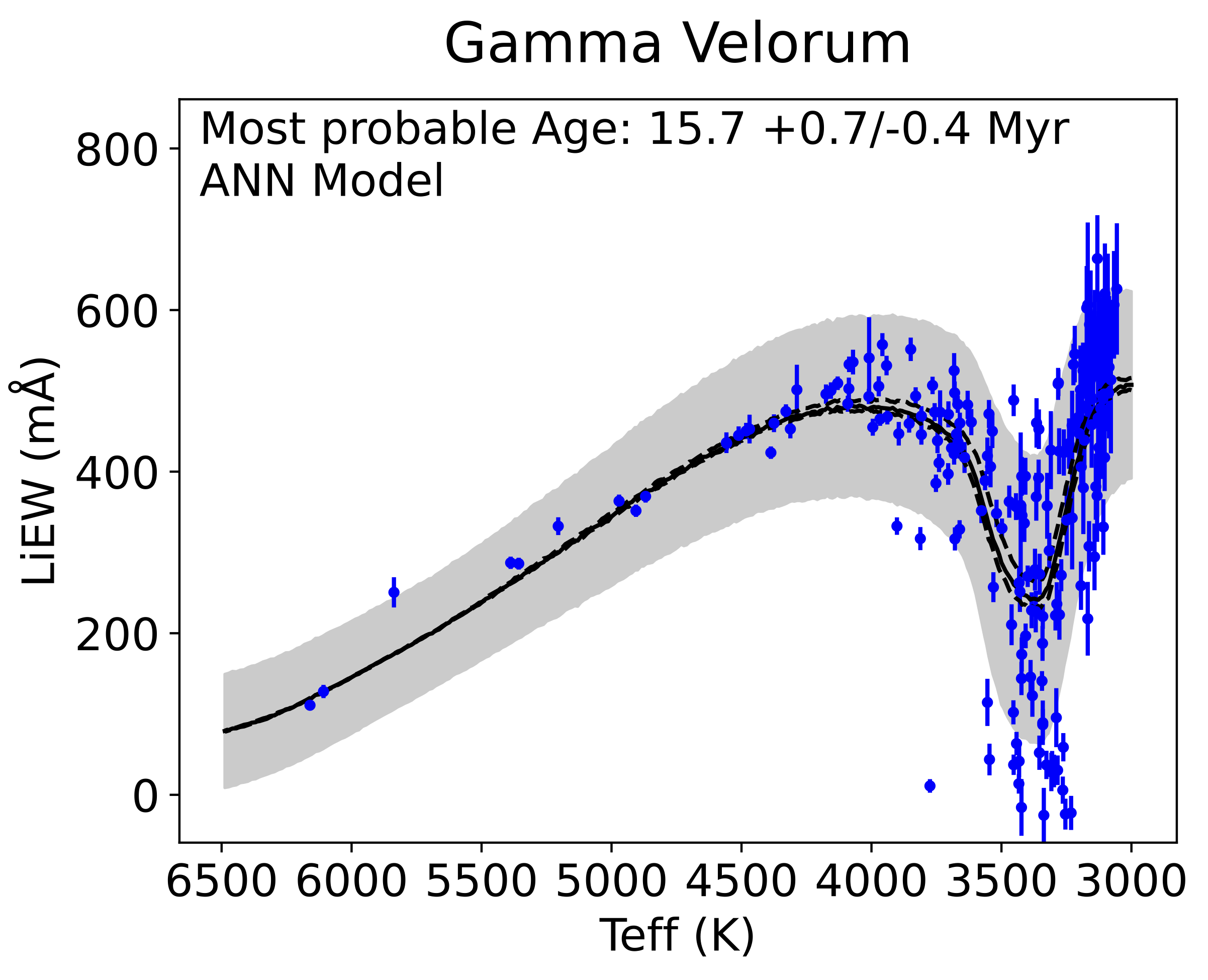}
        }
    \subfigure{
    \centering
    \includegraphics[width=0.328\linewidth]{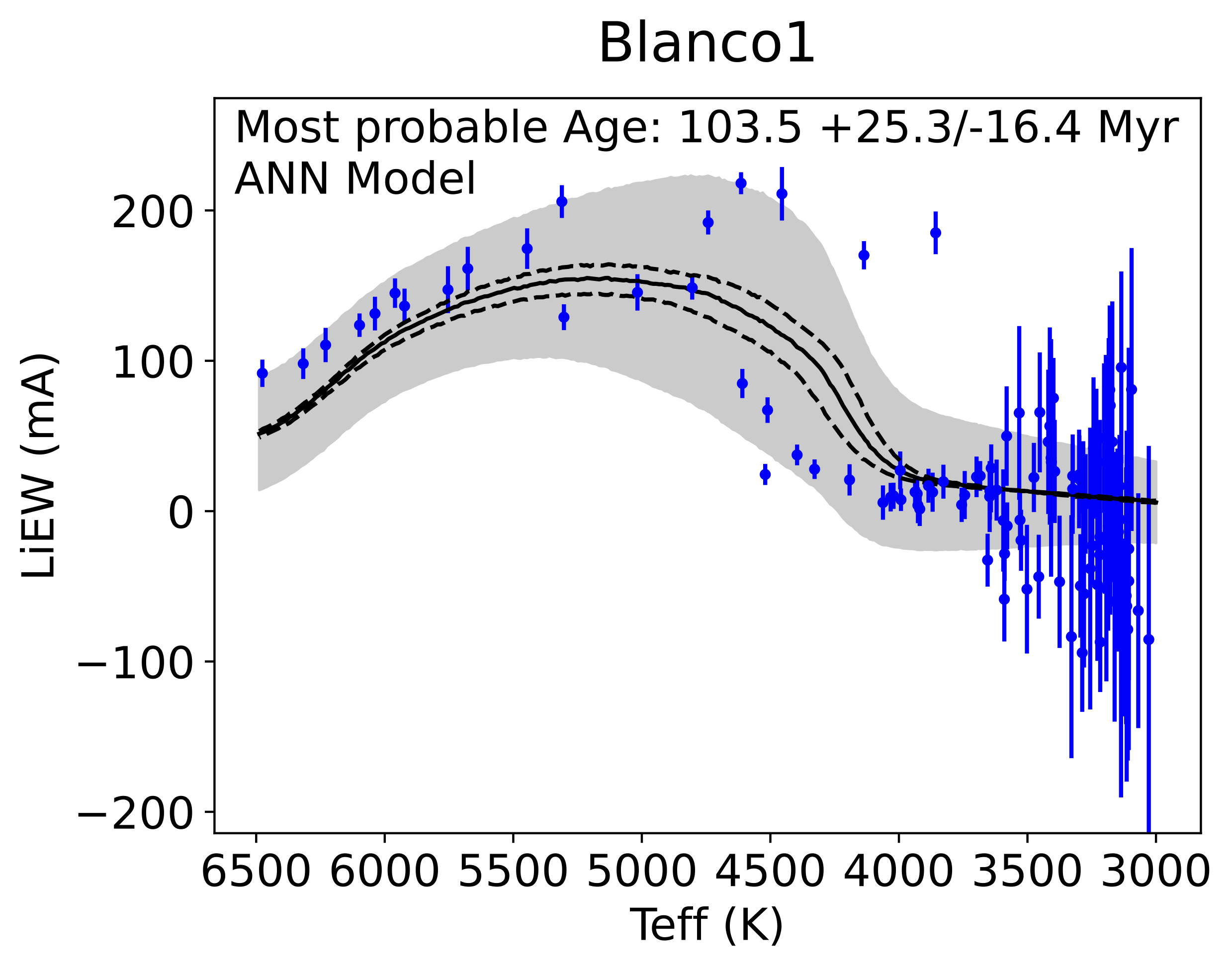}
        }
    \subfigure{
    \centering
    \includegraphics[width=0.321\linewidth]{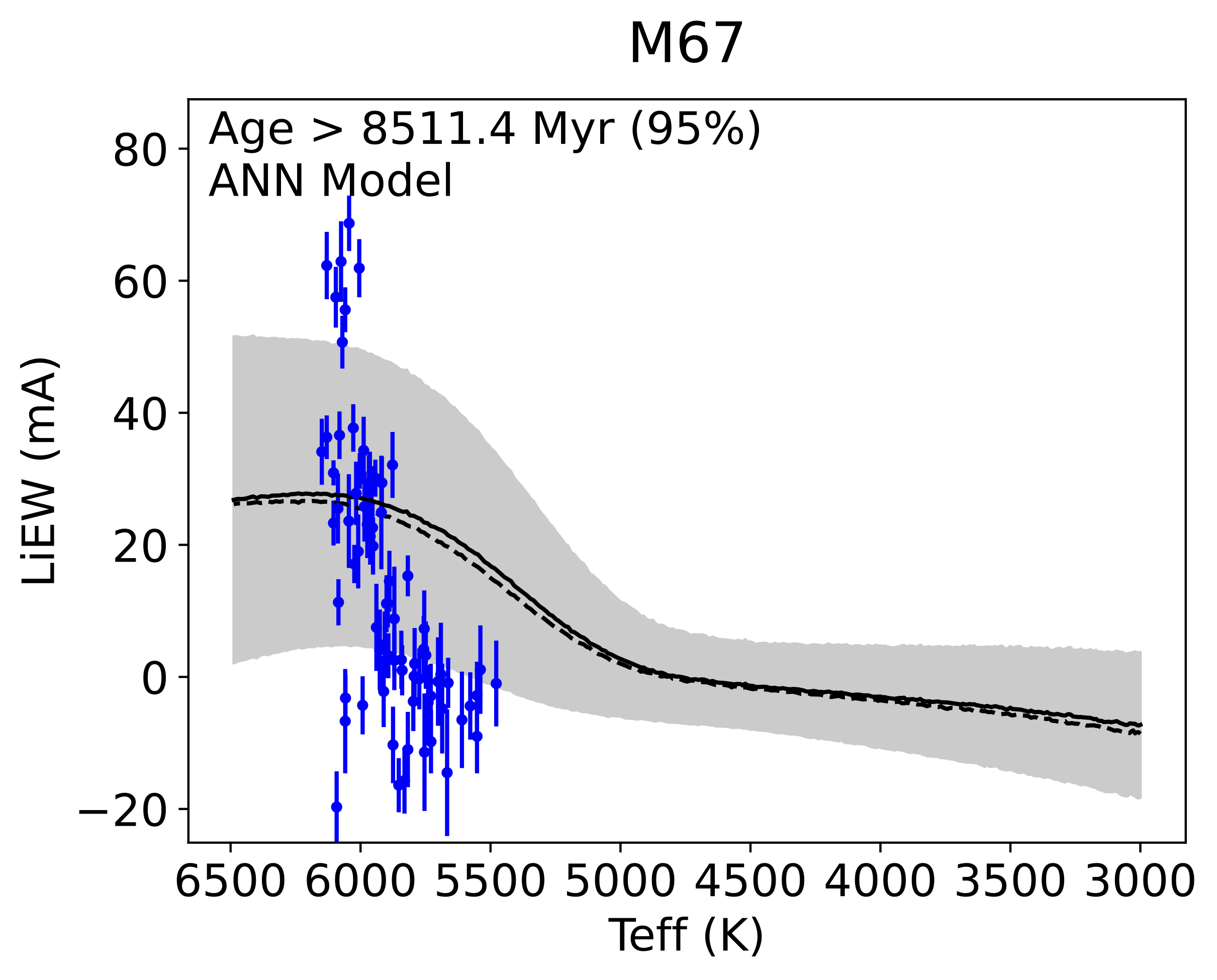}
    }
    
    \subfigure{
    \centering
    \includegraphics[width=0.32\linewidth]{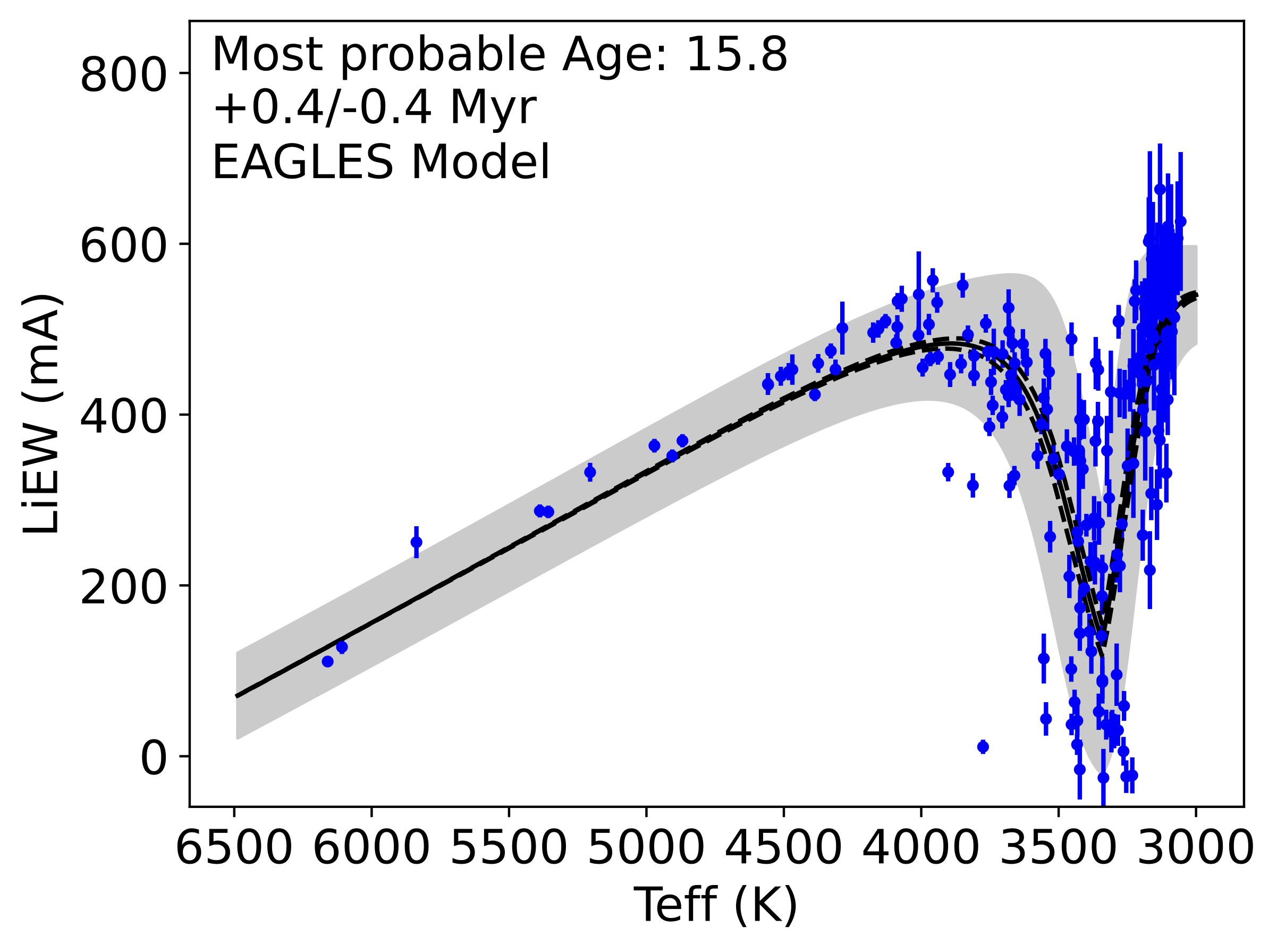}
        }
    \subfigure{
    \centering
    \includegraphics[width=0.328\linewidth]{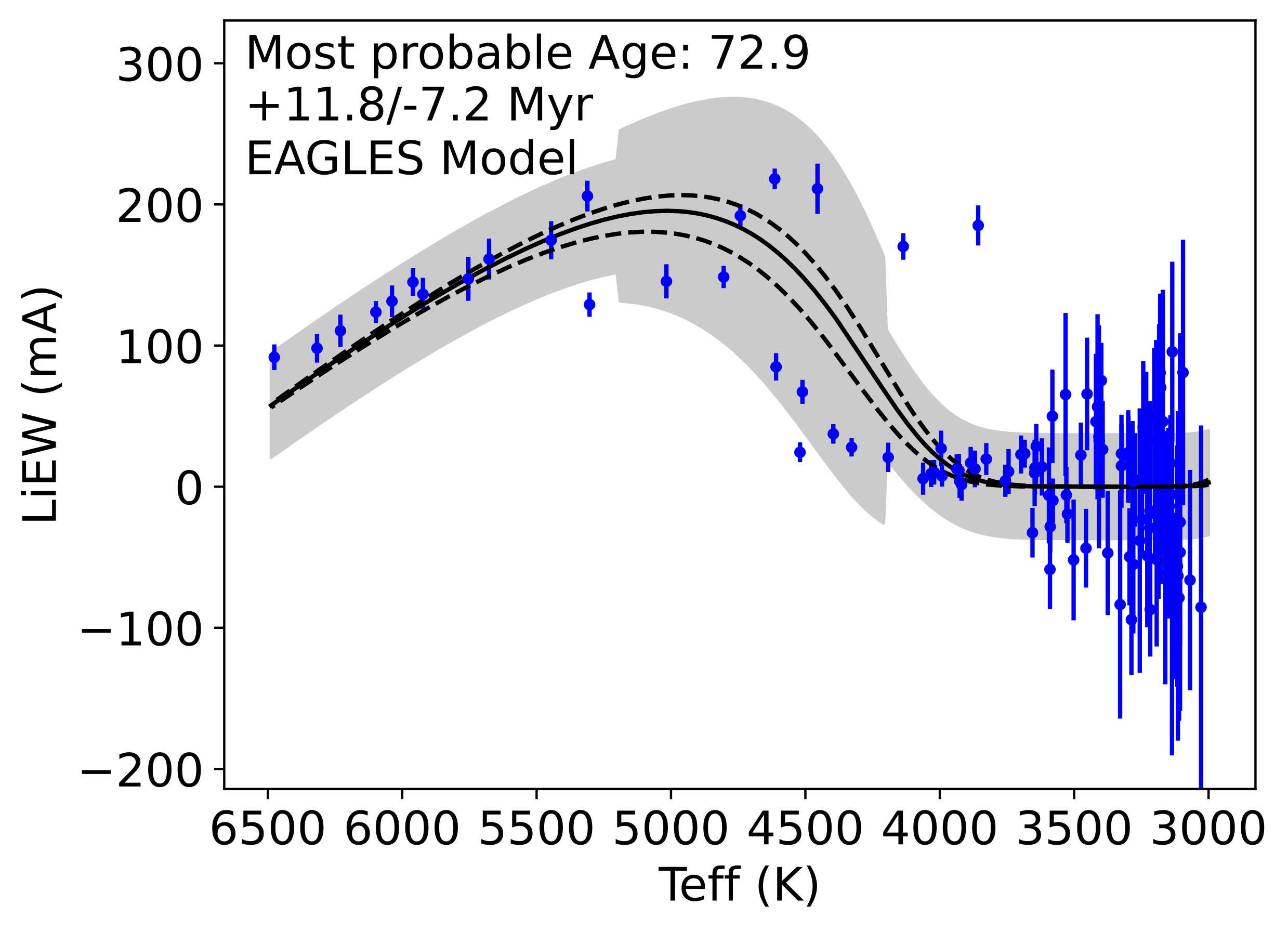}
        }
    \subfigure{
    \centering
    \includegraphics[width=0.321\linewidth]{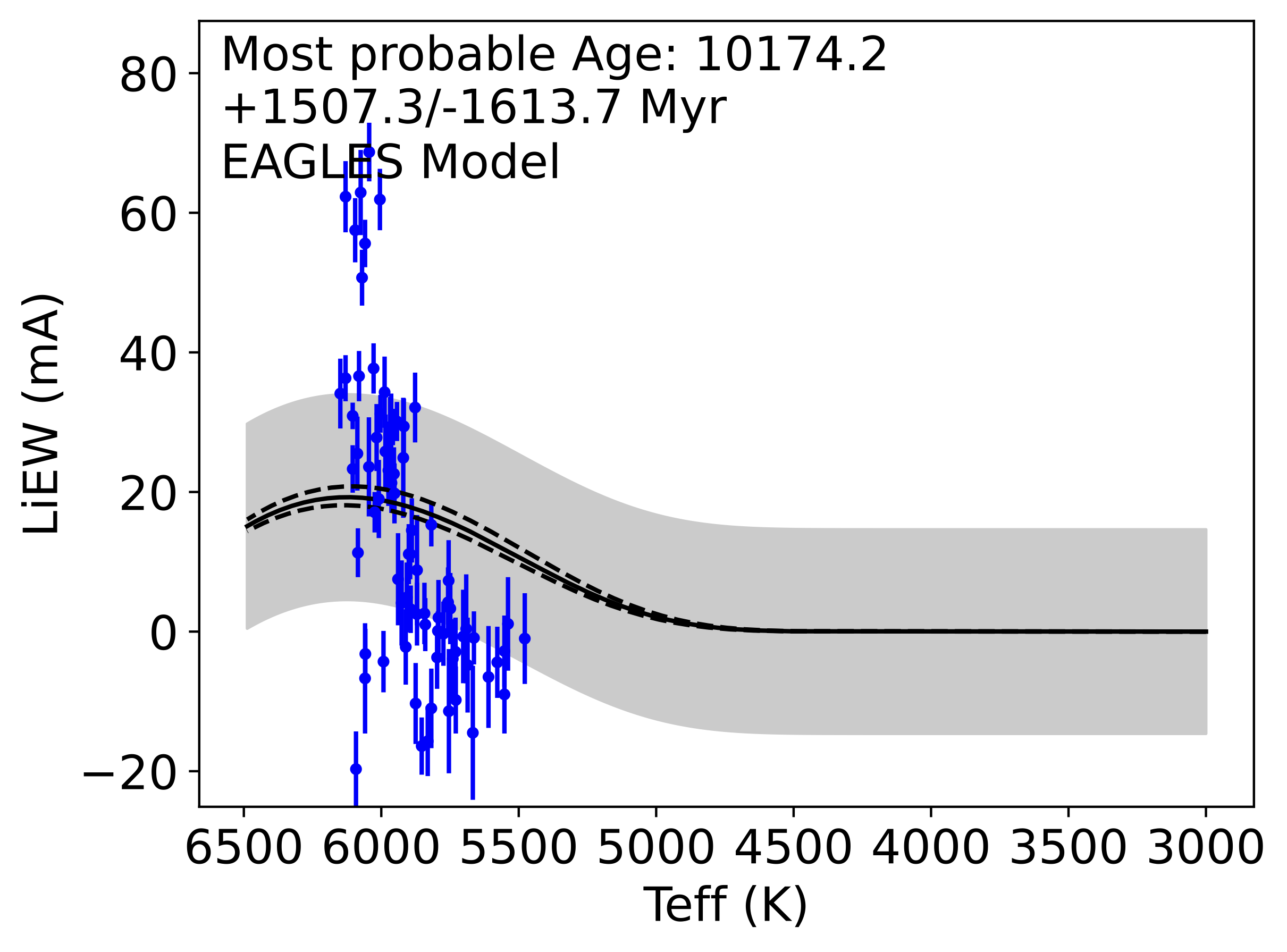}
        }
        
    \caption{ANN (top) and {\sc eagles} (bottom) model fits to three clusters, Gamma Velorum (literature age = $16.2^{+4.7}_{-3.6}$ Myr), Blanco 1 (literature age = $109.7^{+13.4}_{-11.9}$ Myr), and M67 (literature age = $3550^{+170}_{-160}$ Myr).}
    \label{fig:gammavel}
\end{figure*}

\subsubsection{Younger Clusters}
\label{youngerclusters}

The youngest clusters now have well-defined most probable ages, but which are consistent with the upper limits placed on them by the {\sc eagles} estimates.
For clusters with $\log$ (age/yr) $\lesssim$ 7.1, the ANN model's age estimates are almost all within the error bars of the literature age and age estimate, although there is significantly less precision in this age range compared with slightly older clusters where the Li dip formed.
The most notable outlier in Collinder 197, has an age estimate from the ANN model of $6.5^{+2.2}_{-1.6}$ Myrs, compared with the assigned training age of $12.0^{3.1}_{-2.5}$ Myrs.

\subsubsection{Intermediate Aged Clusters}
\label{intermediateclusters}

Between 10 Myrs and 1 Gyr, and particularly in the areas of high Li-depletion sensitivity ($7 < \log$ (age/yr)$ < 8$) the, often very, precise ages determined from the ANN model have small residuals compared with the training ages and agree closely with the {\sc eagles} results.
Using the 33 clusters of all ages which produce a definite most-probable age estimate in both models, {\sc eagles} estimates 14 of these non-limit clusters to within 0.1 dex of the assigned training age and 24 to within 0.2 dex, while the ANN model predicts ages for 21 of the 33 clusters within 0.1 dex, and 26 clusters within 0.2 dex.
We believe this reduction in outliers is due to the increased complexity of the ANN model in both the LiEW and $\sigma$LiEW predictions. Obviously, complexity comes with the risk of overfitting, but we believe the procedures adopted in \S\ref{training} are sufficient for us to say that his is not the case and the ANN model is not generalises well to both the validation dataset and for non-GES clusters and moving groups that are not part of the training set (see \S\ref{nonges}).
Along with some possible age discrimination identified below 10~Myr (\S\ref{subsec:isochrones}), there is some evidence that in the clusters which have a large number of data points falling in the Li dip (\Teff < 4000K and 7 < $\log$ (age/yr) < 7.8), the ANN model fits look better than those from {\sc eagles} (e.g., Gamma Velorum in Fig.~\ref{fig:gammavel}). 
The more complex dispersion pattern (\S\ref{subsec:isochrones}, Fig. \ref{fig:dispersiondifferences}) also appears to better fit some clusters towards the end of the Lithium dip ($\log$ (age/yr) $\sim$ 8) (see Blanco 1 in Fig. \ref{fig:gammavel}).

\subsubsection{Older Clusters}
\label{olderlusters}

Larger disagreements compared with the training ages are seen in many of the ANN model age estimates for older clusters, with $\log$ (age/yr) $> 8.7$.
The ANN model age estimates are quite similar to those of {\sc eagles} but tend to be slightly lower ($\sim 0.1$ dex) with larger error bars.

\begin{figure}
    \centering
    \includegraphics[width = \linewidth]{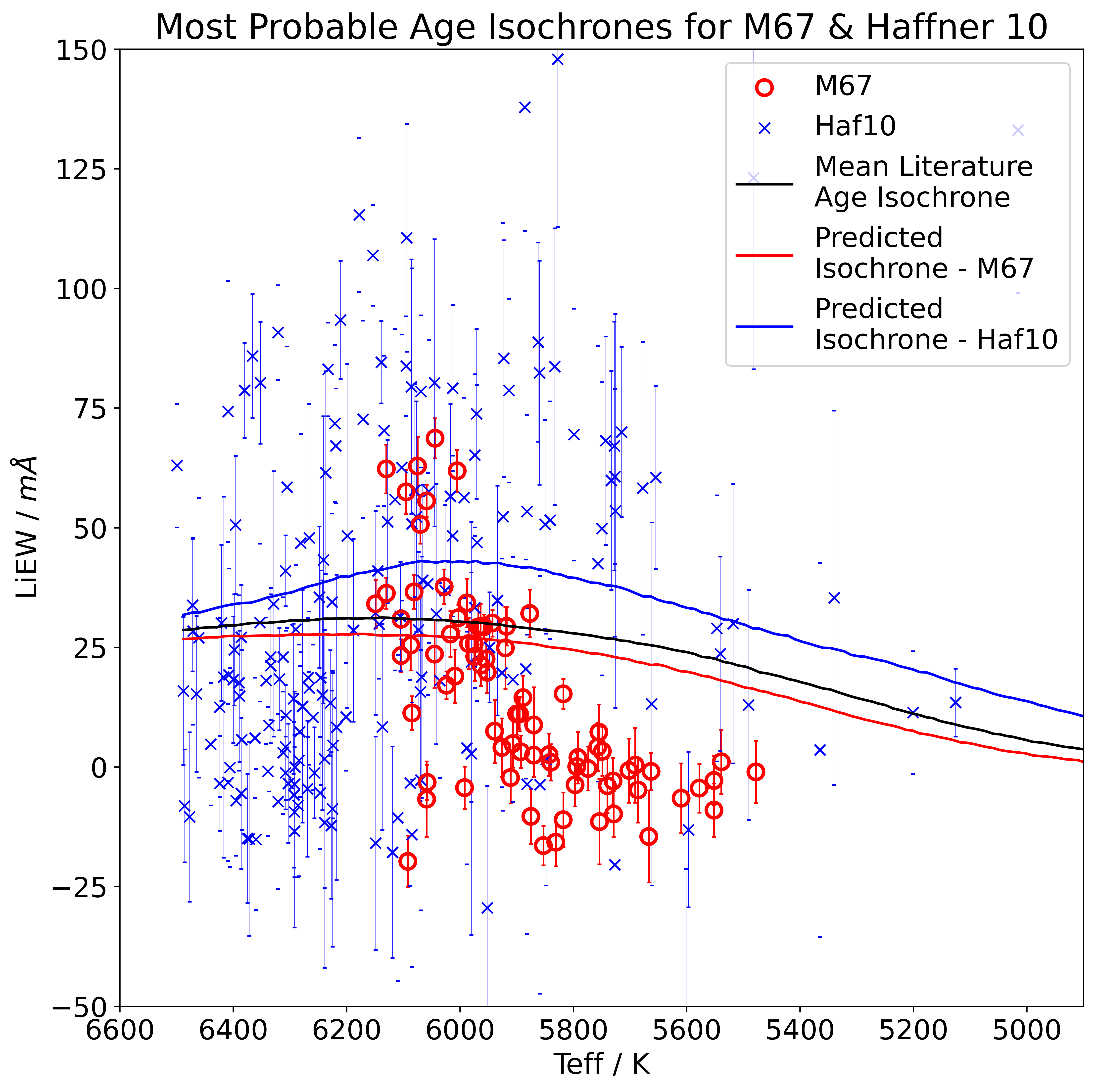}
    \caption{The training data from M67 (literature $\log$ (age/yr) = 9.53) and Haffner 10 (literature $\log$ (age/yr) = 9.55). The coloured isochrones are drawn at the ANN model's most probable $\log$ (age/yr) estimate for Haffner 10 (9.06), and the lower limit predicted age for M67 (> 9.93), and the black isochrone is drawn at the mean literature $\log$ (age/yr) of the two clusters ($\log$ (age/yr) = 9.54).
    Despite being very similar in literature age, the age for Haffner 10 is highly underestimated, and the age for M67 is highly overestimated.}
    \label{fig:m67haf10isochrones}
\end{figure}

In \citet{Rob2023} it was noted that the {\sc eagles} age estimates were quite discrepant with the training ages for some of the older clusters, and the shapes of the model isochrones did not fit the cluster data very well in some cases (e.g. M67). 
The ANN model has more flexibility and offers a test of whether the cause of these discrepancies is the assumed analytical form of the EAGLES model.
However, as can be seen in Fig.~\ref{fig:gammavel}, this is still a problem for some older clusters, including M67, in the ANN model, particular those with an apparent bimodality in their Li depletion.
Neither the ANN model or {\sc eagles} can fit both the high LiEW data at the warmest temperatures and the very low LiEW data at solar temperatures, leading to an age prediction that is significantly higher than the mean literature age.
In principle, the ANN model would be able to adjust the evolution of the isochronal shape to match the M67 data, even if this included allowing an unphysical increase in LiEW with time within certain \Teff ranges -- an impossibility in {\sc eagles}.
However, the ANN model does not do this because other clusters of similar ages show quite different behaviour.
This is seen in Fig.\ref{fig:m67haf10isochrones}, where, despite both clusters having a similar literature/training age, the ANN-estimated age for Haffner 10 is  underestimated, and the age of M67 is greatly overestimated, apparently due to the very different behaviours of Li depletion at high \Teff in these clusters.
The problem is therefore not due to insufficient complexity in the {\sc eagles} model functional form; the additional complexity of the ANN model fit does not solve this problem for the older clusters.

\subsection{Non-GES Data}\label{nonges}

\begin{table*}
 \caption{Age estimates from the ANN and {\sc eagles} models for clusters and groups that were not part of the training set. Both the most probable ages and median estimates are given, alongside the literature ages taken from the {\sc eagles} paper \citep{Rob2023}.}
 \label{tab:nonges}
  \begin{tabular}{cccccc}
   \hline
   Cluster/ & Literature & ANN Most & {\sc eagles} Most & ANN Median & {\sc eagles} Median\\
   Group & Age &Probable Age /&Probable Age /& Age /& Age /\\
   & Myr & Myr & Myr & Myr & Myr \\
   \hline\\[-5pt]
    $\eta$ Cha&$12^{+6}_{-6}$&$2.6^{+4.4}_{-1.0}$&< 10.23&3.31&3.78\\[3pt]
    TW Hydrae &$12^{+8}_{-8}$&$10.2^{+1.1}_{-8.1}$&$11.8^{+0.84}_{-1.0}$&3.51&11.7\\[3pt]
    32 Ori&18 - 25&$27.2^{+3.0}_{-1.2}$&$26.4^{+1.9}_{-1.5}$&28.5&26.4\\[3pt]
    
    $\beta$ Pictoris&$23^{+1}_{-1}$&$25.7^{+1.2}_{-0.87}$&$25.2^{+1.2}_{-0.86}$&25.7&25.5\\[3pt]
    Tuc-Hor Assc&41 - 51&$37.6^{+3.2}_{-2.5}$&$41.4^{+2.5}_{-1.9}$&38.0&41.4\\[3pt]
    IC2391&$42^{+16}_{-12}$&$51.3^{+13}_{-7.6}$&$51.5^{+6.3}_{-4.0}$&53.7&52.1\\[3pt]
    IC2602&$42.6^{+6.2}_{-5.5}$&$37.6^{+5.4}_{-3.3}$&$38.2^{+2.7}_{-2.6}$&38.0&38.2\\[3pt]
    Pleiades&$118^{+6}_{-10}$&$98^{+16}_{-15}$&$95^{+10}_{-10}$&96.6&94.0\\[3pt]
    AB Doradus&100 - 125&$81^{+22}_{-18}$&$74.0^{+12}_{-8.0}$&80.4&74.5\\[3pt]
    Psc-Eri Stream &100 - 125&$121^{+23}_{-19}$&$115^{+11}_{-15}$&123&113\\[3pt]
    M35&$140^{+15}_{-15}$&$117.5^{+9.9}_{-7.8}$&$114.0^{+5.4}_{-5.1}$&120&114\\[3pt]
    Hyades&635&$623^{+84}_{-81}$&$729^{+120}_{-94}$&609&729\\[3pt]
    Praesepe&670&$749^{+140}_{-82}$&$897^{+130.0}_{-110}$&776&905\\[3pt]
   \hline
  \end{tabular}
\end{table*}

Further validation of the ANN model is achieved by estimating ages using lithium datasets from clusters and moving groups {\it not} used in training the ANN. The reference ages, data sources and {\sc eagles} age estimates for these clusters are the same as those in \cite{Rob2023}. The results are summarised in Table \ref{tab:nonges} and the LiEW-\Teff isochrones (from the ANN model) at the most probable ages are shown in Appendix \ref{sec:appendixfits}.

\begin{figure}
    \centering
    \begin{minipage}{.5\linewidth}
        \centering
        \textbf{Eta Chamaeleontis}
        \includegraphics[width=\linewidth]{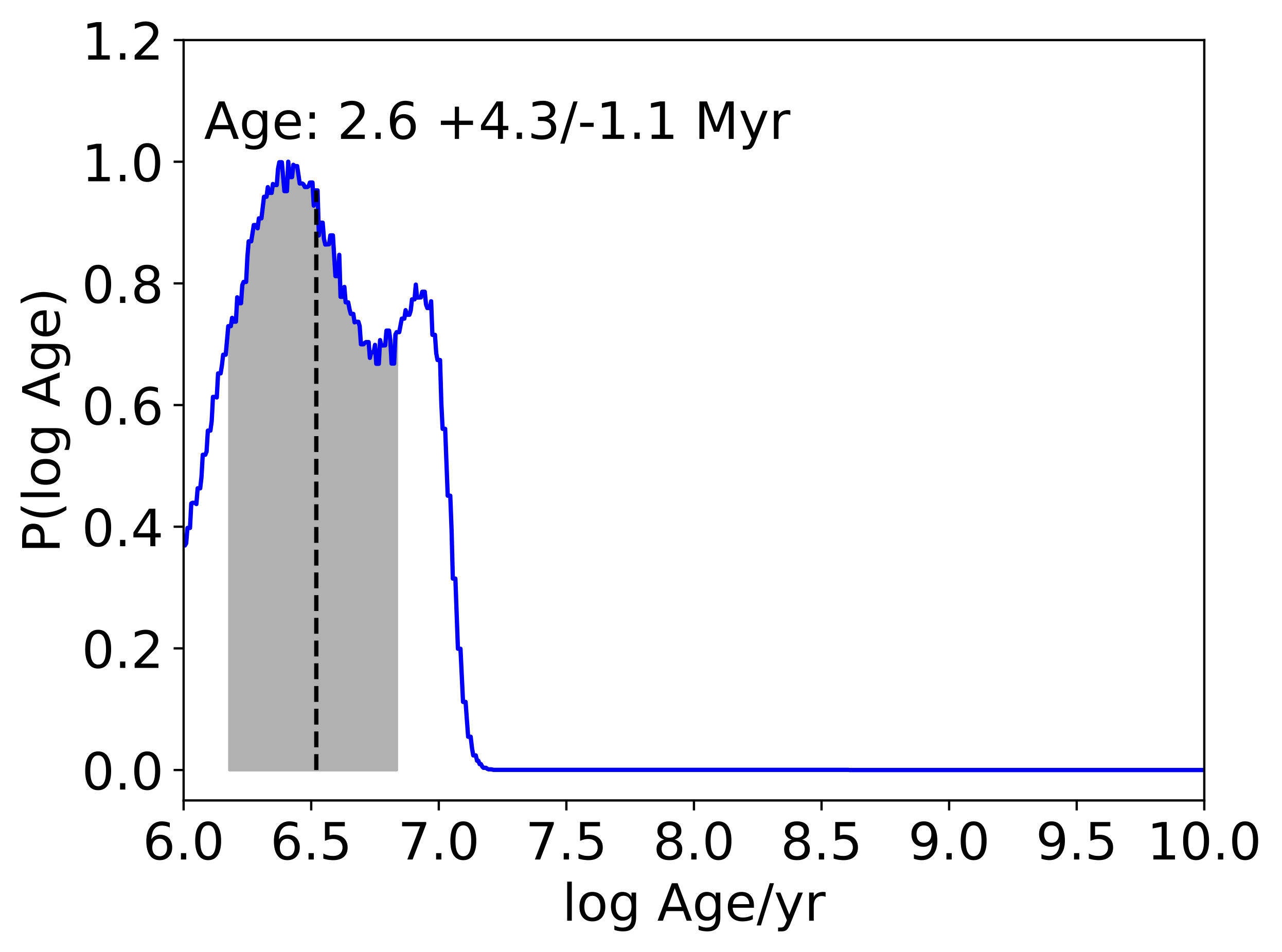}
    \end{minipage}%
    \begin{minipage}{.5\linewidth}
        \centering
        \textbf{TW Hydrae}
        \includegraphics[width=\linewidth]{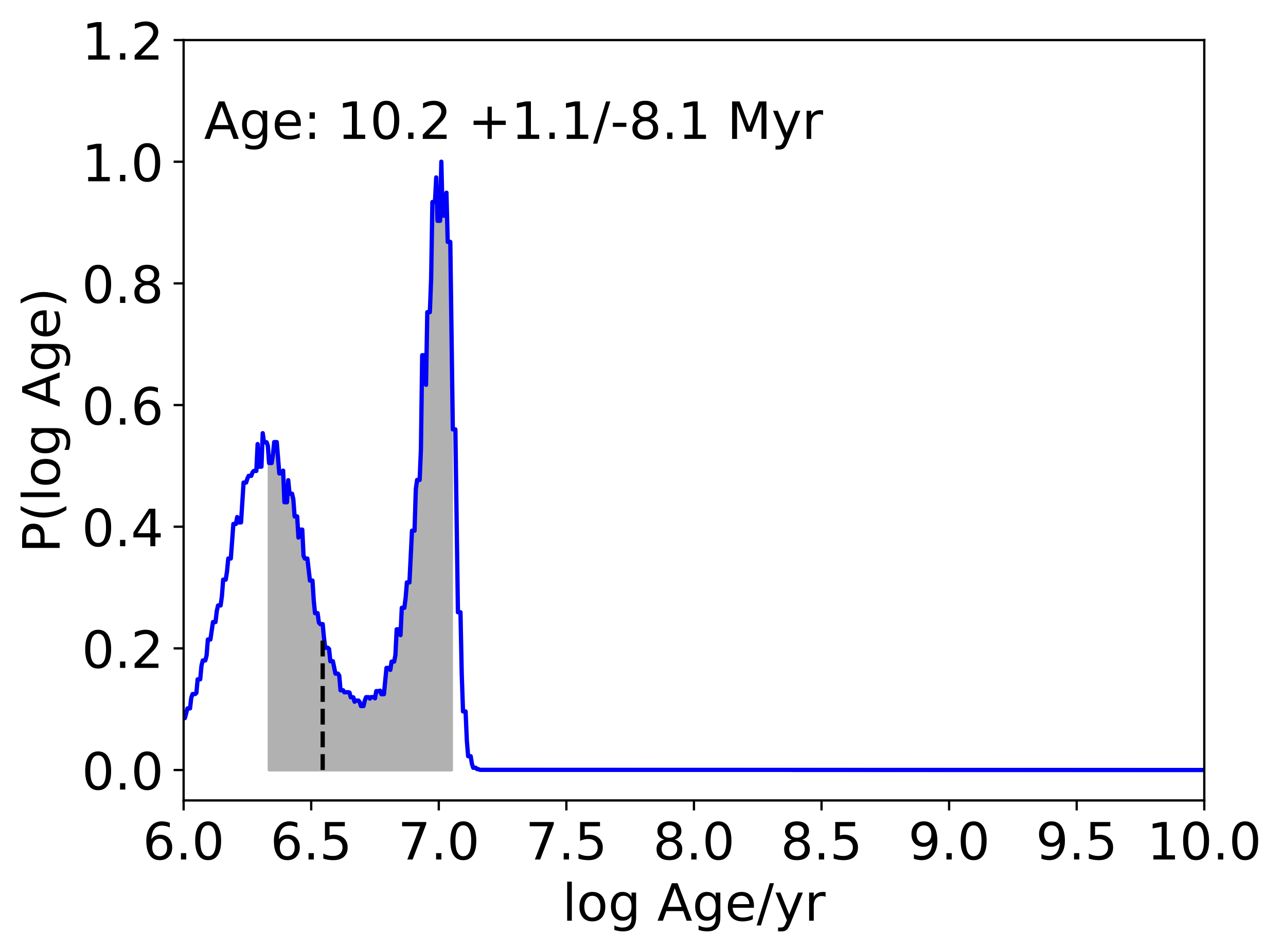}
    \end{minipage}
    \caption{Probability distributions for Eta Chamaeleontis (left) and TW Hydrae (right). Both show bimodal distributions indicating the likelihood of a group having either started to deplete Lithium or being undepleted with lower LiEW at a younger age.}
    \label{fig:probsnonges}
\end{figure}

Table \ref{tab:nonges} shows that the ANN model and {\sc eagles} yield very similar most probable and median ages for almost all the clusters and moving groups. They also provide most probable age estimates that are quite consistent with ages from the literature.
The exception is eta Cha, in which the ANN model has given a resolved most probable age estimate, though this is entirely consistent with the upper limit given by {\sc eagles}.
Additionally, although the age estimate for TW Hydrae is similar, the lower error is significantly larger in the ANN model's estimate.
These differences appear to be a result of the isochronal crossover allowed within the ANN model, as shown in the probability distributions of both TW Hydrae and eta Cha (Fig. \ref{fig:probsnonges}), wherein both distributions have bimodality as a result of the non-monotonic LiEW - age relationship at $\leq 10$ Myr.
In {\sc eagles}, stars with slightly lower LiEW in the association must be attributed to the beginnings of Li depletion and the Li dip, whereas in the ANN model, these could either be stars beginning to deplete Li, or younger stars with lower LiEW, as discussed in \S \ref{subsec:isochrones}.

\begin{figure*}
    \centering
    \begin{minipage}{0.49\textwidth}
        \centering
        \includegraphics[width=\textwidth]{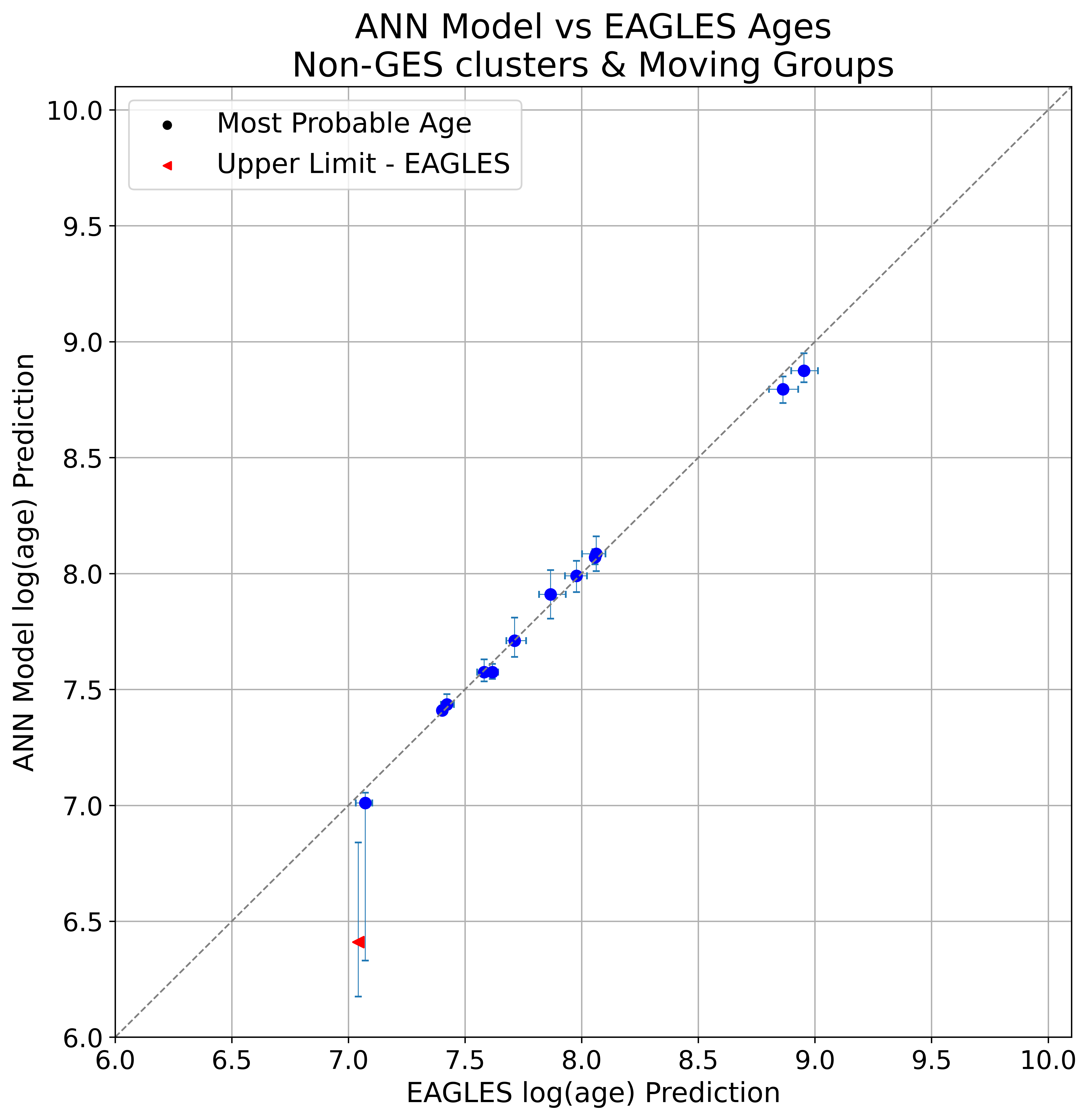}   
    \end{minipage}
    \hfill
    \begin{minipage}{0.49\textwidth}
        \centering
        \includegraphics[width=\textwidth]{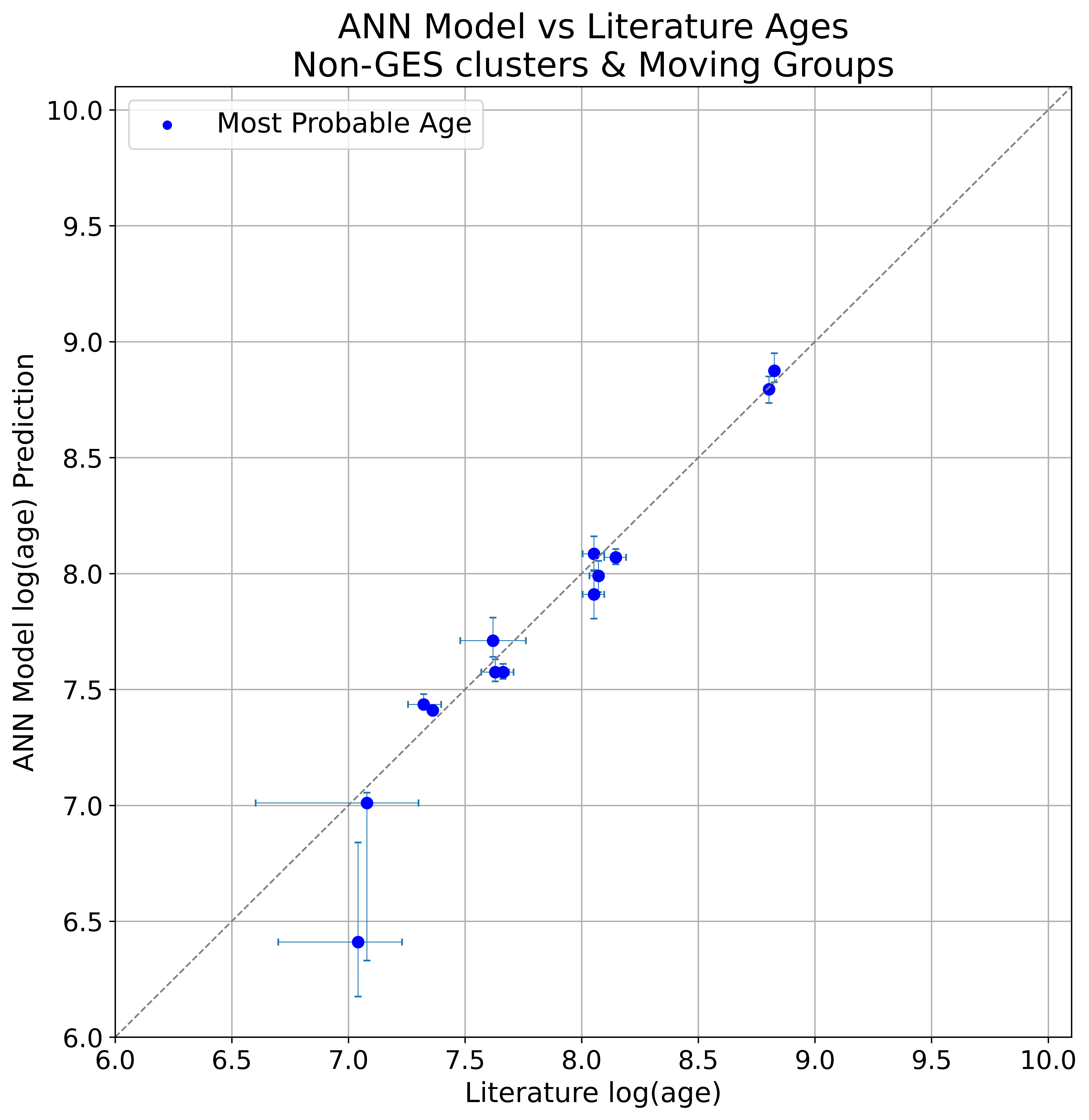}
    \end{minipage}
    \caption{\textit{Left:} Estimated most probable $\log$ (age/yr) from the ANN for 12 clusters/groups not used in training versus estimates using the {\sc eagles} model. \textit{Right:} The ANN model age estimates $\log$ (age/yr) versus reference ages taken from \citet{Rob2023}.}
    \label{fig:nonges}
\end{figure*}

The agreement between the two models and the literature ages for the two oldest clusters (the Hyades and Praesepe) is somewhat misleading.
In common with what was found in \cite{Rob2023}, the isochrones matched to the data by the ANN model are systematically poor fits  (see Fig. \ref{fig:nongesfull}). Both clusters show very low LiEW objects at the hottest \Teff, followed by a group of high-LiEW objects at slightly cooler \Teff, and then a significantly steeper decrease in LiEW with decreasing \Teff than the isochrones can fit (see Fig. \ref{fig:nongesfull}).
The slightly improved accuracy of the ANN age estimates is therefore serendipitous. These large systematic residuals may be related to the similar issue identified in M67 (seen in Figures \ref{fig:gammavel} and \ref{fig:m67haf10isochrones}), in which the isochrones fit by the model do not appear to appropriately follow the shape of the data.
In the 500-1000\,Myr range, there is however a dearth of training data with which to probe this further -- only NGC6633, with 17 members and a literature age of $617^{+60}_{-54}$ Myr.

\section{Discussion}\label{sec:discussion}

\subsection{Comparison with {\sc eagles}}\label{subsec:advantages}

In terms of similarities, the ANN model provides age estimates (both most probable and median) that are in good agreement with those provided by the original {\sc eagles} model and the literature ages for almost all clusters and moving groups up to $\sim$ 1 Gyr.
At older ages, a bimodality in LiEW within some individual clusters and differences in Li depletion patterns between clusters at the same age, suggest additional physical effects contribute and are not fully represented by the LiEW-\Teff-age relationship utilised by either model.

In terms of differences,
the more complex ANN model predicts dispersions that are either higher or lower compared with the rather simply formulated {\sc eagles} representation in different parts of the \Teff - $\log$(age) plane.
In older stars, these are split into predictions of larger dispersions than {\sc eagles} for warmer stars and smaller dispersions for cooler stars.
In younger stars, the dispersion predicted by the ANN model is larger than {\sc eagles} at all \Teff; this is probably because of a lack of an explicit \Teff-dependence in the {\sc eagles} dispersion model.
Where the dispersion is increased, this leads to larger error bars on the age estimate, and as such the influence of the choice of prior is stronger on the final age estimate for the star.
Another point of difference between the models is that the LiEW-\Teff isochrones from the ANN models evolve in a non-monotonic way at $\leq 10$ Myr (Fig \ref{fig:overlapisochrones}), whereas the {\sc eagles} LiEW predictions are constrained by the analytic model to always decrease with age.
This can result in bimodality in the age probability distributions (see Fig. \ref{fig:probsnonges}) in the youngest clusters and therefore differences in the most probable ages (despite often having quite similar median ages).

\subsection{Additional Flexibility Inherent to a Data-Driven Model}\label{disc:constraints}

A key advantage in adopting an ANN as the basis of a data-driven model is avoiding the more rigid constraints imposed by an analytical approach, allowing the discovery of features in the data and insights into important parameters that might otherwise be obscure. 
The ANN model exhibits different behaviour to the original {\sc eagles} model at young ages and in the intrinsic dispersion of LiEW at most ages, but has similar shortcomings when it comes to estimating the ages of older stars and clusters.

\subsubsection{The youngest stars}\label{youngeststars}

\begin{figure}
    \centering
    \subfigure{\includegraphics[width=0.4\textwidth]{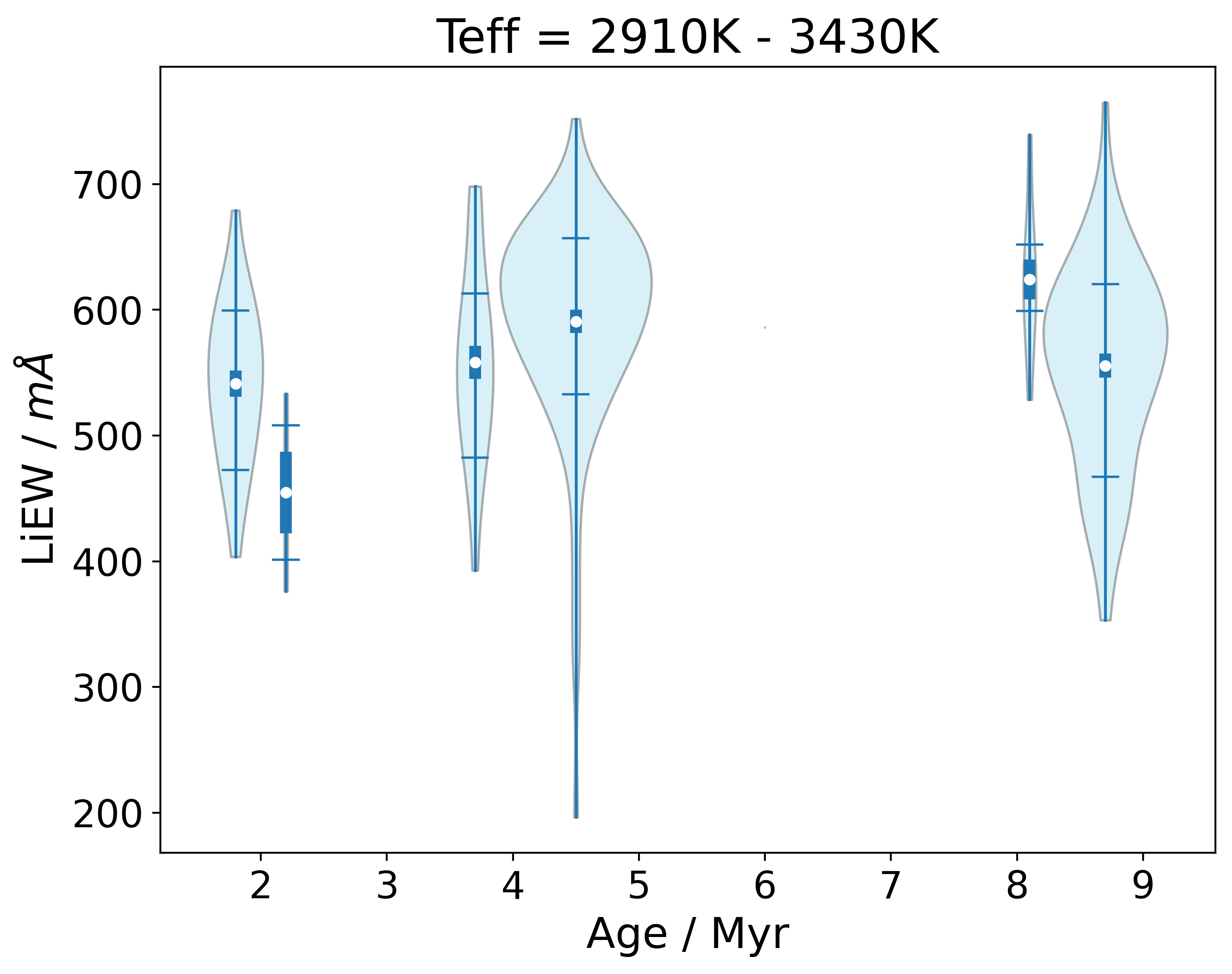}}
    \subfigure{\includegraphics[width=0.4\textwidth]{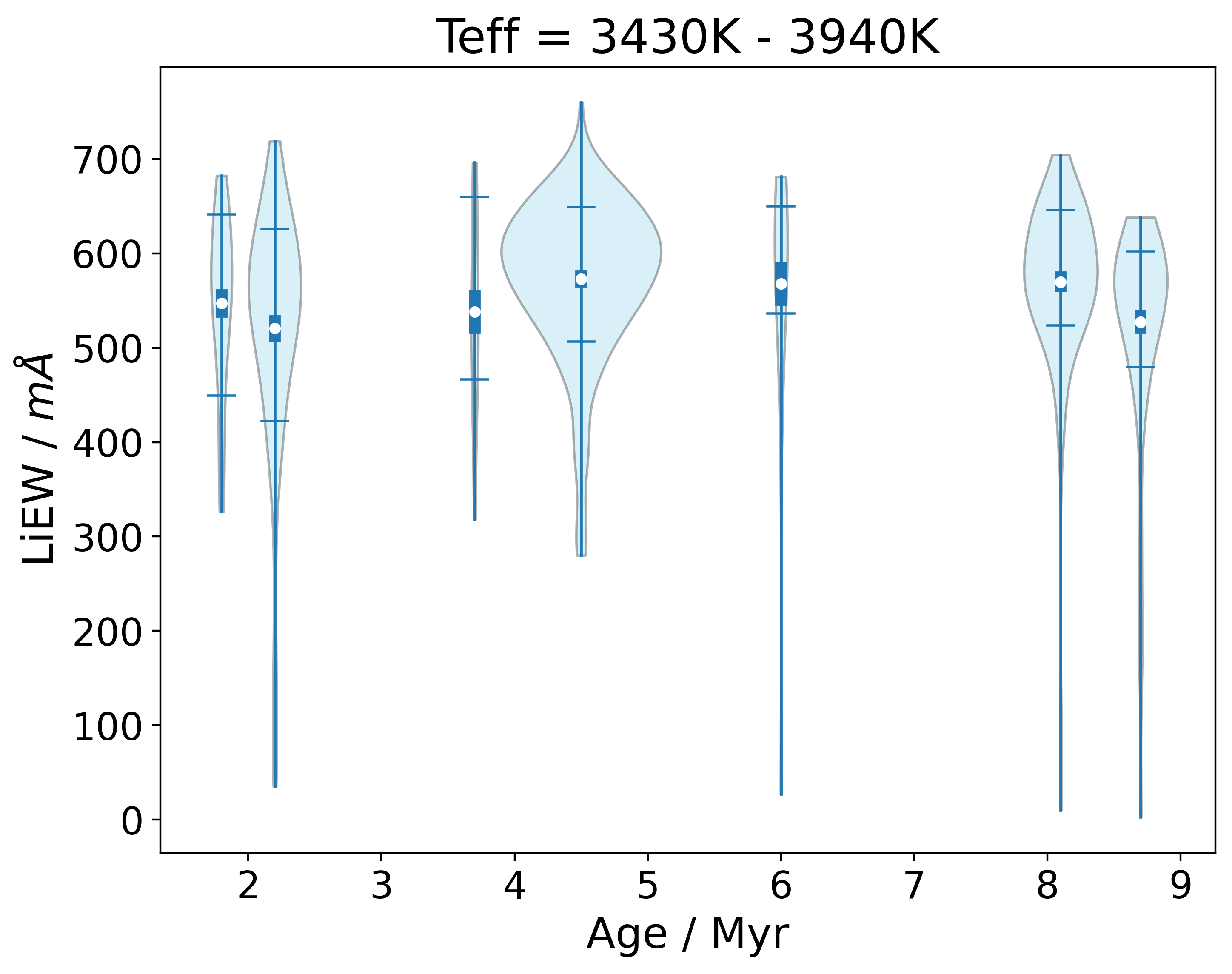}}
    \caption{Violin plots showing the distribution of LiEW values for clusters at ages < 10 Myr in the training dataset. \textit{Left:} stars 2910K $\leq$ \Teff < 3430K. \textit{Right:} stars 3430K $\leq$ \Teff < 3940K. The white point shows the mean LiEW of the stars in each cluster, and the upper and lower horizontal lines represent the 16th to the 84th percentile value of LiEW. The thick vertical line represents the standard error in the mean.}
    \label{fig:youngclusters}
\end{figure}

As noted in \S\ref{subsec:isochrones}, the LiEW predicted by the ANN model increases for the first $\sim 6$ Myr at \Teff $\lesssim$ 5500K (Fig.~\ref{fig:overlapisochrones}).
To investigate whether the non-monotonicity of the isochrone evolution could be a case of over-fitting or random noise, we have created histograms of LiEW for stars belonging to the youngest clusters in the relevant temperature range (See Fig. \ref{fig:youngclusters}). 
These reveal that the LiEW rise modelled by the ANN can be seen in the raw data at 3000-3430\,K but is much more marginal or absent at higher temperatures, and is in any case much smaller than the dispersion in LiEW at these ages.
However, if genuine, then this feature could not be reproduced by the LiEW-\Teff model adopted by {\sc eagles}, which declined monotonically with age and has been revealed by the extra flexibility of the ANN.

Whilst this effect appears statistically marginal and will have little effect on the age determination for individual stars, it is worth exploring whether there are physical reasons to expect an increase in LiEW over the first few Myr of evolution in a low-mass PMS object, not least because the opposite behaviour has been observed in other recent work \citep[][see below]{saad2024}. Most standard evolutionary models (those featuring only convective mixing - e.g. \citealt{siess2000,dotter2008,Baraffe2015}) predict little to no Li depletion over the first few Myr of the PMS and then accelerating depletion over a narrow \Teff range centred on stars with \Teff$\simeq 4000$ K.
More recent models including the effects of star spots and magnetic fields predict an even later onset of Li depletion starting at cooler temperatures \citep[e.g.,][]{Feiden2016, Somers2020}. All of these models predict that Li abundance decreases monotonically with age.

One possibility for increasing abundance within a star at early ages is via accretion.
To alter the abundance of a fully or largely convective star requires an increase in the entire mass of the star appropriately. 
For example, to increase abundance by 0.3 dex (roughly what is needed to increase the LiEW by 10 per cent at these abundances and temperatures) would require a doubling of the mass, with the accretion needing to be exclusively of volatile-depleted material (i.e, no H, He). 
Whilst this might be a possible explanation for some surprisingly Li-rich early G-star or F-stars with shallow convection zones (see \citet{ashwell2005,spina2015,Tognelli2021}) it seems implausible for fully-convective M-type PMS stars at 1--10 Myr, where accretion rates are generally $\leq10^{-8}$\Msun/year.

Assuming no early Li depletion, then another scenario that might lead to the ANN model predicting an increasing Li abundance between 1 and $\sim 10$ Myr is if the training clusters at 3-10 Myr had a higher initial Li abundance than those at immediately younger ages.
Using curves of growth from \citet{Franciosini2022}, with evolutionary modelling from \citet{Baraffe2015}, the observed increase in LiEW in these ages would require an increase in Li abundance of $\sim 0.2-0.4$ dex.
There is no reason to expect this; all the young ($<$ 100 Myr) GES clusters share very similar chemical abundances \citep{randich2022} and the non-monotonic behaviour seems restricted only to cooler temperatures.

\begin{figure}
    \centering
    \includegraphics[width=\linewidth]{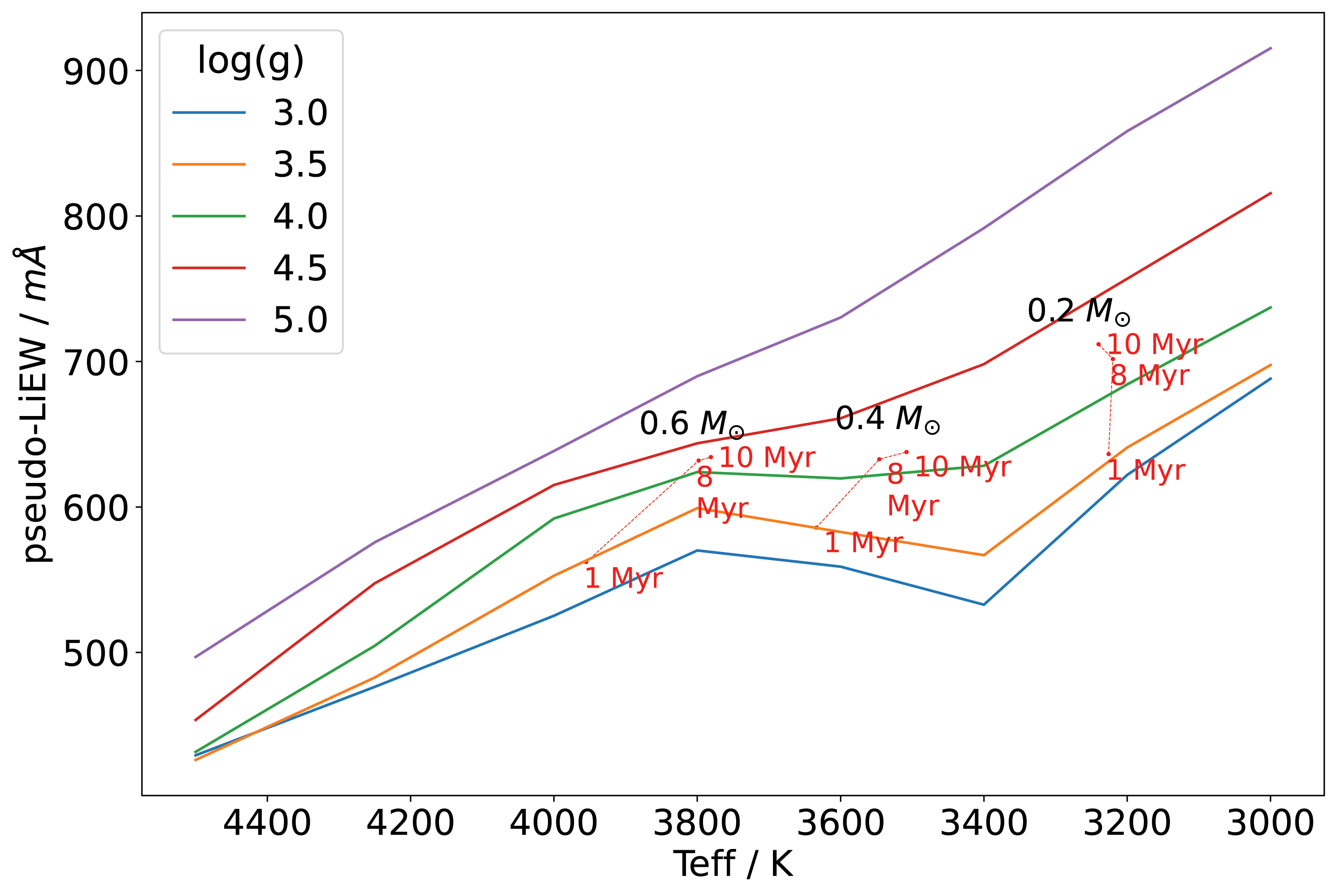}
    \caption{L.T.E. curves of growth for LiEW at various $\log (g)$ values for a fixed $A_{\rm Li} = 3.2$, taken from \citet{Franciosini2022}. Evolutionary tracks are shown for 0.2\Msun, 0.4\Msun and 0.6\Msun stars at 1, 8, and 10 Myrs, demonstrating an expected increase in LiEW if their initial Li remains undepleted. \citep{Baraffe2015}.}
    \label{fig:cog}
\end{figure}

Instead, it could be that the relationship between LiEW and Li abundance changes at early ages, or some observational effect leads to an LiEW underestimation in the very youngest objects. Stars descend Hayashi tracks at almost constant \Teff in this age and mass range \citep{Hayashi61}. Their rapidly decreasing radii lead to increasing surface gravities and may lead to an increased equivalent width for a fixed abundance in the saturated Li~I line due to pressure broadening. To investigate this, Fig. \ref{fig:cog} shows theoretical curves of growth plotted at a fixed Li abundance (roughly the value expected for undepleted stars), for a series of different surface gravities \citep[from][]{Franciosini2022}.
Superimposed are tracks for contracting PMS stars of different mass at ages from 1 to 10 Myr \citep[from][]{Baraffe2015}. 
The curves of growth predict LiEW should become stronger by $\sim 10$ per cent between 1 and 10 Myr (but mostly between 1 and 8 Myr) in rough agreement with the ANN model predictions over this age range.

Another observational bias could be due to the veiling effect of an accretion disc continuum around the youngest stars. This can cause a large decrease in LiEW for heavily veiled objects, decreasing the average LiEW at that age and increasing the observed dispersion. Then, since accretion discs disperse on an e-folding time of $\sim$ 2-3 Myr \citep{Fedele2010} we expect, as Li depletion has not begun, that the {\it average} LiEW would rise and the dispersion would fall over the course of the first 10 Myr. This might be apparent in the data as a `tail' of accreting stars with low LiEW, that then disappears along with the accretion discs.

\begin{figure}
    \centering
    \includegraphics[width = \linewidth]{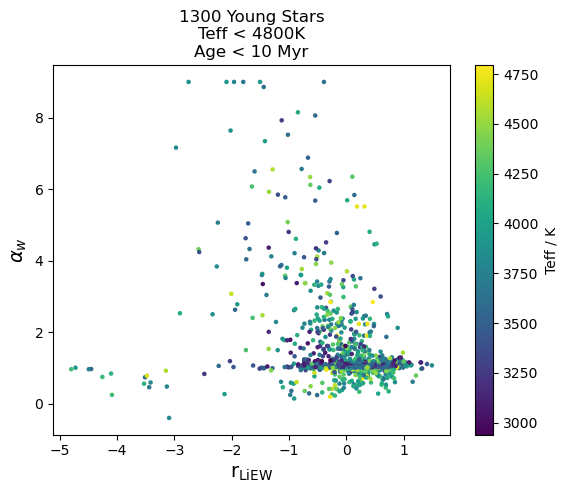}
    \caption{$\alpha_{w}$ vs ${\rm r_{LiEW}}$, the standardised residual from model predictions, normalised by the observed errors and predicted dispersion. The `tail' of negative ${\rm r_{LiEW}}$ stars indicate an observed LiEW that is far below the prediction. The plot shows a weak anti-correlation with activity, indicating that there may be a veiling effect from accretion on the LiEW observed in these youngest stars.}
    \label{fig:youngaccretion}
\end{figure}

Figure \ref{fig:youngaccretion} shows some evidence of this effect, where a standardised residual LiEW discrepancy from the ANN model, defined by
\begin{equation}
    \label{eq:standLiew}
    {\rm r_{LiEW}} = \frac{{\rm LiEW}_{\rm obs} - {\rm LiEW}_{\rm pred}}{\sqrt{\sigma_s^{2} + 
    \Delta^{2}}}\,
\end{equation}
with LiEW$_{\rm obs}$ the observed LiEW, LiEW$_{\rm pred}$ the LiEW predicted by the ANN and the other symbols as defined in Eqn.~1, is correlated with $\alpha_w$, an index that measures how much excess H$\alpha$ emission is found in any extended wings of the Balmer line in the GES spectra \citep{damiani2014}. 
This was calculated from the GES iDR6 spectra.
The strength of H$\alpha$ emission, particularly in the wings at $>270$\,km/s, correlates with veiling and is a well-used means of identifying strong accretion \citep{muzerolle98,alencarbatalha,whitebasri}.
There is indeed a weak correlation between ${\rm r_{LiEW}}$ and $\alpha_w$ and the asymmetry in the distribution of ${\rm r_{LiEW}}$ for cool stars (shown in the righthand panel of Fig.~\ref{fig:dispersions}) appears to be attributable to likely accreting objects with $\alpha_w>2$.
This correlation with $\alpha_w$ strongly implicates an accretion-related continuum in reducing the average LiEW in groups of young, cool stars and suggests an improvement to age estimation could be to include empirical accretion-indicators (such as $\alpha_w$) as features in the ANN model (see \S\ref{sec:expansion}).

Our results for the youngest stars disagree with those of \citet{saad2024}, who found that LiEW fell by 10-15 per cent in the first $\sim 3$ Myr, remained roughly constant over 3-10 Myr, and then continued to fall thereafter in a much larger sample of cool young PMS star candidates with isochronal ages.
Whilst our sample of stars is much smaller and restricted to relatively few clusters, the trend found by Saad et al. could not be reproduced in our data by reassigning the ages assumed for these clusters within their likely error bounds (see Fig.~\ref{fig:youngclusters}).
It is worth noting that the spectral resolution of the GES data used here is an order of magnitude higher than the LAMOST and SDSS V data used by \citet{saad2024}, and their LiEW determination has not accounted for the significant blending of the Li line with other metal lines and molecular features that become increasingly important in cooler stars.
This may account for the significantly lower peak LiEW found by Saad et al. ($\sim 450$\,m\AA\ compared with $\sim 570$\,m\AA\ in our sample) and perhaps the different early PMS evolution of LiEW.

\subsubsection{Dispersion and Single Star Age Estimation}\label{disp}

The ANN and original {\sc eagles} model differ in their evaluation of the intrinsic dispersion of LiEW as function of \Teff and age.
The {\sc eagles} dispersion calculation includes two main terms, a decreasing exponential with time, and another that depends on the derivative of LiEW with respect to time. There is no explicit \Teff dependence.
As a result, in areas where LiEW is not rapidly decreasing, the {\sc eagles} dispersion appears to not be complex enough to fully describe the dispersion across all areas of the age - \Teff - plane.
As an example, older stars ($\gtrsim 1$\,Gyr) in the training dataset show a smaller dispersion in cooler regions, where the dataset is largely made up of entirely Li-depleted field stars, and a larger dispersion in warmer regions where cluster members are the dominant data source.
As shown in Fig. \ref{fig:dispersiondifferences}, the lack of \Teff dependence in {\sc eagles} leads to underestimates of the dispersion compared to the ANN model in warmer regions, and overestimates in the cooler regions.

In younger stars, the difference approximates to the additional epistemic uncertainty in the ANN model contributed by the model's variation across the 2000 iterations.
The variation in dispersion in the Li dip area is a result of the differing shape of the isochrones between the two models.
The sharp uptick of the Li dip isochrones in {\sc eagles} is a result of the hard cut-off value in the functional fit.
As this fit is not constrained by a cut-off in the ANN model, the isochrones can take a shape that better represents the data for stars at these ages, and as such the isochrones are more `rounded' (Fig. \ref{fig:isochrones}).
Through the main area of the dip (between 10-20 Myrs), {\sc eagles} must inflate the dispersion to account for the poorer fit to the data from the sharp dip.
At ages > 20Myrs, at the bottom of the Li dip, the {\sc eagles} model again underestimates the dispersion as it predicts much lower LiEW than the ANN model, again due to the 'pointed' nature of the isochrones.
Fits to clusters in this age, LiEW and \Teff range show that the dispersion appears to be better described by the ANN model (See Fig. \ref{fig:gammavel}).

As a result, the intrinsic dispersion around the best-fitting LiEW-age-\Teff relation is more complex than assumed in {\sc eagles}. 
The additional freedom inherent to a data-driven model has allowed a more faithful reproduction of this dispersion, in particular, describing \Teff dependencies that are not explicitly allowed in the {\sc eagles} model and avoiding the need to introduce ad-hoc \Teff-dependent dispersion among ZAMS K-type stars.

\begin{figure}
    \centering
    \begin{minipage}{.5\linewidth}
    \centering
        \includegraphics[width=\linewidth]{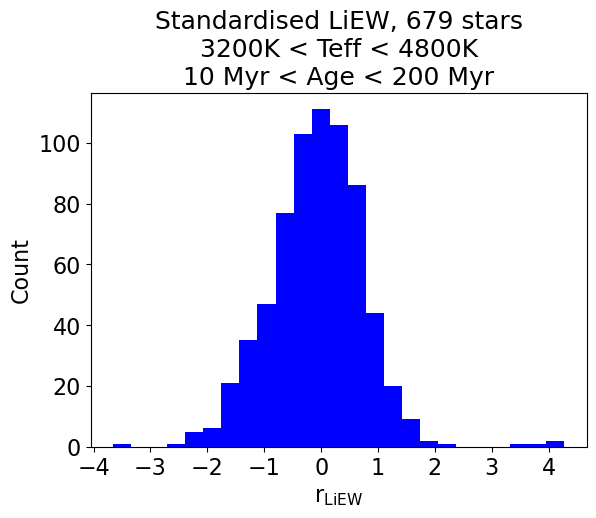}
    \end{minipage}%
    \begin{minipage}{.5\linewidth}
        \centering
        \includegraphics[width=\linewidth]{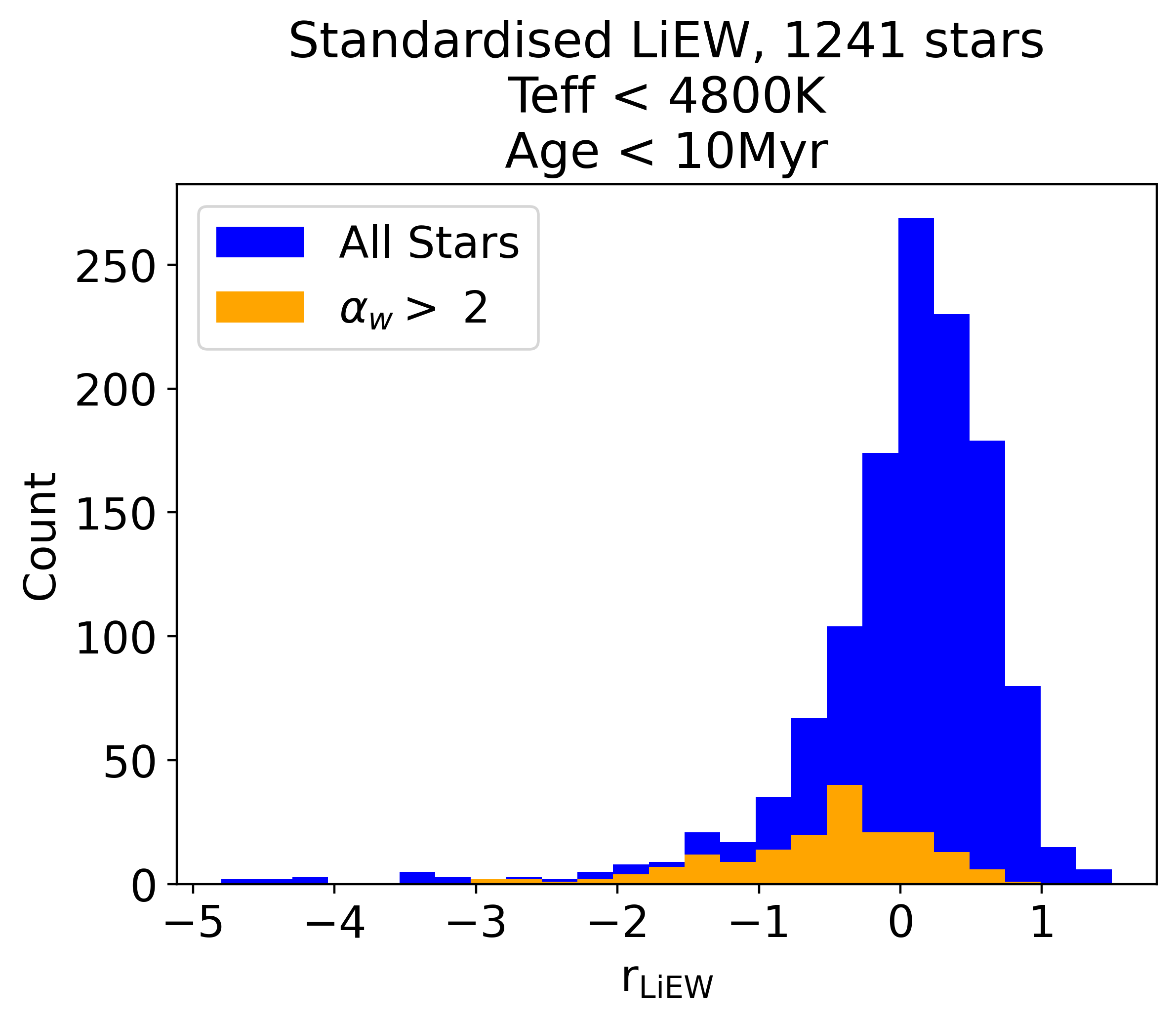}
    \end{minipage}
    \caption{Histograms of standardised residual ${\rm r_{LiEW}}$. \textit{Left:} Stars at intermediate temperatures of 3200K-4800K and age between 10 Myr - 200 Myr. These stars show a near-Gaussian dispersion about the mean. \textit{Right:} Younger stars of age < 10 Myr, with temperatures below 4800K. These stars show a distinct skew in the Gaussian shape, with a tail of observed LiEW well below prediction by the model. Stars with a Gaia $\alpha_{w} \geq 2$ are shown in orange.}
    \label{fig:dispersions}
\end{figure}

The intrinsic dispersion in LiEW is assumed to have a normal distribution in both models.
Comparisons of the residuals in single-star estimates from the validation dataset using ${\rm r_{LiEW}}$ (Eqn.~\ref{eq:standLiew}) at young, intermediate, and old ages in a range of \Teff bins show that the dispersion is approximately Gaussian in the vast majority of the LiEW - \Teff plane (Fig. \ref{fig:dispersions}), and so this assumption seems largely satisfactory.
Although in one area, at \Teff $< 4800$\,K and ages $\leq 10$ Myr, there is evidence for a non-Gaussian `tail' of stars in the distribution of ${\rm r_{LiEW}}$. These stars, seen in Fig.\ref{fig:dispersions}, have an observed LiEW much lower than predicted by the model, and the distribution of stars suggests that the intrinsic dispersions are more likely to become skewed at the boundary of lower age. As discussed in \S\ref{youngeststars}, some of this skew in the form of the dispersion may be due to veiling from accretion, shown by the orange histogram bins in Fig. \ref{fig:dispersions}.

\subsubsection{Differences in Age Estimation of Clusters}

The discrepancies between the ANN model estimates and literature ages in some older clusters (> 1 Gyr, see \S\ref{olderlusters}) is similar to deficiencies identified in the original {\sc eagles} model.
In the training data, clusters of the same age and of similar metallicity show differing Li depletion for stars of the same \Teff.
This phenomenon was similarly found in earlier work by \citet{sestito2005} and \citet{randich2010}.
Even with the additional flexibility of the ANN model, it cannot produce isochrones to fit all these data simultaneously, and so the modelled intrinsic dispersion is inflated.
This means that single star age estimates will have appropriately large uncertainties, even with precise measurements.
For clusters, the results are likely to give misleadingly small age error bars, with a `mean isochrone' that may not be a good fit to the cluster (See \S\ref{olderlusters}, Figs. \ref{fig:gammavel}, \ref{fig:m67haf10isochrones}).
Given that the ANN model has effectively removed any restriction on the form of the LiEW - \Teff - age relationship, and greatly increased the complexity with which this relationship is fitted, it appears that these deficiencies, common to both models, are likely astrophysical issues that require additional parameters to solve (See \S\ref{sec:expansion}).
It is still unclear which parameters may reduce the issues, and this problem is still unsolved.

\begin{figure}
    \centering
    \includegraphics[width = \linewidth]{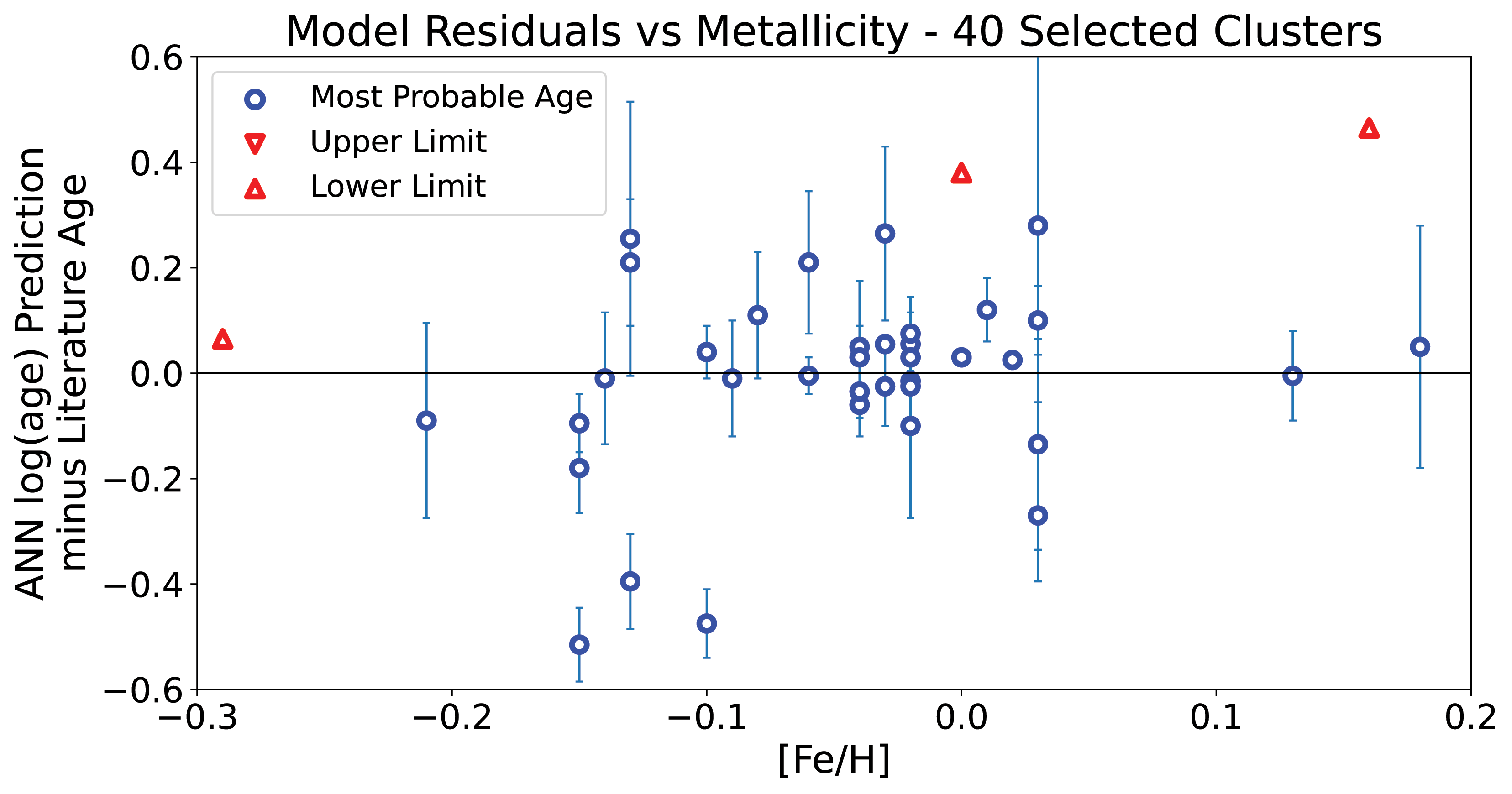}
    \caption{ANN predicted age residuals vs Metallicity.}
    \label{fig:metal}
\end{figure}

Similarly to tests of the {\sc eagles} model, the ANN-model residuals for each cluster to the literature ages were compared with the mean cluster [Fe/H] metallicity taken from GES \citep{randich2022}. 
These residuals show no apparent trend with metallicity (Fig. \ref{fig:metal}), appearing to rule out this parameter as an explanation for the this age estimate problem.

Within the Lithium dip (\Teff < 4000K and 7 < $\log$ (age/yr) < 7.8), the ANN model shows some variation in isochrone shape compared with {\sc eagles} (Fig. \ref{fig:gammavel}).
As a result, there is some evidence that the additional flexibility of the ANN leads to slightly improved results and this can be seen in better-looking fits for Gamma Velorum and the much better age estimation for Blanco 1 in Fig.\ref{fig:gammavel}. The lack of constraints in the ANN model also allow for greater complexity in the rate of change of LiEW, as seen in the sharp variations in the spacing of isochrones between 20 Myrs and 300 Myrs at intermediate \Teff in Fig.~\ref{fig:isochrones}.
The more rapid rate of change in LiEW in some areas of the LiEW-\Teff plane leads to increased age uncertainties (Table~1 and Fig.\ref{fig:nonges}). An example of this is AB Doradus, where age uncertainties from the ANN model are approximately doubled from {\sc eagles} because there is less separation in LiEW between isochrones with a fixed separation in age (See Appendix \ref{sec:appendixfits}).
Increased age uncertainties are also expected in the youngest clusters because non-monotonicity in the LiEW-age relation leads to a bimodal age probability distribution as seen in Fig. \ref{fig:probsnonges}).
This represents another clear indication for the need to include further parameters in the model in order to break the degeneracy between older stars with lower LiEW due to the commencement of Li depletion and younger, undepleted stars that have a similar LiEW, due to veiling, lower surface gravity, or other astrophysical effects.

\subsection{Development Areas}

\subsubsection{Modelling the Dispersion}

The intrinsic dispersion in LiEW is a proxy for the unknown parameters driving the LiEW-\Teff relationship aside from age.
The aim of models such as those discussed in this paper is to minimise this dispersion (see \S\ref{sec:expansion}). The assumed form of the intrinsic dispersion distribution is unimportant if it is smaller than observational errors. But if not, then it may be important to provide a more realistic description of the distribution of residuals.
As shown discussed in \S\ref{disp} and shown in Fig.~\ref{fig:dispersiondifferences} in some areas of the \Teff-age plane, the intrinsic dispersion may not be Gaussian, and so introducing more flexible functions, such as a skewed Gaussian or beta distribution, might be considered, at the expense of introducing another free empirical parameter.

\subsubsection{Expansion}\label{sec:expansion}

The ANN model has made some demonstrable improvements over {\sc eagles} in terms of the fidelity of LiEW predictions and a more accurate modelling of dispersion. However, there are still shortcomings and it is clear from \S\ref{disc:constraints} that there are still improvements that could be made if the causes of intrinsic dispersion can be identified and incorporated into the model.

The highly-adaptable nature of machine learning modelling, particularly ANNs, leads to the possibility of expansion to use other physical variables to improve the accuracy of LiEW prediction and hence age estimation, or to include other age-sensitive features.
An expansion in this way would be difficult with the empirical analytic methods used in the original {\sc eagles} model, but requires much simpler changes within an ANN model. Including these additional features is beyond the scope of this initial paper but the following should be considered in future work:
\begin{itemize}
    \item Gravity indicators: surface gravity measurements are sensitive to age and likely have an influence on Li line formation.
    The inclusion of spectroscopic $\log$(g) or gravity-sensitive indices may tighten age constraints for contracting PMS stars, or break the degeneracy in LiEW vs age, as discussed in \S\ref{youngeststars}.
    \item Accretion: In \S\ref{youngeststars} it was shown that there is a correlation between low-LiEW objects and strong accretion activity, presumably as a result of veiling.
    Accretion is also a crude age indicator in the youngest PMS stars ($\lesssim$10 Myr).
    \item Rotation: the correlation between LiEW and rotation is well established in young PMS and ZAMS stars and much of the intrinsic dispersion in LiEW at these ages is correlated with rotation \citep{Bouvier2018, Jeffries2021}.
    Inclusion of rotation rates (either periods or projected rotation velocities) could improve LiEW predictions. 
    In addition, rotation is itself age-sensitive (gyrochronology) though most of that sensitivity is for post-ZAMS stars.
    \item Magnetic activity: activity is strongly correlated with, and may be degenerate with, or serve instead of, rotation, but may also have a direct influence on LiEW through changes to the effective temperature and temperature structure of the photosphere \citep{King2010,Barradoactivity2016}.
    Activity is also age-dependent, though similarly to rotation, largely for post-ZAMS stars.
    \item Metallicity: differences in composition may have an effect on LiEW \citep{cummings2017} as higher metallicity PMS stars should deplete Li faster, due to their deeper convective zones \citep{Pinsonneault1997,piau2002,Tognelli2012,Tognelli2021}, and possibly in the Main Sequence phase too \citep{chaboyer1995}.
    In addition, the Li I 6708 $\AA$ line is usually blended with a weak Fe I 6707.4 $\AA$ line. The contribution of the blend to the observed LiEW ought to become important when considering older clusters with smaller LiEW values. However, the evidence for any metallicity effects is weak or absent (see Fig.~\ref{fig:metal}).
    \item Galactocentric radius: Currently, the ANN model implicitly assumes that clusters of the same age in different parts of the galaxy had a common initial Li abundance and a common Li depletion history.
    This may not be a valid assumption \citep{Romano2021,magrini2023}, and we may expect some dependence of the initial Li on galactocentric radius alongside metallicity. The current dataset has a limited range of metallicity and galactocentric radius for older clusters, but a much smaller range in the younger clusters and that should be borne in mind if applying the model to stars or clusters outside these ranges.
\end{itemize}

\subsubsection{Expanding the Training Set}\label{sec:expandtraining}

The dataset used in model training in this work could be expanded in order to provide better coverage of the LiEW - \Teff - Age (and other parameters) space and improve the model fitting and prediction as follows.

\begin{itemize}
    \item In several areas of the parameter space, particularly among older stars, the observational errors become similar in magnitude to the intrinsic dispersion of LiEW.
    This means the modelling of the intrinsic dispersion, or its resolution in terms of other parameters, becomes highly contingent on the accuracy and distribution of observational errors.
    Improving the precision of training data in these areas of parameter space could improve the model.
    \item Several areas of the parameter space also have epistemic uncertainties that are a significant component of the uncertainties in the LiEW prediction due to a sparsity of data.
    Greater data coverage in these areas would reduce epistemic uncertainties and improve prediction.
    \item At ages below 10 Myr, and between 300 Myr - 1 Gyr, there are relatively few clusters in the training data.
    The former issue hampers attempts to accurately calibrate the behaviour of LiEW at the youngest ages, particularly in confirming the trend discussed in \S\ref{disc:constraints} that LiEW initially increases, then decreases during this period.
    The latter issue hampers our attempts to understand the factors leading to the poor model fits in the Hyades and Praesepe, and whether this is connected to metallicity or to other parameters.
    \item The dataset is also limited in the range of additional parameters that might be included as a part of any investigation into the minimisation of the intrinsic dispersion.
    In particular, the range of metallicities for younger clusters is very narrow, hampering any attempt to pin down the influence of composition on PMS Li depletion.
\end{itemize}
A big advantage of the training dataset used in this work is the homogeneity of using data solely from GES.
Additional training data would require careful standardisation and cross-calibration with GES in terms of LiEW analysis and \Teff estimation.
Two possible sources of such data are the upcoming WEAVE \citep{Dalton2012} and 4MOST \citep{dejong2019} spectroscopic surveys, both of which have programmes planning to observe and analyse large numbers of star clusters in a homogeneous way and at a resolving power capable of precisely measuring the strength of the Li {\sc{i}} 6708\AA\ line.

\section{Conclusions}\label{sec:conclusions}

We have trained an artificial neural network (ANN) model, similar to the analytical {\sc eagles} model of \citep{Rob2023}, using the same dataset of stars from 52 open clusters, taken from Gaia EDR3 and the Gaia-ESO Survey.
This ANN model predicts the equivalent width of the Li~{\sc i}~6708\AA\ line (LiEW, and its dispersion) and is used to produce age estimates and probability distributions for stars and clusters with Li and \Teff measurements.
The additional flexibility provided by the data-driven approach has provided improvements in some respects but in others it has not, highlighting known, suspected or unknown observational and astrophysical effects that are not yet incorporated into the model and which currently inflate the predicted LiEW dispersions and consequent age uncertainties.
\begin{itemize} 
    \item The model's inherent added flexibility has found new features in the LiEW-age-\Teff relationship, including a possible increase in LiEW for cool PMS stars at ages from 1--6\,Myr.
    \item The increased flexibility has also led to more realistic modelling of the intrinsic dispersion in LiEW, replacing the ad-hoc dispersion that was included in {\sc eagles}. This has led to marginal improvements in age estimate residuals for the training and validation clusters.
    \item The model still gives poor age discrimination at ages $> 1$\,Gyr. The predicted LiEW-\Teff isochrones are systematically poor fits to the data in some older clusters, at any age.
    \item We have confirmed then that adding more flexibility to the functional form of the fit is not sufficient to fully describe the relationship between LiEW, \Teff and age.
    Further astrophysical parameters are required to fully constrain the dispersion in the fits.
    \item The ANN method more easily allows for this expansion, which is the next step, alongside expanding the training dataset, and better representing dispersion with a more flexible functional form.
\end{itemize}
The ANN-model predictions and age estimation methodology and model are provided as an updated version 2.0 of the {\sc eagles} software and described in Appendix \ref{code}.

\section*{Acknowledgements}

We thank Laura Magrini for her useful comments and suggestions.
GW acknowledges receipt of an STFC postgraduate studentship.
This work is based on data products from observations made with ESO Telescopes at the La Silla Paranal Observatory under programme IDs 188.B-3002, 193.B-0936 and 197.B-1074.
These data products have been processed by the Cambridge Astronomy Survey Unit (CASU) at the Institute of Astronomy, University of Cambridge, and by the FLAMES/UVES reduction team at INAF/Osservatorio Astrofisico di Arcetri. 
These data have been obtained from the Gaia-ESO Survey Data Archive, prepared and hosted by the Wide Field Astronomy Unit, Institute for Astronomy, University of Edinburgh, which is funded by the UK Science and Technology Facilities Council.

\section*{Data Availability}

The training and testing datasets were obtained from \citet{Rob2023}.
The reduced stacked spectra underlying that paper can be obtained via the European Southern Observatory (ESO) archive and is identified as the ‘Phase 3 release’ of Gaia-ESO spectroscopic survey DR4.
The full catalogue of stellar parameters derived from these spectra by the various GES working groups is also available in the ‘Phase 3 release’ of Gaia-ESO spectroscopic survey DR5. 
Raw data can also be obtained from the ESO archive.

\bibliographystyle{mnras}
\bibliography{refs} 

\appendix

\section{EAGLES v2.0 Code}\label{code}

The model as described in Section \ref{sec:training} has been packaged into the updated {\sc python} code as Version 2.0 of {\sc eagles}, available at \url{https://github.com/robdjeff/eagles}.
This code works in the same manner as Version 1.0, as a command-line driven script taking an ascii input file of \Teff, blending-corrected LiEW, observed error in LiEW and additional \Teff uncertainty, for one or more stars and returns Bayesian estimates for their age using a prior flat in either age or $\log$(age).
Input stars can be treated either as individuals or a coeval cluster.
Outputs include most probable age, median age, an asymmetric 68\% confidence interval, and the full posterior age probability distribution.

The updated Version 2.0 includes a flag to utilise the ANN model described in this paper, which automatically sets the upper and lower age limits, and the lowest age at which likelihood is calculated, to the recommended values for this model.
Scripts are also included to generate isochrones and a grid of estimated age as a function of LiEW and \Teff for a given level of uncertainty in LiEW.

\section{Custom Loss Function}\label{sec:appendixloss}

As described in \S\ref{subsec:architecture}, the compiled ANN model makes use of a custom loss function to evaluate the quality of the model's fit to the training data in predicting LiEW and its intrinsic dispersion $\sigma$LiEW.
The negative log likelihood is given by
\begin{equation}
    \text{NLL} = \mathlarger{\mathlarger{\sum}}\frac{({\rm LiEW}_{\rm obs} - {\rm LiEW}_{\rm pred})^{2}}{2(\sigma{\rm LiEW}^{2} + \Delta^{2})}
    + \ln\left(\frac{\sigma{\rm LiEW}^{2}+\Delta^{2}}{2}\right)\
\end{equation}
Where LiEW$_{\rm obs}$ and $\Delta$ are the observed LiEW and its error bar, 
whilst LiEW$_{\rm pred}$, and $\sigma$LiEW are the predicted LiEW and its intrinsic dispersion that are the ``targets" for the ANN model.

\section{Supplementary Plots and Data}\label{sec:appendixfits}

Table \ref{table:data} below shows a sample of the training and validation data (see \S\ref{dataset}) used in the ANN model fitting \citep{Rob2023}. The full version of the table is available online as supplementary material.

\begin{figure*}
    \centering
    \subfigure[]{\includegraphics[width=0.27\textwidth]{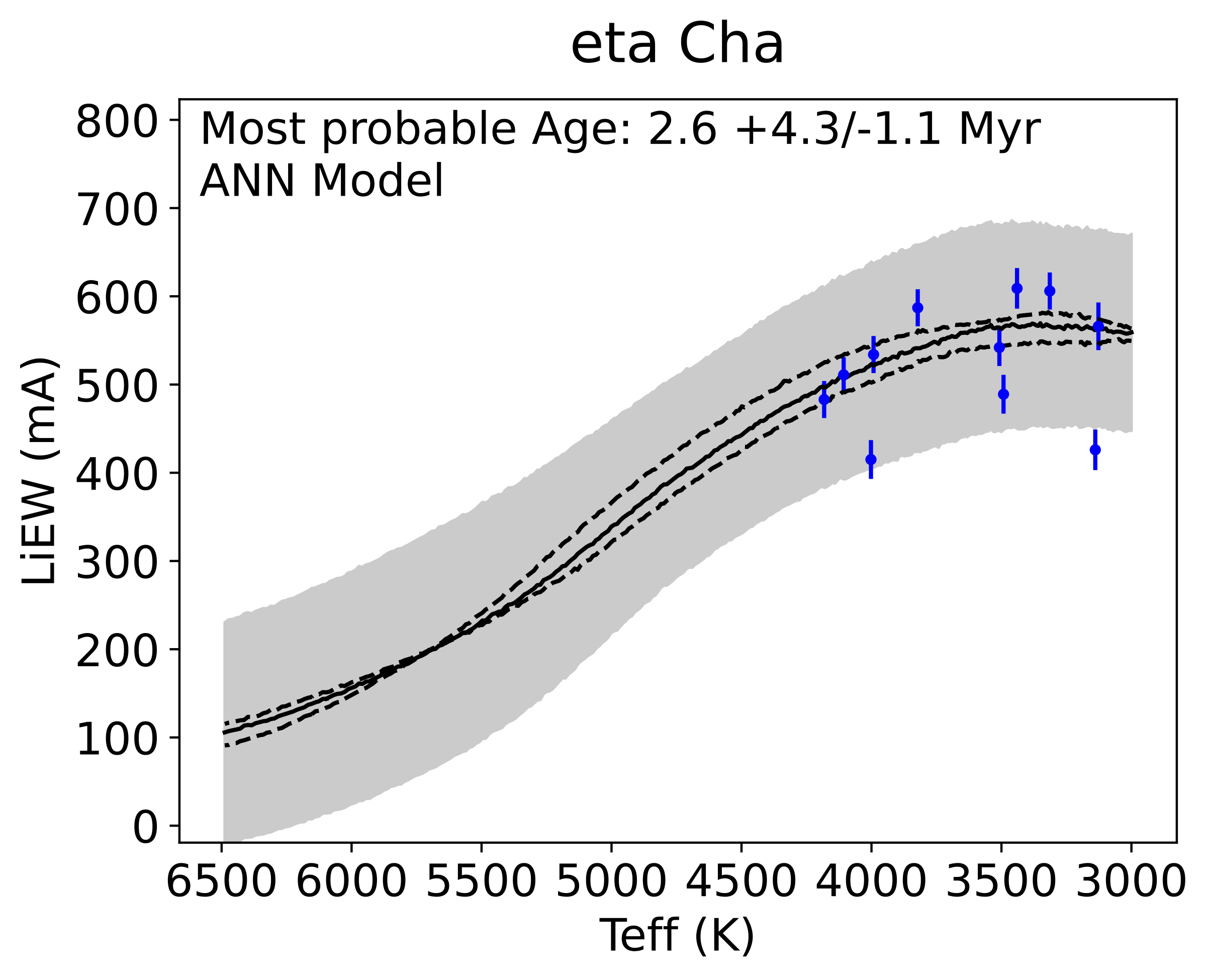}}
    \subfigure[]{\includegraphics[width=0.27\textwidth]{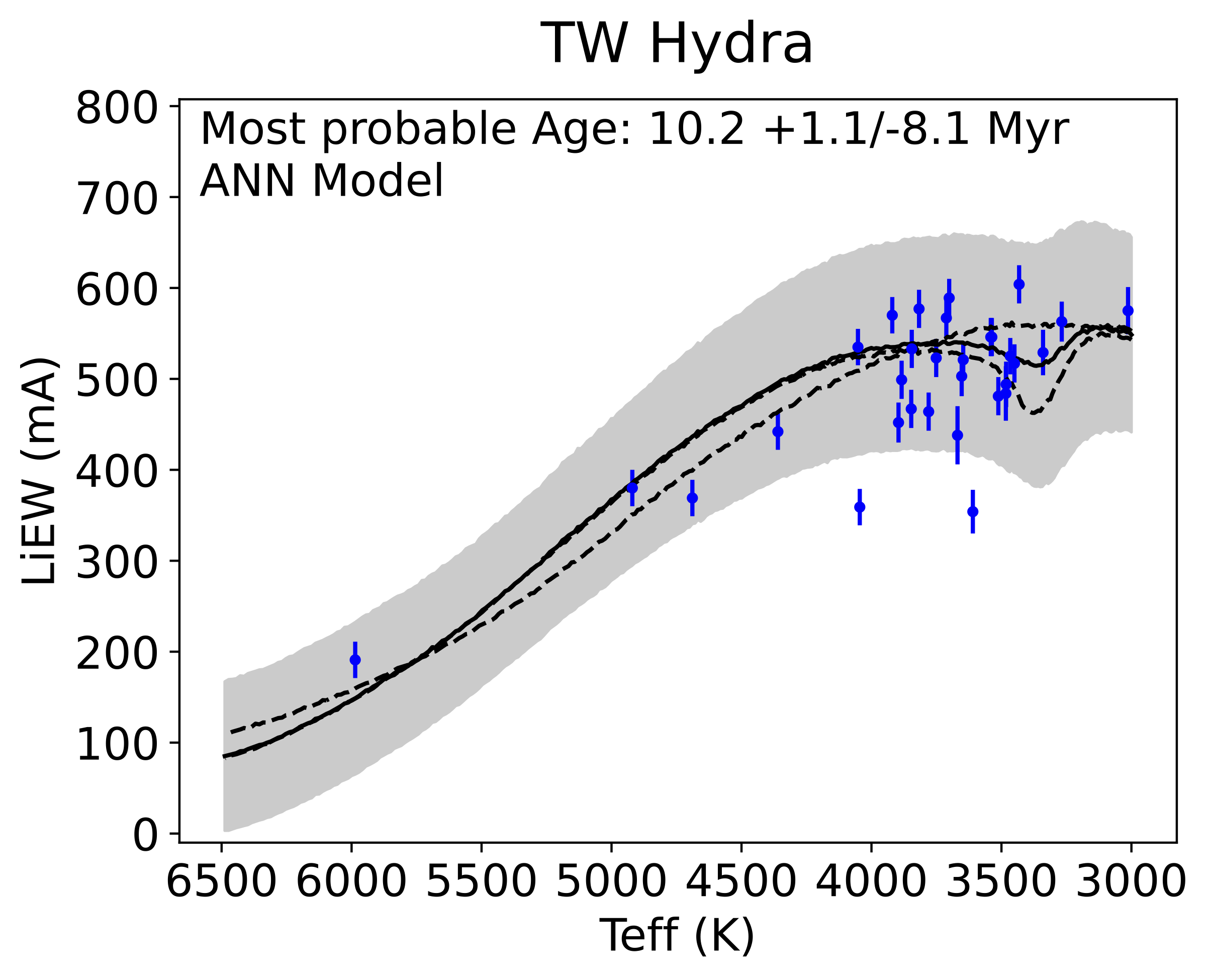}}
    \subfigure[]{\includegraphics[width=0.27\textwidth]{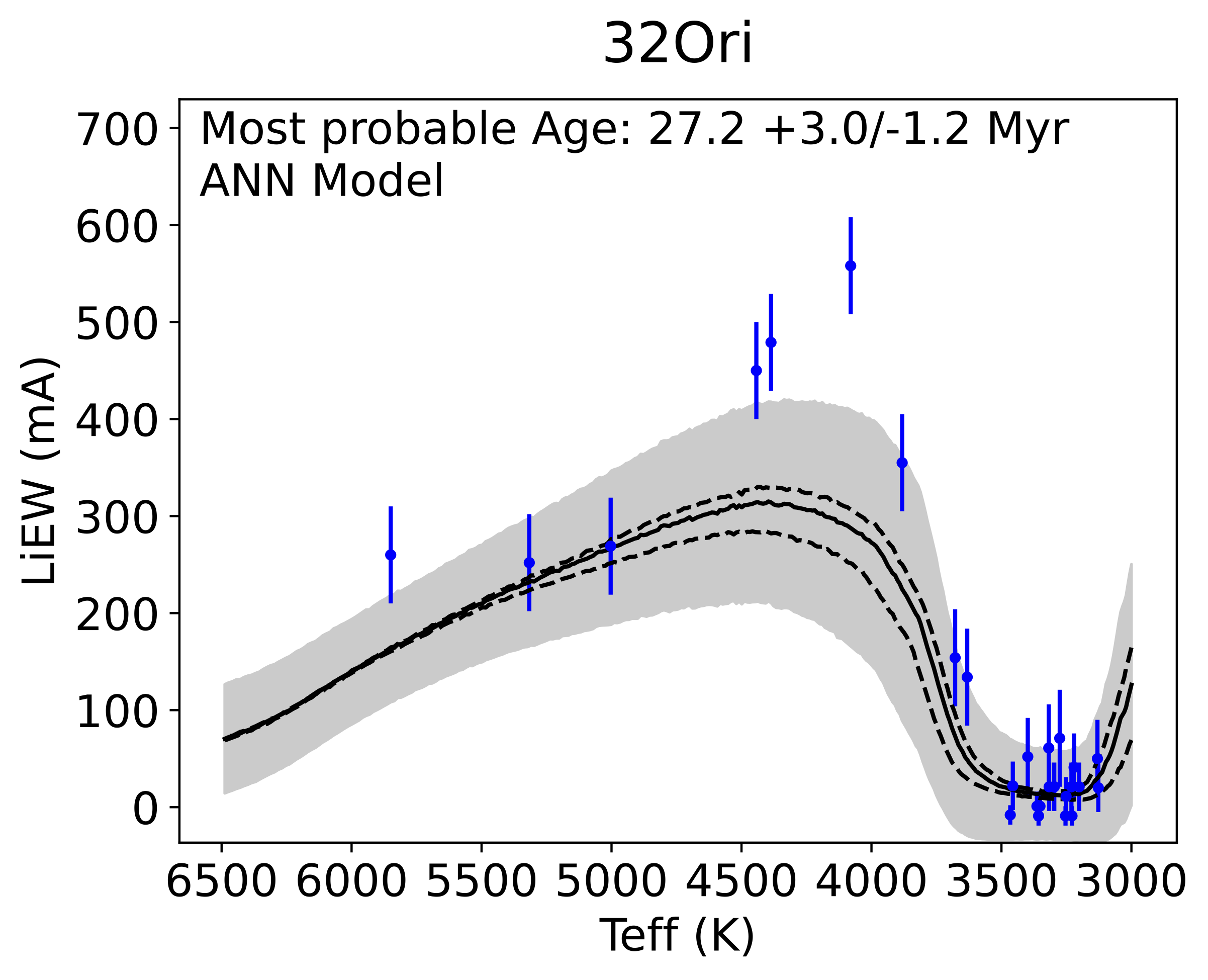}}
    \subfigure[]{\includegraphics[width=0.27\textwidth]{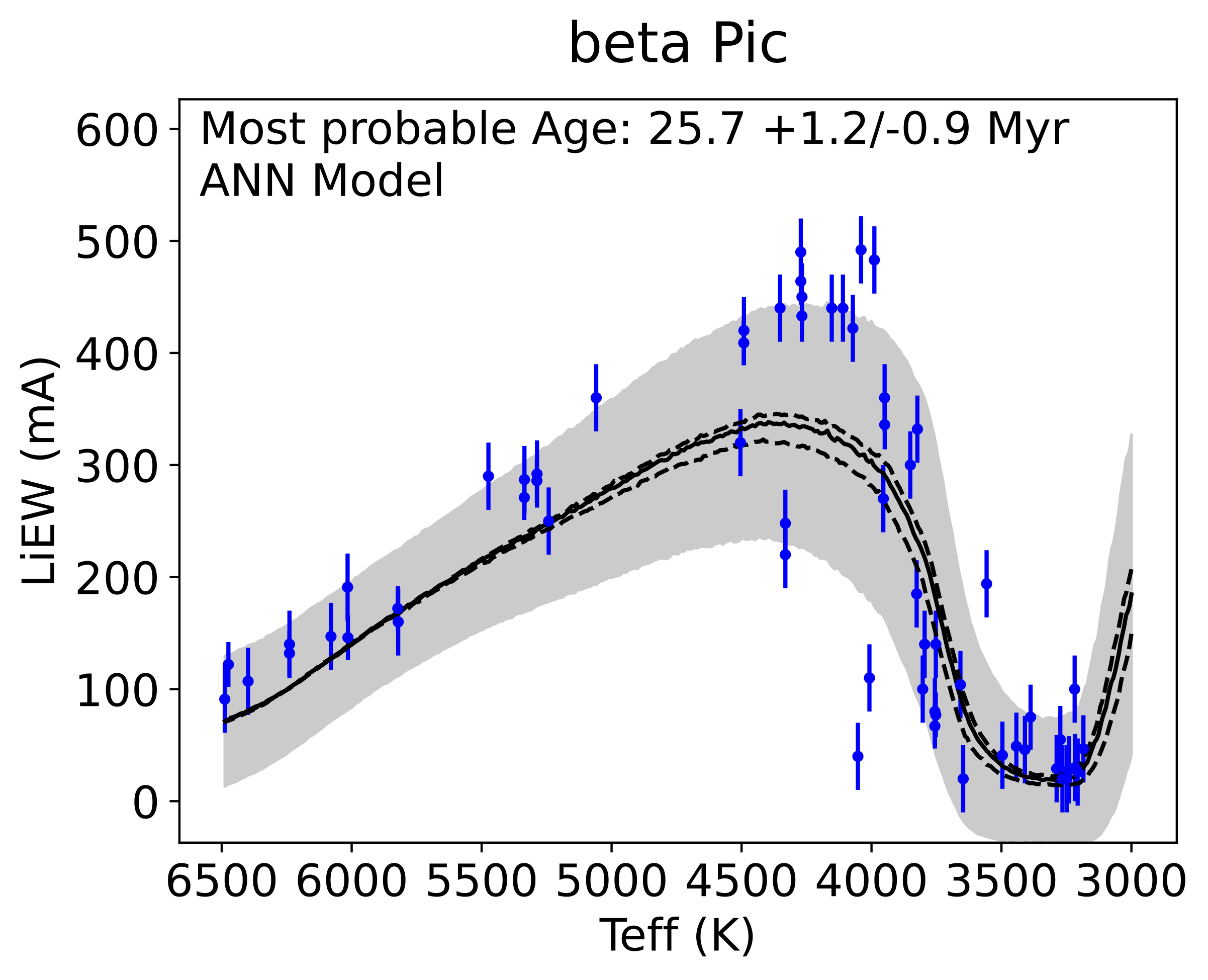}}
    \subfigure[]{\includegraphics[width=0.27\textwidth]{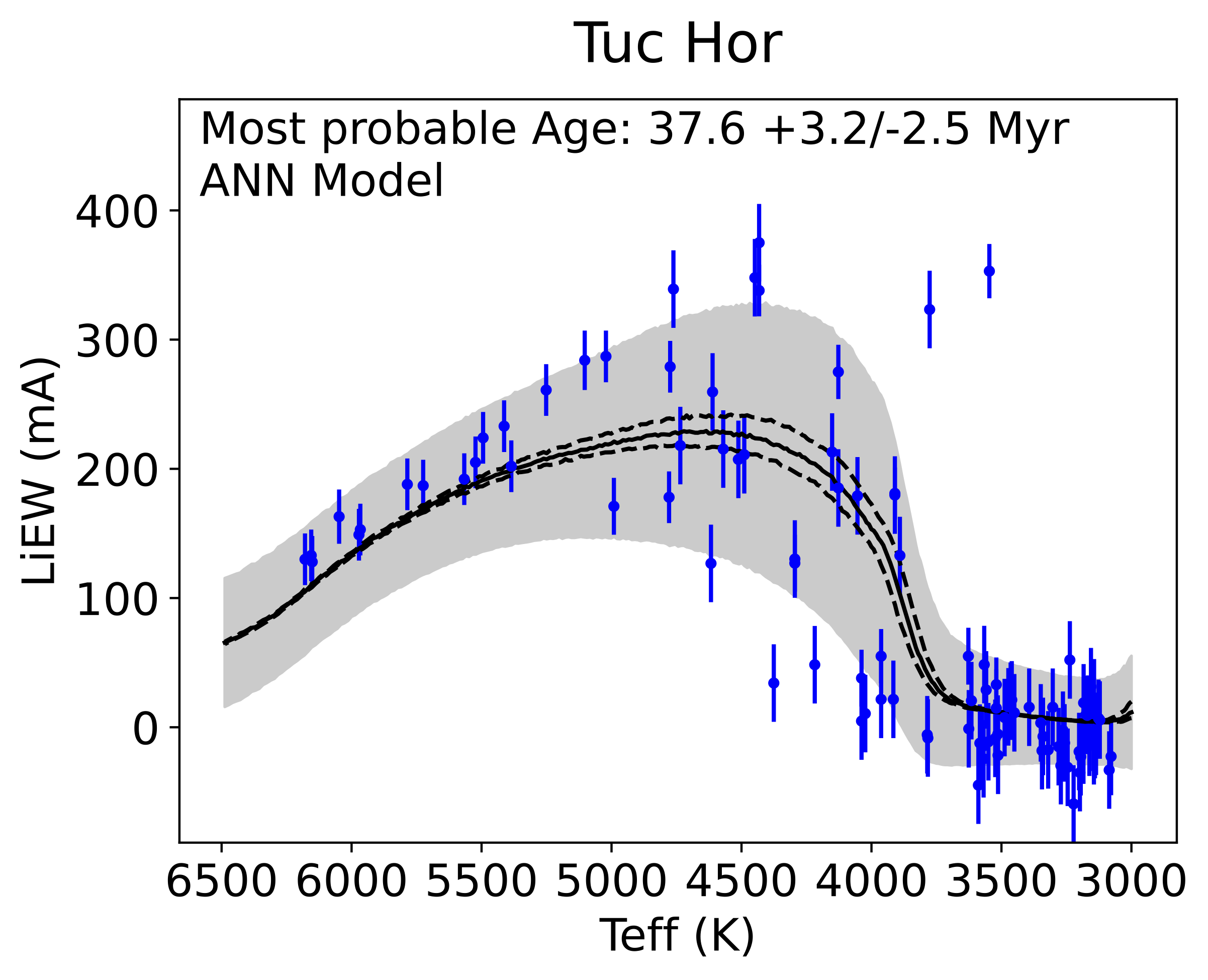}}
    \subfigure[]{\includegraphics[width=0.27\textwidth]{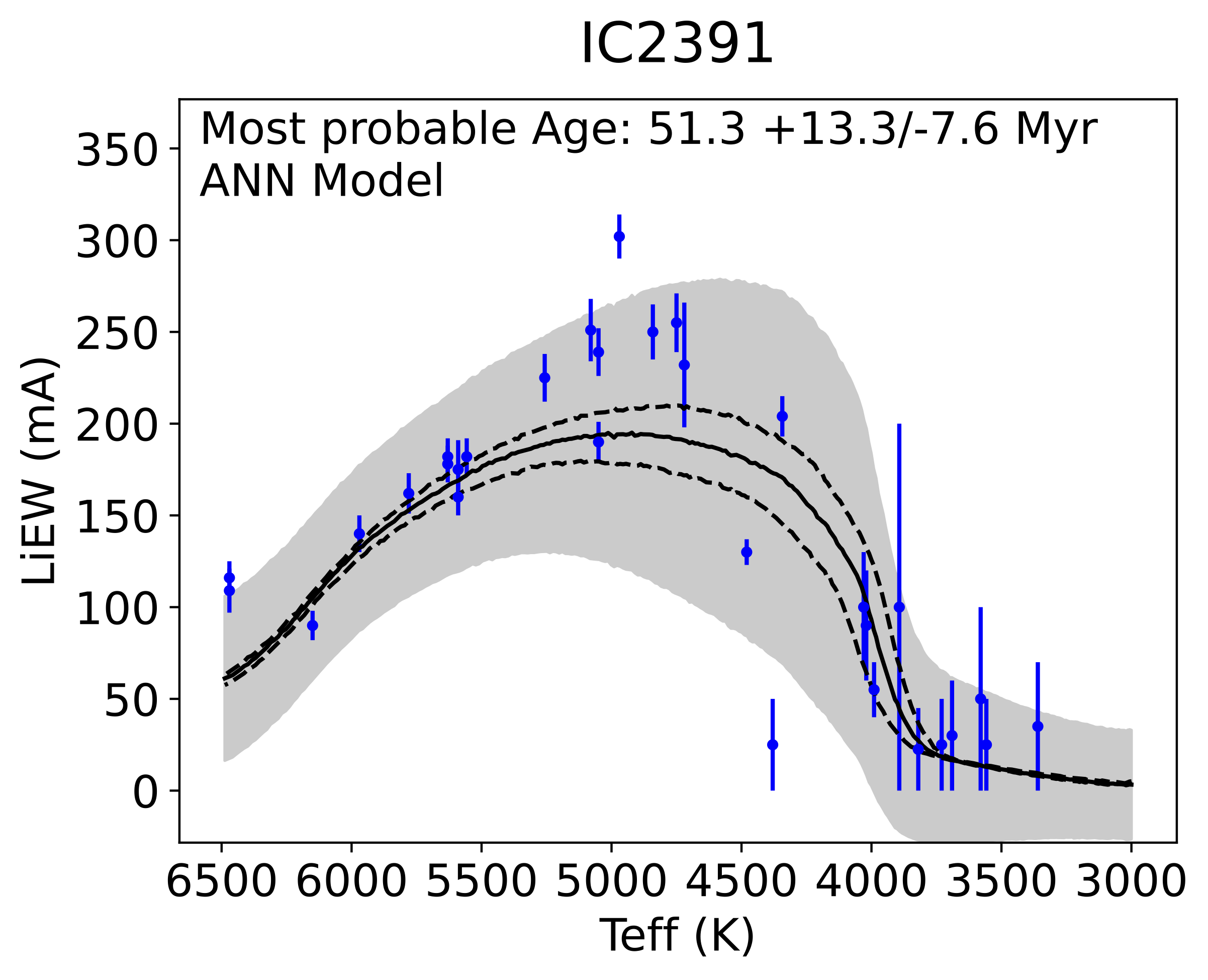}}
    \subfigure[]{\includegraphics[width=0.27\textwidth]{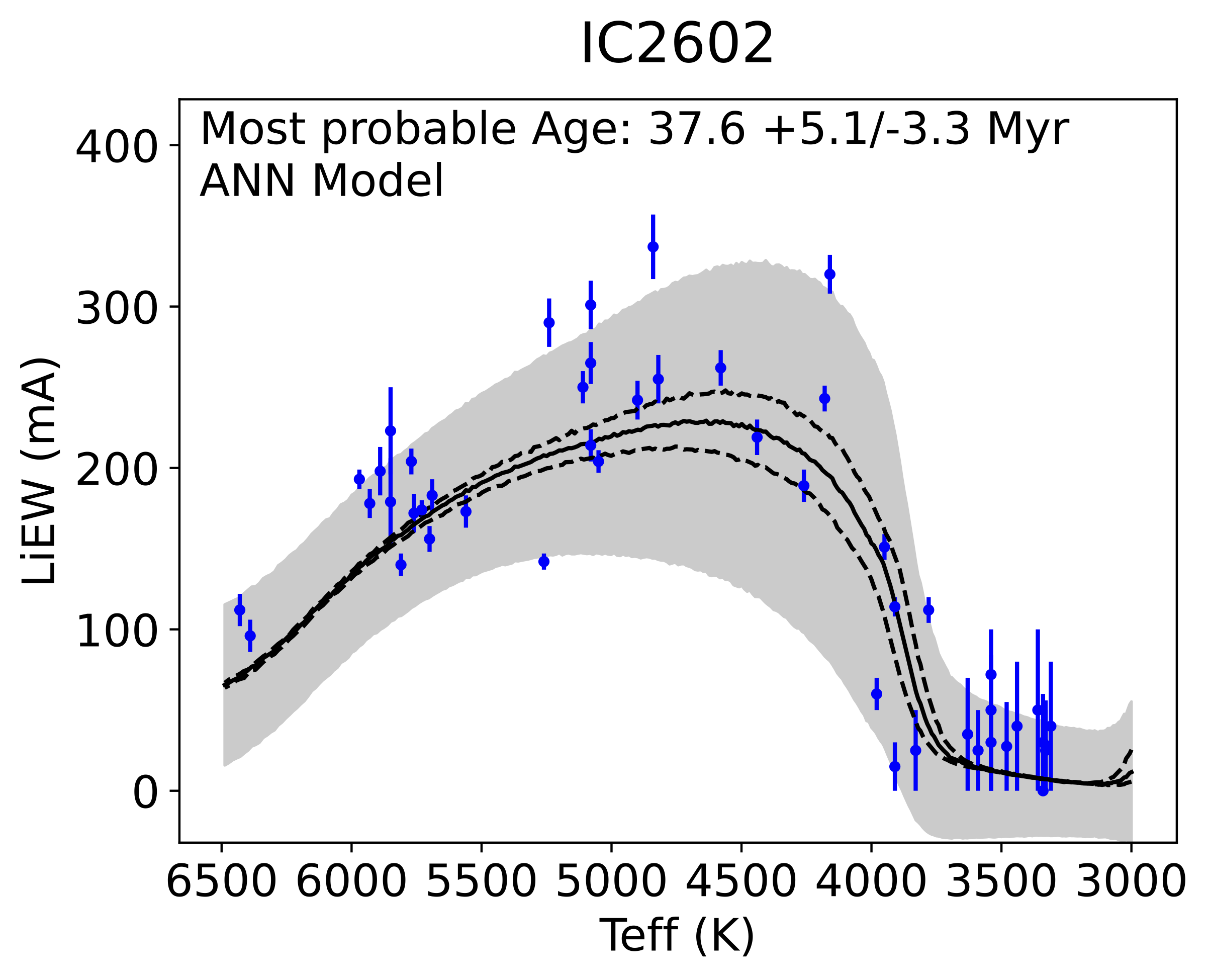}}
    \subfigure[]{\includegraphics[width=0.27\textwidth]{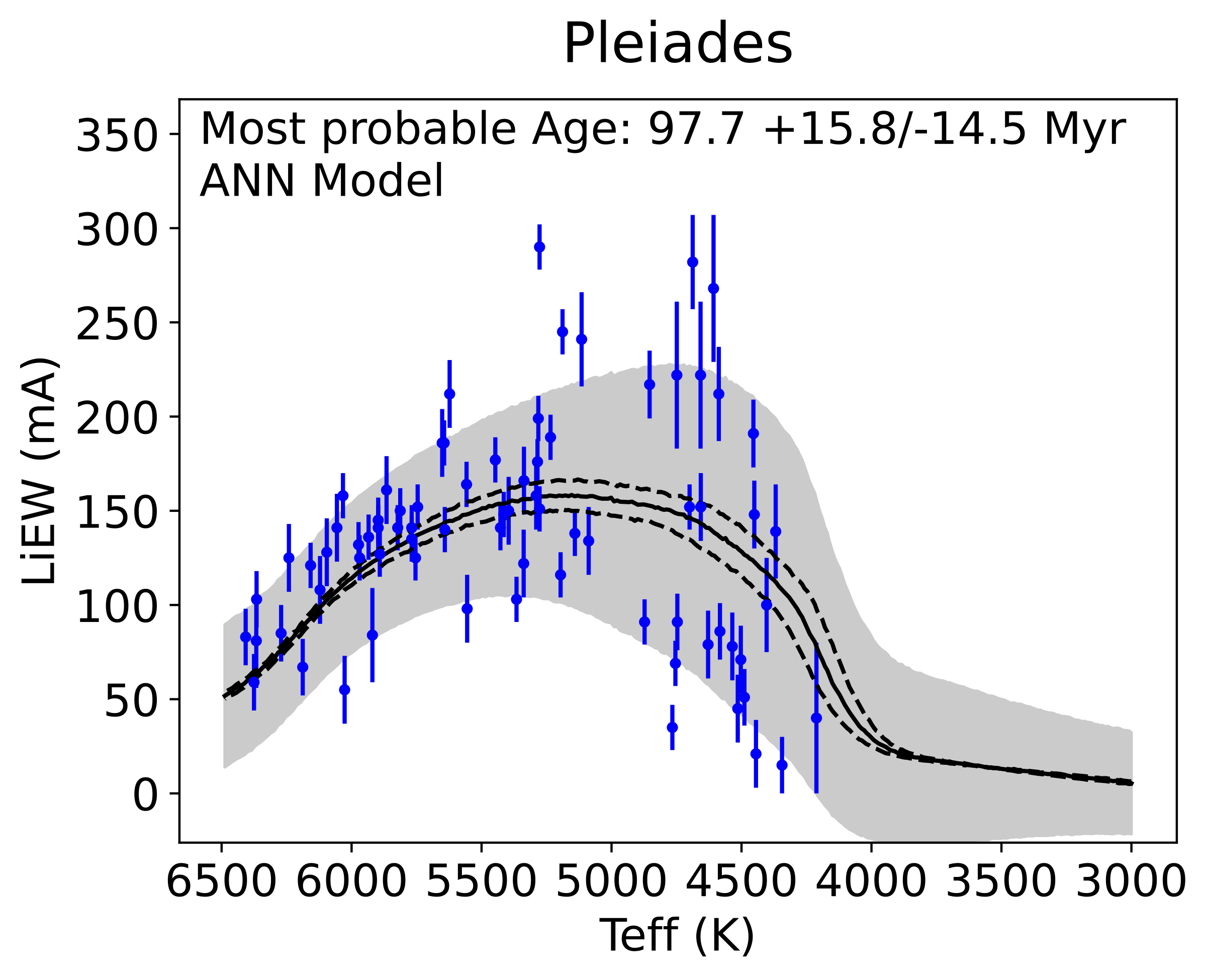}}
    \subfigure[]{\includegraphics[width=0.27\textwidth]{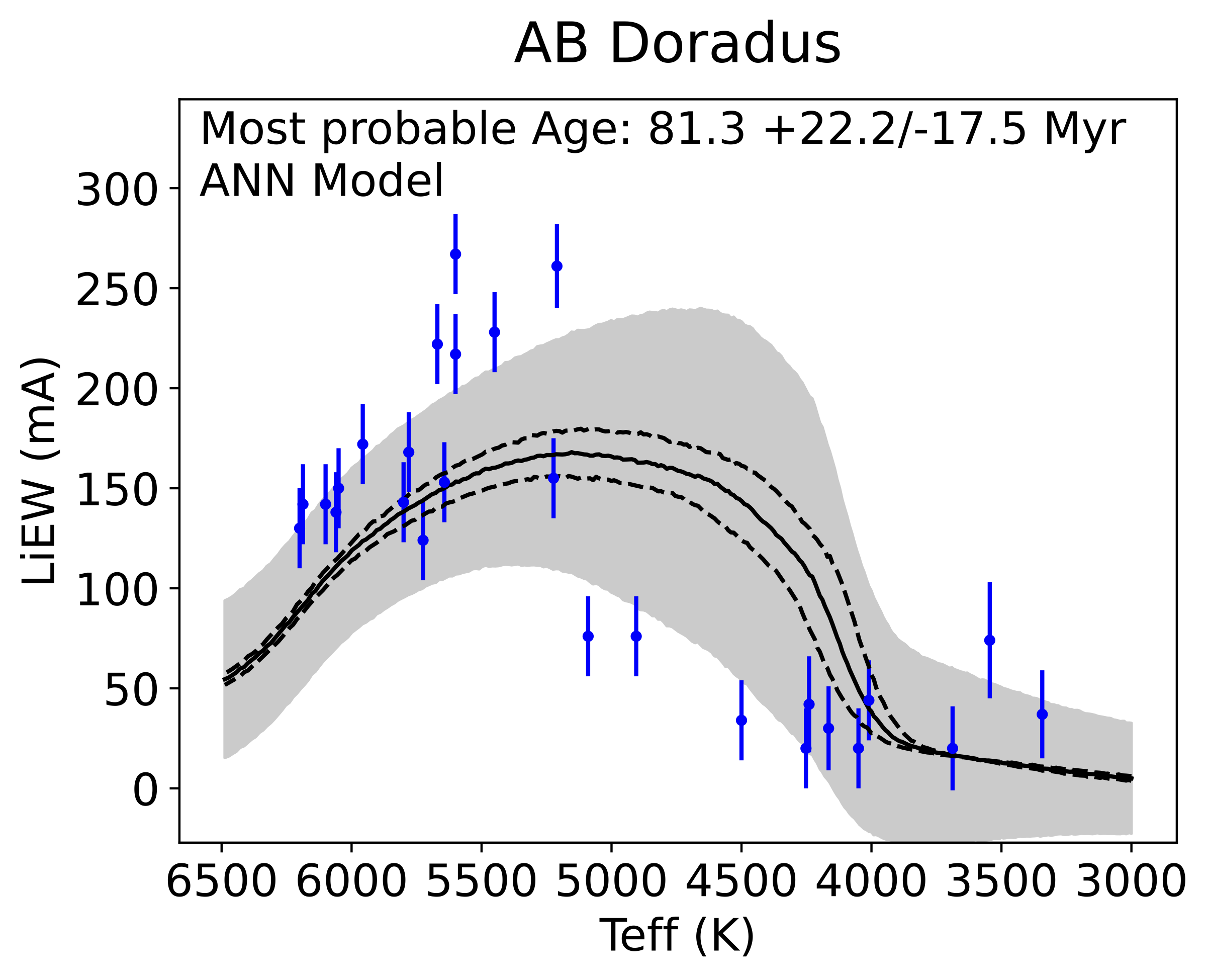}}
    \subfigure[]{\includegraphics[width=0.27\textwidth]{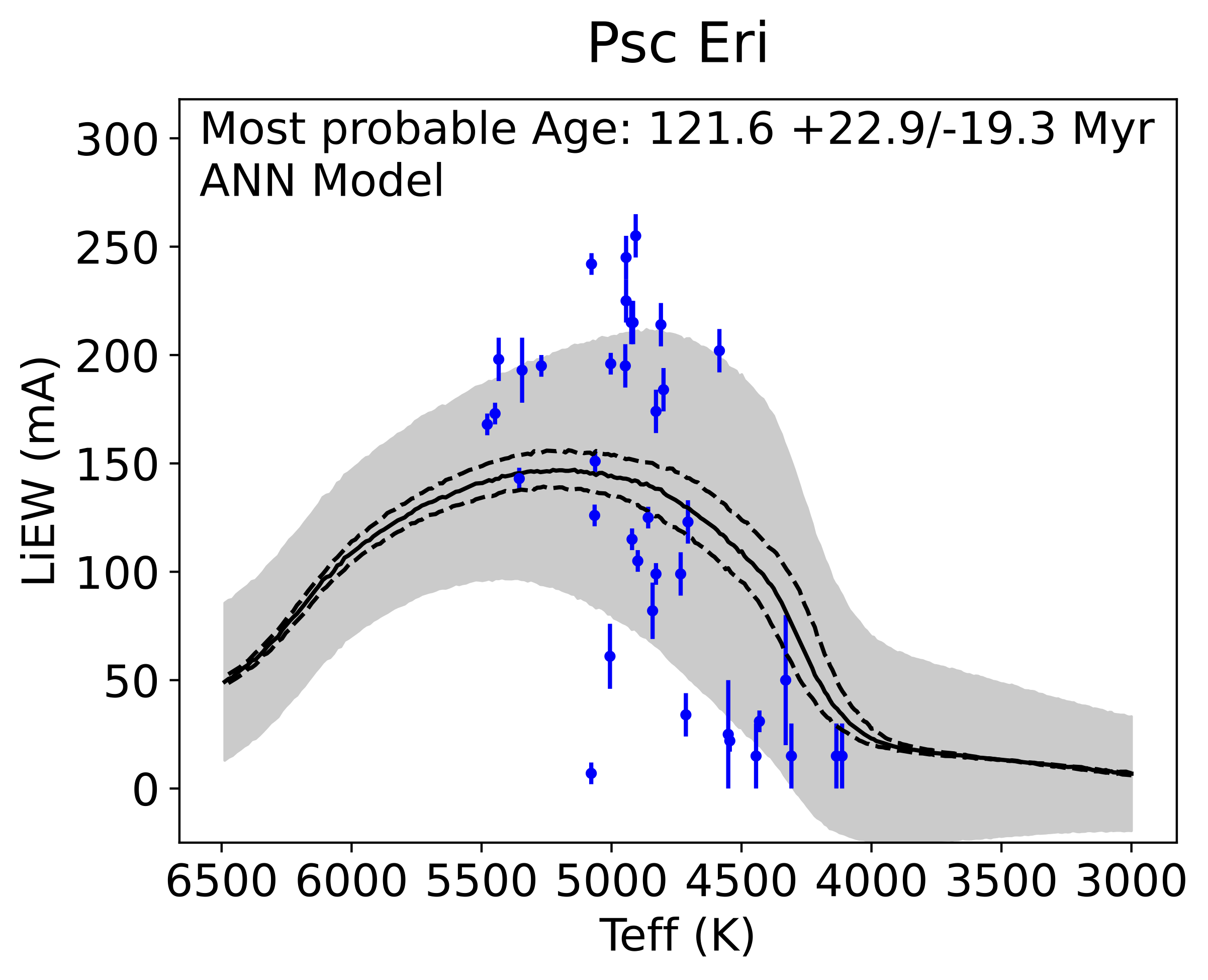}}
    \subfigure[]{\includegraphics[width=0.27\textwidth]{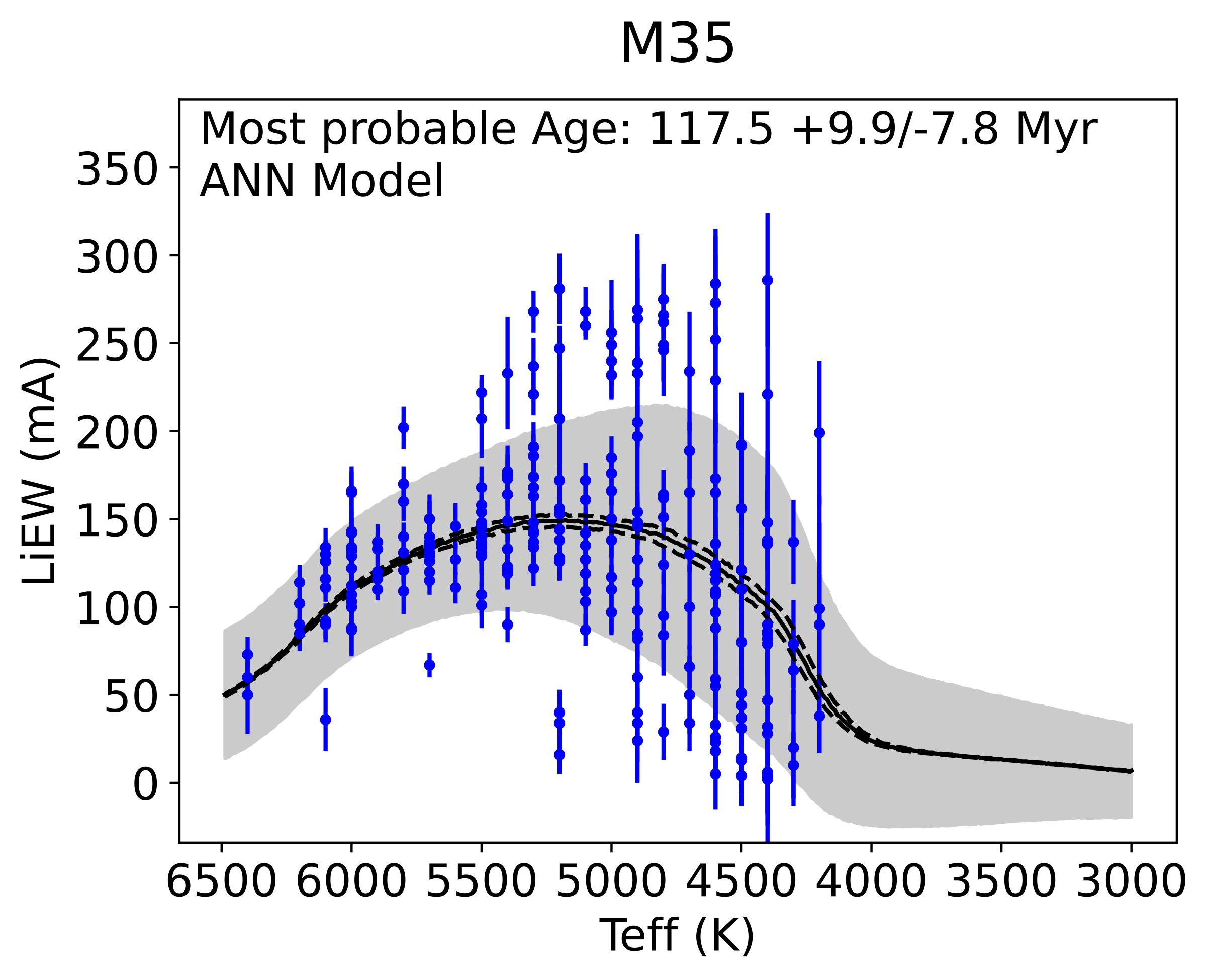}}
    \subfigure[]{\includegraphics[width=0.27\textwidth]{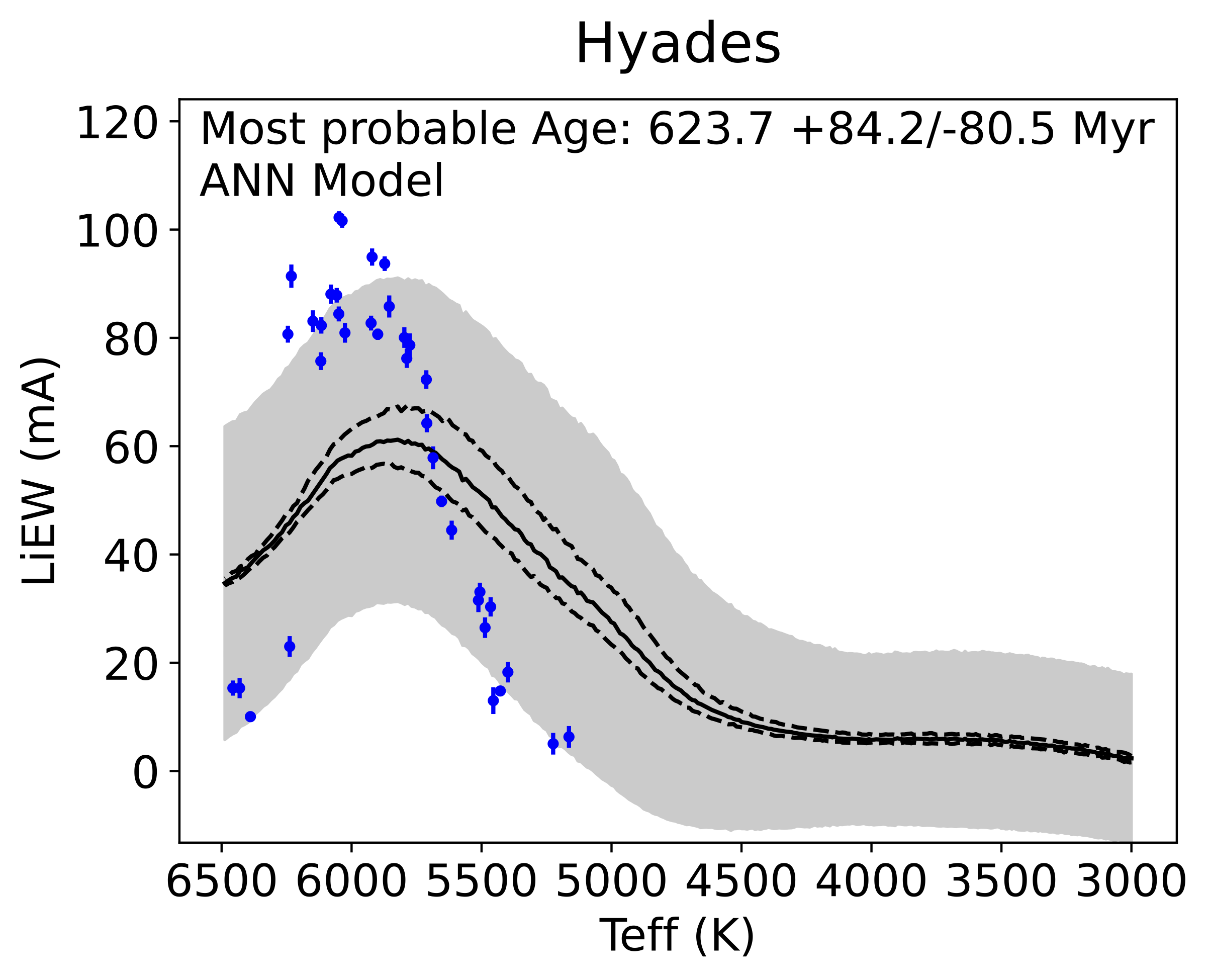}}
    \subfigure[]{\includegraphics[width=0.27\textwidth]{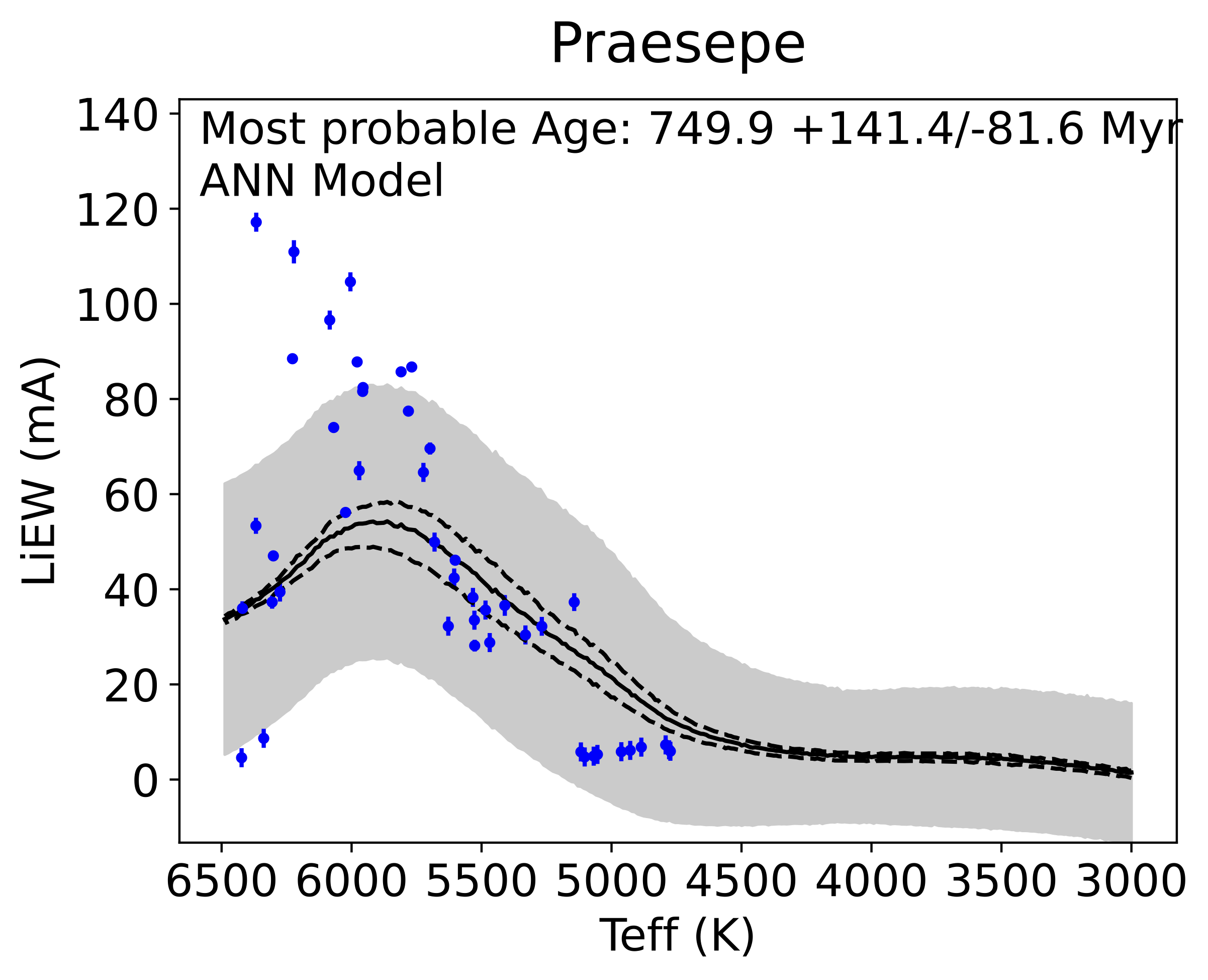}}
    \caption{Fits for 13 non-GES Clusters, moving groups and associations from the ANN models as in Table \ref{tab:nonges}.}
\label{fig:nongesfull}
\end{figure*}

Also included below are the fits for additional testing data not used in training, including non-GES clusters, associations and moving groups (Fig. \ref{fig:nongesfull}).

\begin{table*}
    \centering
    \begin{tabular}{cccccccccccc}
    \hline
    Cluster & Target & Filter & RA & DEC & Age & Probability & $(G_{\rm{BP}} - G_{\rm{RP}})_{0}$ & \Teff & LiEW & eLiEW & $\alpha_{w}$ \\ 
     &  & Centre $\lambda$ & (dec) & (dec) & (Myr) & member & (mag) & (K) & (m\AA) & (m\AA) & \\
    \hline
    25 Ori & 05224842+0140439 & 665.0 & 80.70175 & 1.67886 & 14.6 & 0.998 & 3.013 & 3148 & 633.7 & 38.6 & 1.0330 \\ 

    25 Ori & 05225186+0145132 & 665.0 & 80.71608 & 1.75367 & 14.6 & 1.000 & 3.173 & 3203 & 544.4 & 41.2 & 1.0283 \\ 

    25 Ori & 05225609+0136252 & 665.0 & 80.73371 & 1.60700 & 14.6 & 1.000 & 2.727 & 3333 & 12.0 & 19.5 & 1.0659 \\ 

    25 Ori & 05225678+0147404 & 665.0 & 80.73658 & 1.79456 & 14.6 & 0.992 & 2.821 & 3299 & 296.4 & 26.8 & 1.1467 \\ 

    25 Ori & 05225889+0145437 & 665.0 & 80.74538 & 1.76214 & 14.6 & 1.000 & 2.857 & 3320 & 463.6 & 34.6 & 1.0426 \\
    \hline
    \end{tabular}
    \caption{Training and validation data (see \S\ref{dataset}) used in the ANN model training, and the original calibration of the {\sc eagles} model \citep{Rob2023}. Details are shown for 6200 cluster members observed as part of the Gaia-ESO survey and a further 1503 low-probability members defined as field stars. This data includes the calculated $\alpha_{w}$ value (see \S\ref{youngeststars}). A sample of the table is shown here, with the full version available online as supplementary material.}
    \label{table:data}
\end{table*}

\bsp	
\label{lastpage}
\end{document}